RICE UNIVERSITY

# Thermodynamic Perturbation Theory for Associating Fluids: Beyond First Order

by

**Bennett D. Marshall**

A THESIS SUBMITTED
IN PARTIAL FULLFILLMENT OF THE
REQUIRMENTS FOR THE DEGREE

**Doctor of Philosophy**

APPROVED, THESIS COMMITTEE

Dr. Walter G. Chapman, Chair
William W. Akers Professor
Chemical and Biomolecular Engineering

Dr. Rafael Verduzco
Lewis Owen Assistant Professor
Chemical and Biomolecular Engineering

Dr. Kenneth R. Cox
Professor in the Practice
Chemical and Biomolecular Engineering

Dr. Anatoly Kolomeisky
Professor
Chemistry

HOUSTON, TEXAS
April 2014

# Abstract

Thermodynamic Perturbation Theory for Associating Fluids: Beyond First Order

by

Bennett D. Marshall

Association interactions such as the hydrogen bond are a key component in many physical and biological systems. For this reason accurate theories are needed to describe both the thermodynamics and self – assembly of associating species. Applications of these theories range from those of industrial importance, such as equations of state for process simulations, to the realm of materials science where these theories can be used to predict how molecular structure determines the self – assembly of associating species into advanced supramolecular materials. Semi – empirical equations of state based on chemical or lattice theories do not contain the molecular level detail to make predictions on how molecular structure affects self – assembly of associating species. For this, one needs a theory whose starting point is the interaction potential between two associating species which includes this molecular level detail.

Development of accurate molecular theories for associating fluids is hampered by the strength, directionality and limited valence of the association interaction. This has proven particularly true in the extension of Mayer's cluster theory to these associating fluids. This problem was largely solved by Wertheim in the 1980's who developed an exact cluster expansion using a multi – density formalism. Wertheim's cluster theory incorporates the geometry of the association interaction at an early point in the derivation. This allowed Wertheim to develop the theory in such a way that accurate and simple approximation methods could be applied such as thermodynamic perturbation theory (TPT). When treated at first order in perturbation (TPT1), Wertheim's theory gives a simple and general equation of state which forms the basis of the statistical associating fluid theory (SAFT) free energy model. SAFT has been become a standard in both academia and industry as an equation of state for associating (hydrogen bonding) fluids. While simple in form and widely applied, the development of TPT1 rest on a number of, sometimes severe, simplifying assumptions: no interaction between associated clusters beyond that of the reference fluid, association sites are singly bondable, no double bonding of molecules, no cycles of association bonds, no steric hindrance between association sites, association is independent of bond angle and there is no bond cooperativity. The purpose of this dissertation is to relax these assumptions.

Chapters 2 – 4 extend TPT to allow for multiple bonds per association site. Chapter 2 focuses on the case of associating spheres with a doubly bondable association site as a model for patchy colloids with a multiply bondable patch. Chapters 3 – 4 extend TPT to associating mixtures of spheres where the first component has directional association sites and the second component has spherically symmetric association sites. This theory is applicable as a model for mixtures of patchy and spherically symmetric colloids and ion – water association.

Chapters 5 – 6 extend TPT to account for the effect of relative association site location. In chapter 5 the case of associating hard spheres with two association sites is considered. For this case the angle between the centers of the association sites (bond angle) is treated as a independent variable. This is the first application of TPT which has included the effect of bond angle. It is shown that as bond angle is decreased the effects of steric hindrance, ring (cycle) formation and eventually double bonding of molecules must be accounted for. The developed theory accounts for each of these higher order interactions as a function of bond angle. The resulting theory is shown to be accurate over the full bond angle range for both the distribution of cluster types (chains, rings, double bonded) as well as the equation of state. In chapter 6 this theory is extended to the case there are more than two association sites.

In chapter 7 TPT is extended to account for hydrogen bond cooperativity for the case of molecules with two association sites. The derived theory is shown to be highly accurate for molecules which exhibit positive or negative hydrogen bond cooperativity. Finally, in chapter 8 the case of associating fluids in spatially uniform orienting external fields is considered. An example system for this case would be associating dipolar molecules in a uniform electric field. By employing classical density functional theory in the canonical ensemble exact results are obtained for the orientational distribution function. These exact results contain the monomer fraction which must be approximated in TPT1. The resulting theory is in good agreement with simulation data for the prediction of the effect of a linear orienting field on the chain length and orientation of associated chains of spheres with two association sites.

# Acknowledgments

First, I would like to thank my advisor Dr. Walter Chapman for his guidance, in both research and life, throughout my time at Rice. I am grateful to Dr. Kenneth Cox for his guidance and constant reminders to not enter the theoretical abyss. I thank all of my group members for their support throughout my research endeavors; especially Deepti Ballal and Dr. Kai Gong with whom I shared an office and spent many hours discussing complex research problems.

I thank Dr. Virginia Davis for mentoring me when I was an undergraduate at Auburn University. After all, it was at her suggestion that I applied to the graduate program at Rice. I would also like to thank Dr. Margarida Telo da Gama for supporting me during my stay in Lisbon, Portugal.  I am grateful to Dr. Rafael Verduzco and Dr. Anatoly Kolomeisky for being on my thesis committee.

Most importantly, I would like to thank my family, whose unwavering support and steadfast dedication have given me the confidence and ability to pursue my goals. I especially thank my wife Karen and son Benny who are the source of my happiness. I dedicate this thesis to you.

# Contents

# CHAPTER 1

# Introduction

This dissertation develops new theoretical approaches to describe the self – assembly and equation of state for associating fluids. This chapter introduces the background and terminology needed for the remainder of the dissertation. After a brief introduction to associating fluids, some existing theoretical methods are reviewed. In this work we focus on theories of association which are derived through approximations of exact cluster expansions. An extensive background on cluster expansions is not required; section 1.2 gives an overview of the necessary features of these expansions. After an introduction to cluster expansions, general theories for associating fluids based on cluster expansions are reviewed. We obtain approximate solutions to these exact expansions in perturbation theory after a number of simplifying assumptions. At the end of the chapter, we outline the assumptions made in the development of these theories, and the remainder of the dissertation is focused on relaxing these assumptions.

# 1.1: What is an associating fluid?

In this dissertation an associating fluid is meant to describe a fluid which is composed of molecules which interact with association interactions. Here an association interaction is an attractive interaction between two molecules which is both short ranged and highly directional, resulting in saturation of the interaction. In general these molecules will interact with other repulsive and attractive forces such as steric repulsions and Van der Waals attractions. A common example of an association interaction is the hydrogen bond. Hydrogen bonding is responsible for the remarkable properties of water[1], folding of proteins[2] and is commonly exploited in the self – assembly[3] of advanced materials.

Patchy colloids give another example of using association interactions to guide self – assembly. Patchy colloids are colloids with some number of attractive surface patches giving rise to association like anisotropic inter – colloid potentials. [4] These "patchy" colloids have been synthesized by glancing angle deposition[5, 6], the polymer swelling method[7] and by stamping the colloids with patches of single stranded DNA.[8] The form of the anisotropic pair potentials, and resulting fluid behavior, can be manipulated by varying the number, size, shape, interaction range and relative location of attractive patches on the surface of the colloid. Through the tailored design of these patch parameters these colloids can be programmed to self – assemble into predetermined structures such as colloidal molecules[7], the Kagome lattice[9] and diamond structures for photonic applications[10]; as well as form fluid phases such as gels[11], empty liquids[12] and fluids which exhibit reentrant phase behavior[13].

Given the importance of associating fluids, accurate theories are needed to predict their behavior. For instance, simulations of industrial processes require an equation of state which can accurately correlate experimental data for hydrogen bonding fluids and use fitted pure component parameters to predict the properties of mixtures. From a materials science perspective, accurate theories for association are needed to describe association mediated self – assembly of supramolecular materials. The reversible and ordered nature of these materials allows for the bottoms up design of functional materials with applications in microelectronics[14], photonics[15], self – healing materials[16], solar panels[17] etc... With the rapid advance in our ability to synthesize these building blocks to ever increasing complexity, one of the largest challenges we face is determining what to synthesize. If a certain self – assembled structure is required, which must have a predetermined set of properties, how do we design molecules / colloids *a priori* which perform the required task? To sort through the large parameter space, accurate theories for association are needed.

The first models used to describe hydrogen bonding fluids were developed using a chemical approach, where each associated cluster is treated as a distinct species created from the reaction of monomers and smaller associated clusters. [18, 19] The "reactions" are governed by equilibrium constants which must be obtained empirically. This type of approach has been incorporated into a Van der Waals type equation of state[20], the perturbed anisotropic chain theory equation of state (APACT)[21] and the Sanchez – Lacombe[22] equation of state.

Alternatively, lattice theories may be employed to model hydrogen bonding fluids. These approaches generally follow the method of Veytsman[23] who showed how the free energy contribution due to hydrogen bonding could be calculated in the mean field by enumerating the number of hydrogen bonding states on a lattice. Veytsman's approach was incorporated into the Sanchez – Lacombe equation of state by Panayiotou and Sanchez[24] who factored the partition function into a hydrogen bonding contribution and a non – hydrogen bonding contribution. The lattice approach has also been applied to hydrogen bond cooperativity[25] and intramolecular[26] hydrogen bonds.

Both the chemical and lattice theory approaches to hydrogen bonding yield semi - empirical equations of state which are useful for several hydrogen bonding systems. [18] The drawback of these approaches is a result of their simplistic development. As discussed above, we wish to predict the self – assembly behavior of associating systems by considering the molecular details of the associating species. This cannot be accomplished using a lattice or chemical theory. For this we must incorporate these molecular details from the outset. The starting place for any molecular theory of association is the definition of the pair potential energy $\phi(12)$ between molecules (or colloids). Molecules are treated as rigid bodies with no internal degrees of freedom. In total 6 degrees of freedom describe any single molecule, three translational coordinates represented by the vector $\vec{r}_1$ and three orientation angles represented by $\Omega_1$. These six degrees of freedom are represented as $1=\{\vec{r}_1,\Omega_1\}$. It is assumed that the intermolecular potential can be separated as

$$\phi(12)=\phi_{ref}(12)+\phi_{as}(12) \tag{1.1}$$

Where $\phi_{as}(12)$ contains the association portion of the potential and $\phi_{ref}(12)$ is the reference system potential, which contains all other contributions of the pair potential including a harsh short ranged repulsive contribution.

Considering molecules (or colloids) which have a set of association sites $\Gamma=\{A,B,C,\cdots,Z\}$, where association sites are represented by capital letters, the association potential is decomposed into individual site – site contributions

$$\phi_{as}(12) = \sum_{A\in\Gamma}\sum_{B\in\Gamma}\phi_{AB}(12) \tag{1.2}$$

The potential $\phi_{AB}(12)$ represents the association interaction between site $A$ on molecule 1 and site $B$ on molecule 2. One of the challenges in developing theoretical models for associating fluids stems from the short ranged and directional nature of the association potential $\phi_{AB}$, which results in the phenomena of bond saturation. For instance, considering molecules which consist of a hard spherical core of diameter $d$

$$\phi_{ref}(12) = \phi_{HS}(r_{12}) = \begin{cases} \infty & r < d \\ 0 & r \geq d \end{cases} \tag{1.3}$$

and a single association site $A$ (see Fig. 1.1), bond saturation arises as follows. When spheres 1 and 2 are positioned and oriented correctly such that an association bond is formed, the hard cores of these two spheres may, depending on the size and range of the association site, prevent sphere 3 from approaching and sharing in the association interaction. That is, if $\phi_{as}(12) < 0$ and $\phi_{as}(13) < 0$ then $\phi_{HS}(r_{23}) = \infty$, meaning that each association site is singly bondable (has a valence of 1). In hydrogen bonding it is usually the case that each association bond site is singly bondable, although there are exceptions. For the case of patchy colloids, the patch size can be controlled to yield a defined valence controlling the type of self-assembled structures which form.

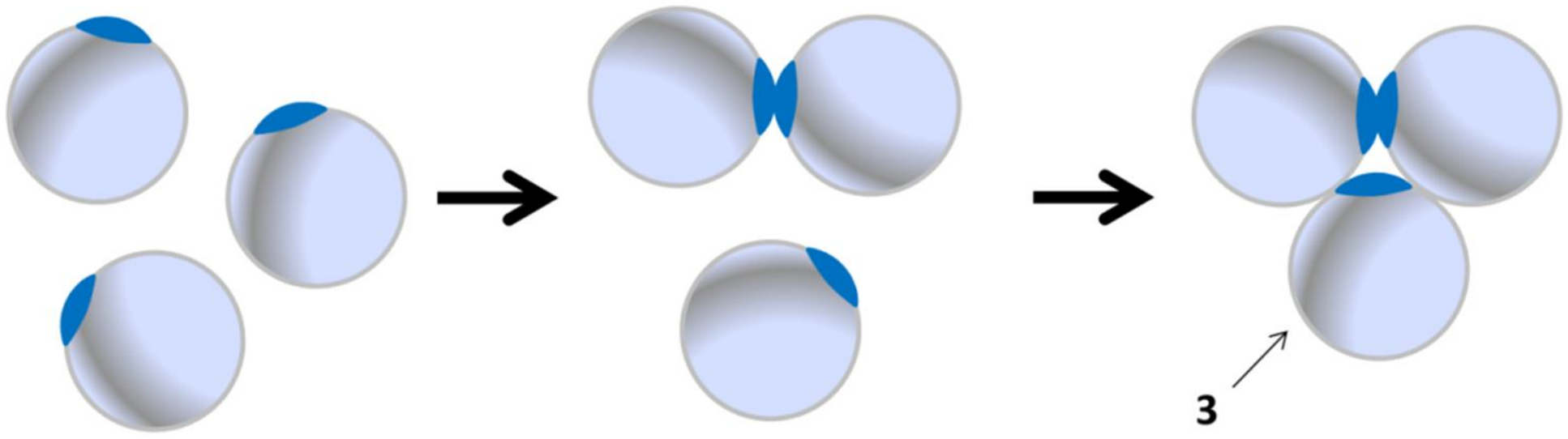

**Figure 1.1:** Illustration of bond saturation for hard spheres with a single association site of valence 1

The development of accurate theories to model potentials of type Eqns. (1.1) – (1.2) is complicated by the limited valence of the interactions as well as the strength of the association bonds. In the transitioning between microscopic pair potential to macroscopic theory these basic features of the association interaction must be retained. In the following sections we review some of the existing theories to model associating fluids with potentials of the form of Eqns. (1.1) – (1.2), but first a brief introduction to cluster expansions is required.

## 1.2: A brief introduction to cluster expansions

In this section we give a very brief overview of cluster expansions. For a more detailed introduction the reader is referred to the original work of Morita and Hiroike[27] and the reviews by Stell[28] and Andersen[29]. Cluster expansions were first introduced by Mayer[30] as a means to describe the structure and thermodynamics of non – ideal gases. In cluster expansions Mayer $f$ functions are introduced

$$f(12) = \exp(-\phi(12)/k_B T) - 1 \tag{1.4}$$

The replacement $\exp(-\phi(12)/k_B T) = f(12) + 1$ in the grand partition function and the application of the lemmas developed by Morita and Hiroike[27] allows for the pair correlation function $g(12)$ and Helmholtz free energy $A$ to be written as an infinite series in density where each contribution is an integral represented pictorially by a graph. A graph is a collection of black circles and white circles with bonds connecting some of these circles. The bonds are represented by two molecule functions such as Mayer functions $f(12)$ and the black circles are called field points represented by single molecule functions such as fugavity $z(1)$ or density $\rho(1)$ integrated over the coordinates (1). The white circles are called root points and are not associated with a single molecule function, and the coordinates of a root point are not integrated. Root points are given labels 1, 2, 3 etc… The value of the diagram is then obtained by integrating over all coordinates associated with field points and multiplying this integral by the inverse of the symmetry number $S$ of the graph. Figure1.2 gives several examples. The volume element $d(1)$ is given by $d(1) = d\vec{r}_1 d\Omega_1$ .

**A)** 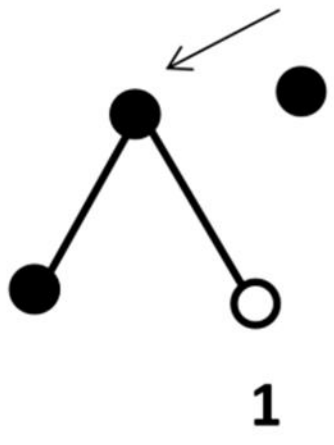

$$= \int \rho(2)\rho(3)\rho(4) f(12) f(23) d(2) d(3) d(4)$$

**B)** 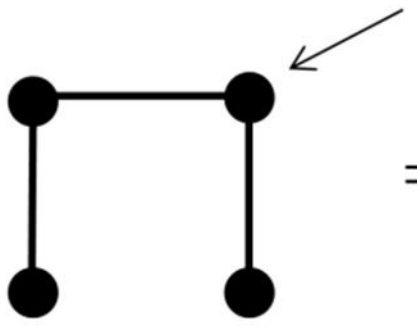

$$= \frac{1}{2}\int \rho(1)\rho(2)\rho(3)\rho(4) f(12) f(23) f(34) d(1) d(2) d(3) d(4)$$

**C)** 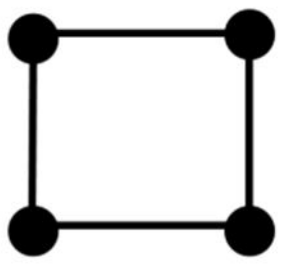

$$= \frac{1}{8}\int \rho(1)\rho(2)\rho(3)\rho(4) f(12) f(23) f(34) f(14) d(1) d(2) d(3) d(4)$$

**Figure 1.2:** Examples of integral representations of graphs. Arrows point towards articulation circles

Before giving graphical representations of the pair correlation function $g(12)$ and the Helmholtz free energy $A$, a few definitions must be given

1) A graph is connected if there is at least one path between any two points. Graph **A** in Fig. 1.2 is disconnected and graphs **B** and **C** are both connected.

2) An articulation circle is a circle in a connected graph whose removal makes the graph disconnected, where at least one part contains no root point and at least one field point Arrows in Fig. 1.2 point to articulation circles.

3) An irreducible graph has no articulation circles. Graph **C** in Fig. 1.2 is an example of an irreducible graph.

Using these definitions, the pair correlation function and Helmholtz free energy are given as

$$g(12)=\left\{\begin{array}{l}\text{sum of topologically different irreducible graphs that have two root points}\\ \text{labelled 1 and 2, any number of field points } \rho \text{, and at most one } f \text{ bond}\\ \text{between each pair of points}\end{array}\right\} \tag{1.5}$$

and

$$\frac{A}{k_bT}=\int \rho(1)\left(\ln\left(\rho(1)\Lambda^3\right)-1\right)d(1)-c^{(o)} \tag{1.6}$$

Where $\Lambda$ is the de Broglie wavelength, $\rho(1)$ is the density where $\rho(\vec{r})=\int \rho(1)d\Omega$, and $c^{(o)}$ is the graph sum given by

$$c^{(o)}=\left\{\begin{array}{l}\text{sum of topologically different irreducible graphs that have no root points}\\ \text{any number of field points } \rho \text{, and at most one } f \text{ bond between each pair}\\ \text{of points}\end{array}\right\} \tag{1.7}$$

Equations (1.5) – (1.7) are rigorous and exact mathematical statements. Unfortunately, the exact evaluation of these infinite sums cannot be performed and numerous approximations must be made to obtain any usable result. In these approximations only some subset of the original graph sum is evaluated.

Performing these partial summations in hydrogen bonding fluids is complicated by both the strength of the association interaction and the limited valence of the interaction. Hydrogen bond strengths can be many times that of typical Van der Waals forces giving Mayer functions which are very large. If the entire cluster series were evaluated for $g(12)$ and $c^{(o)}$ these divergences would cancel; however, when performing partial summations, care must be taken to eliminate divergences if meaningful results are to be obtained. Similarly, in most hydrogen bonding fluids, each hydrogen bonding group is singly bondable. Hence, any theory for hydrogen bonding fluids must account for the limited valence of the attractions. Again, if the full cluster series were evaluated for $g(12)$ this condition would be naturally accounted for; however, when performing partial summations care must be taken to ensure this single bonding condition holds. There have been two general methods to handle these strong association interactions using cluster expansions. The first was the pioneering work of Andersen[31, 32] who developed a cluster expansion for associating fluids in which the divergence was tamed by the introduction of renormalized bonds, and the second is the method of Wertheim[33-37] who used multiple densities.

## 1.3: Single association site: Bond renormalization

Before discussing the more general case of associating fluids with multiple association sites, we will discuss the simpler case of molecules with a single association site *A*. For a single association site the Mayer function is decomposed as

$$f(12) = f_{ref}(12) + F_{AA}(12) \tag{1.8}$$

where

$$\begin{aligned} F_{AA}(12) &= e_{ref}(12) f_{AA}(12) \\ e_{ref}(12) &= \exp(-\phi_{ref}(12)/k_B T) = 1 + f_{ref}(12) \\ f_{AA}(12) &= \exp(-\phi_{AA}(12)/k_B T) - 1 \end{aligned} \tag{1.9}$$

In Eq. (1.9) the $f_{AA}(12)$ accounts or the anisotropic / short ranged attraction of the association interaction and the function $e_{ref}(12)$ prevents the overlap of the cores of the molecules. It is the functions $e_{ref}(12)$ which give rise to the single bonding condition. Now inserting Eq. (1.8) into Eq. (1.5) and simplifying

$$g(12) = \left\{ \begin{array}{l} \text{sum of topologically different irreducible graphs that have two root points} \\ \text{labelled 1 and 2, any number of field points } \rho,\ f_{ref} \text{ and } F_{AA} \text{ bonds, and at most} \\ \text{one bond between each pair of points} \end{array} \right\} \tag{1.10}$$

Andersen[31, 32] defines a renormalized association Mayer function $\tilde{F}_{AA}(12)$ as the sum of the graphs in Eq. (1.10) which are most important in the determination of $g(12)$. Since the Mayer functions $F_{AA}$ may take on very large numerical values in the bonding region, the most important graphs in the calculation of $g(12)$ are the ones whose root points are connected by an $F_{AA}$ bond. Hence, it is natural to define $\tilde{F}_{AA}$ as

$$\tilde{F}_{AA}(12) = \{\text{sum of graphs in (1.10) whose root points are connected by an } F_{AA} \text{ bond}\} \tag{1.11}$$

Andersen assumes that the intermolecular potential was such that the association site was singly bondable. This single bonding condition was exploited in the cluster expansion by use of the *cancelation theorem* as described by Andersen, who was able to sum the diagrams in Eq. (1.11) as

$$\tilde{F}_{AA}(12) = F_{AA}(12) Y_p(12) \frac{1 + 2\Delta_{AA} - \sqrt{1 + 4\Delta_{AA}}}{2\Delta_{AA}^2} \tag{1.12}$$

Where the term $\Delta_{CD}$ is given by (where for a homogeneous fluid $\rho = \int \rho(1) d\Omega_1$)

$$\Delta_{CD} = \frac{\rho}{\Omega} \int Y_p(12) F_{CD}(12) d(2) \tag{1.13}$$

and $\Omega = \int d\Omega'$ is the total number of orientations. The function $Y_p(12)$ is given by

$$Y_p(12) = \left\{ \begin{array}{l} \text{sum of graphs in (1.10) which have no } F_{AA} \text{ bond attached to either} \\ \text{root, and no bond between the roots} \end{array} \right\} \quad (1.14)$$

It is easily shown that $\tilde{F}_{AA}(12)$ is bounded as

$$0 \leq \frac{\rho}{\Omega} \int \tilde{F}_{AA}(12) d(2) \leq 1 \quad (1.15)$$

Equation (1.15) shows that the renormalized association bond remains finite even when the association potential $\phi_{AA}$ takes on infinitely large negative values. Using this renormalized bond the average number of hydrogen bonds per molecule is calculated as

$$N_{HB} = \frac{\rho}{\Omega} \int \tilde{F}_{AA}(12) d(2) \quad (1.16)$$

Comparing Eqns. (1.15) and (1.16) it is easy to see

$$0 \leq N_{HB} \leq 1 \quad (1.17)$$

Equation (1.17) demonstrates that the single bonding condition is satisfied and that the method of Andersen was successful. Unfortunately, the function $Y_p(12)$ must be obtained through the solution of a series of integral equations using approximate closures.

To the author's knowledge, this approach has never been applied for numerical calculations of the structure or thermodynamics of one site associating fluids. Here we will show how a single simple approximation allows for the calculation of $N_{HB}$. To approximate $Y_p(12)$ we note that this function can be decomposed into contributions from graphs which contain $k$ association bonds $F_{AA}$

$$Y_p(12) = \sum_{k=0}^{\infty} Y_p^{(k)}(12) \quad (1.18)$$

The terms $Y_p^{(k)}$ give the contribution to $Y_p$ from graphs which contain *k* association bonds. The simplest possible case is to keep only the first contribution *k* = 0 and disregard all $Y_p^{(k)}$ for *k* > 0. For this simple case

$$Y_p(12) = y_{ref}(12) \tag{1.19}$$

where $y_{ref}$ is the cavity correlation function of the reference fluid; meaning association is treated as a perturbation to the reference fluid. Combining Eqns. (1.12), (1.13), (1.16) and (1.19) the monomer fraction (fraction of molecules which do not have an association bond) can be written as

$$X_o = 1 - N_{NB} = \frac{-1 + \sqrt{1 + 4\Delta_{AA}}}{2\Delta_{AA}} \tag{1.20}$$

where $\Delta_{AA}$ is now given by $\Omega\Delta_{AA} = \rho \int y_{ref}(12) F_{AA}(12) d(2)$. Equation (1.20) gives a very simple relationship for the monomer fraction. This same equation became famous, one decade later, after it was derived by Wertheim using a very different cluster expansion. Equation (1.20) has been shown to be highly accurate in comparison to simulation data.[38] Now we will introduce Wertheim's two density formalism for 1 site associating fluids.

## 1.4: Single association site: Two density approach

In the previous section it was shown that Andersen's formalism can be applied to derive a highly accurate and simple relationship for the monomer fraction. In order to obtain this result the renormalized association Mayer functions $\tilde{F}_{AA}$ were employed. The applicability of Andersen's approach to more complex systems (mixtures, multiple bonds per association site etc…) is limited by the fact that for each case the renormalized Mayer functions must be obtained by solving a rather complex combinatorial problem. A more natural formalism for describing association interactions in one site associating fluids is the two density formalism of Wertheim.[33, 34]

Instead of using the density expansion of the pair correlation function $g(12)$ or Helmholtz free energy *A*, Wertheim uses the fugacity expansion of $\ln\Xi$, where $\Xi$ is the grand partition function, as the starting point. Building on the ideas of Lockett[39], Wertheim then regroups the expansion such that individual graphs are composed of *s* – mer graphs. An *s* – mer represents a cluster of points which are connected by paths of $F_{AA}$ bonds, each pair of points in an *s* – mer, which are not directly connected by a $F_{AA}$ bond, receives an $e_{ref}(12)$ bond. This regrouping serves to include the geometry of association with the $e_{ref}(12)$ bonds enforcing the limited valence of the association interaction. In the *s* – mer representation, graphs which include unphysical core overlap are identically zero. That is, if the association site is singly bondable all graphs composed of *s* – mers of size *s* > 2 immediately vanish due to steric hindrance. This is not the case in Andersen's approach where these unphysical contributions are allowed in individual graphs, with the single bonding condition being exploited with the *cancelation theorem*.

This regrouping of the fugacity expansion allows for the easy incorporation of steric effects. Now, unlike Andersen who tamed the arbitrarily large $F_{AA}$ bonds through the introduction of a renormalized $\tilde{F}_{AA}$, Wertheim uses the idea of a multiple densities, splitting the total density of the fluid as

$$\rho(1) = \rho_o(1) + \rho_b(1) \qquad (1.21)$$

Where $\rho_o(1)$ is the density of monomers (molecules not bonded) and $\rho_b(1)$ is the density of molecules which are bonded. The density $\rho_o(1)$ is composed of all graphs in $\rho(1)$ which do not have an incident $F_{AA}$ bond, and $\rho_b(1)$ contains all graphs which have one or more incident $F_{AA}$ bonds. Performing a topological reduction from fugacity graphs to graphs which contain $\rho_o(1)$ and $\rho(1)$ field points, allowed Wertheim to arrive at the following exact free energy

$$\frac{A}{k_B T} = \int \left(\rho(1)\ln\left(\rho_o(1)\Lambda^3\right) - \rho_o(1)\right) d(1) - c^{(o)} \tag{1.22}$$

where for this case the graph sum $c^{(o)}$ is given as

$$c^{(o)} = \left\{ \begin{array}{l} \text{sum of all irreducible graphs consisting of monomer points carrying factors} \\ \text{of } \rho,\ s\text{-mer graphs with } s \geq 2 \text{ and every point carrying a factor of } \rho_o \text{, and} \\ f_{ref} \text{ -bonds between some sets of points in distinct } s\text{-mers.} \end{array} \right\} \tag{1.23}$$

The first few graphs in the infinite series for $c^{(o)}$ are given in Fig. 1.3

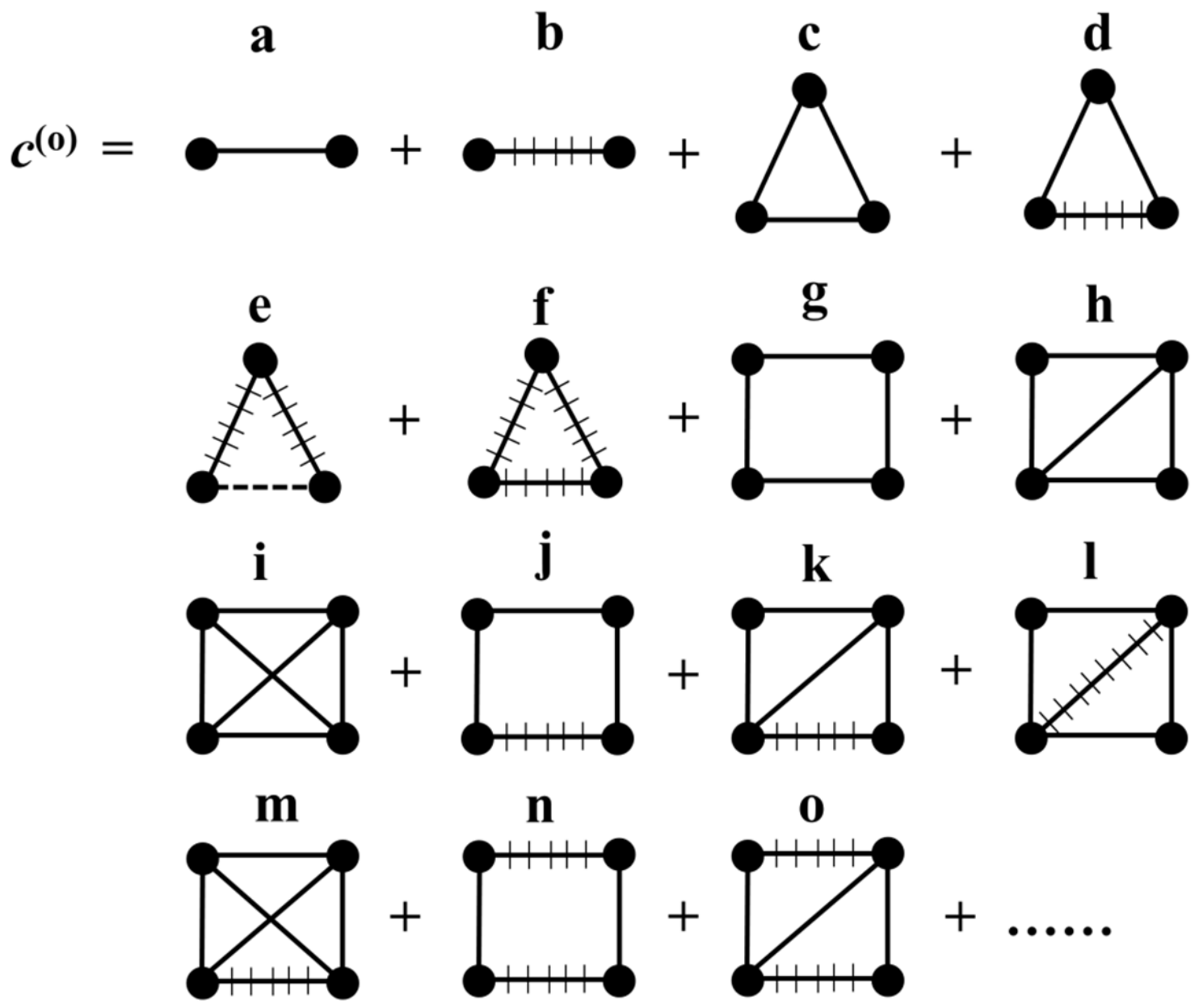

**Figure 1.3:** Graphical representation of Eq. (1.23)

In Fig. 1.3 crossed lines ┼┼┼┼ represent $F_{AA}$ bonds, dashed lines represent $e_{ref}$ bonds and solid lines represent $f_{ref}$ bonds. All points with one or more incident $F_{AA}$ bonds carry a factor $\rho_o(1)$ and each point with no incident $F_{AA}$ bonds carries a factor $\rho(1)$. All graphs without any $F_{AA}$ bonds (graphs **a**, **c**, **g**, **h**, **i** in fig. 1.3) represent the reference system contribution $c_{ref}^{(o)}$. Any point which has 2 incident $F_{AA}$ bonds (graphs **e** and **f** in fig 1.3 are s = 3 – mers) represents a molecule with an association site which is bonded to two other molecules.

If $\phi(12)$ is chosen such that the single bonding condition holds, then all *s* – mer graphs with *s* > 2 vanish and Eq. (1.23) can be summed exactly to yield

$$c^{(o)} = c_{ref}^{(o)} + \frac{1}{2}\int \rho_o(1) f_{AA}(12) g_{oo}(12) \rho_o(2) d(1) d(2) \tag{1.24}$$

The quantity $g_{oo}(12)$ is the monomer / monomer pair correlation function which can be ordered by graphs which contain $k\,F_{AA}$ bonds

$$g_{oo}(12) = \sum_{k=0}^{\infty} g_{oo}^{(k)}(12) \tag{1.25}$$

Similar to the approximation made for $Y_p$ in the formalism of Andersen (Eq. 1.19), only the lowest order contribution is retained, and all contributions with *k* > 0 are neglected. This is the single chain approximation, which yields

$$g_{oo}(12) = g_{ref}(12) = y_{ref}(12) e_{ref}(12) \tag{1.26}$$

Equation (1.26) forms the basis of Wertheim's thermodynamic perturbation theory (TPT), which assumes the monomer – monomer correlation function is the same as that of the reference fluid. This amounts to neglecting all graphs in Eq. (1.23) which contain more than a single $F_{AA}$ bond. To obtain an equation for $\rho_o(1)$, Eq. (1.22) is minimized

$$\frac{\delta A / k_B T}{\delta \rho_o(1)} = \frac{\rho(1)}{\rho_o(1)} - 1 - \int f_{AA}(12) g_{ref}(12) \rho_o(2) d(2) = 0 \tag{1.27}$$

The operator $\delta / \delta \rho_o(1)$ represents the functional derivative. Combining Eqns. (1.22), (1.24) and (1.27) the free energy is simplified as

$$\frac{A - A_{ref}}{k_B T} = \int \rho(1) \left( \ln X_o(1) - \frac{X_o(1)}{2} + \frac{1}{2} \right) d(1) \tag{1.28}$$

where $A_{ref}$ is the Helmholtz free energy of the reference system. Now, assuming a homogeneous fluid $\rho = \rho(1)\Omega$ and solving Eq. (1.27), the monomer fraction Eq. (1.20) is obtained.

As can be seen, under the single bonding condition when treated as a perturbation theory, both Andersen's and Wertheim's approaches give the same result for homogeneous fluids. Indeed, Eq. (1.12) can be rewritten in terms of monomer fractions

$$\tilde{F}_{AA}(12) = F_{AA}(12) y_{ref}(12) X_o^2 = f_{AA}(12) g_{ref}(12) X_o^2 \tag{1.29}$$

and for a homogeneous fluid the renormalized Mayer functions can be used to represent $c^{(o)}$

$$c^{(o)} = c_{ref}^{(o)} + \frac{1}{2} \frac{V \rho^2}{\Omega} \int \tilde{F}_{AA}(12) d(2) \tag{1.30}$$

While identical for singly bondable sites in the single chain approximation (perturbation theory), the two density approach of Wertheim is much more versatile than the approach of Andersen. For instance, for the case that the association site can bond a maximum of $n$ times, there is a clear path forward in the development of a perturbation theory using Wertheim's approach (keep all $s$ – mer graphs with $s \le n$). Attempting to apply Andersen's formalism to this case would be hopelessly complex. Also, Eqns. (1.27) – (1.28) are generally valid for inhomogeneous fluids where the density and monomer fraction vary with position and orientation. In fact, density functional theories based on Wertheim's TPT have proven to be very accurate in the description of inhomogeneous one site associating fluids.[40] It seems unlikely the approach of Andersen could be utilized to derive the inhomogeneous form of the theory.

## 1.5: Multiple association sites: Multiple density approach

Now the case of molecules with a set of association sites $\Gamma = \{A, B, C, \cdots, Z\}$ will be considered. The majority of hydrogen bonding molecules contain multiple association sites: water, alcohols, proteins, hydrogen fluoride etc… Theoretically, this case is more difficult to model than the single site case due to the fact that these molecules can form extended hydrogen bonded structures.

The two density approach of Wertheim allows the development of accurate and simple theories for molecules with a single association site. To extend this idea to the case of multiple association sites $n(\Gamma) > 1$, Wertheim again begins with the fugacity expansion of $\ln \Xi$ which he regroups into the $s$ – mer representation. Where, as for the one site case, an $s$ – mer represents a cluster of $s$ points (hyperpoints here) connected by association bonds $f_{ij}$. However, in contrast to the two density case, all points in an $s$ – mer are not connected by $e_{ref}$ bonds. Only points with bond connected association sites within an $s$ - mer are connected by $e_{ref}$ bonds. Wertheim defines two association sites as bond connected if there is a continuous path of association sites and bonds between these two association sites. Figure 1.4 demonstrates this for the case of a two site $AB$ ($A$ – red association site, $B$ – blue association site) molecule. The wavy lines represent $f_{AB}$ bonds and the dashed lines represents $e_{ref}$ bonds, remembering $f_{AB}(12)e_{ref}(12) = F_{AB}(12)$. All molecules which share $f_{AB}$ bonds are bond connected receiving $e_{ref}$ bonds. Molecules which do not share association bonds (for instance molecules **1** and **3**, **3** and **5** in Fig. 1.4) can only be bond connected if an association site is bonded more than once. This is the only way two association sites not directly connected by a $f_{AB}$ bond can be connected by a continuous path of sites and $f_{AB}$ bonds. For this reason, molecules **1** and **3** receive an $e_{ref}$ bond and molecules **3** and **5** do not. This choice to only fill with $e_{ref}$ bonds between bond connected sites greatly facilitates the formulation of approximation methods.

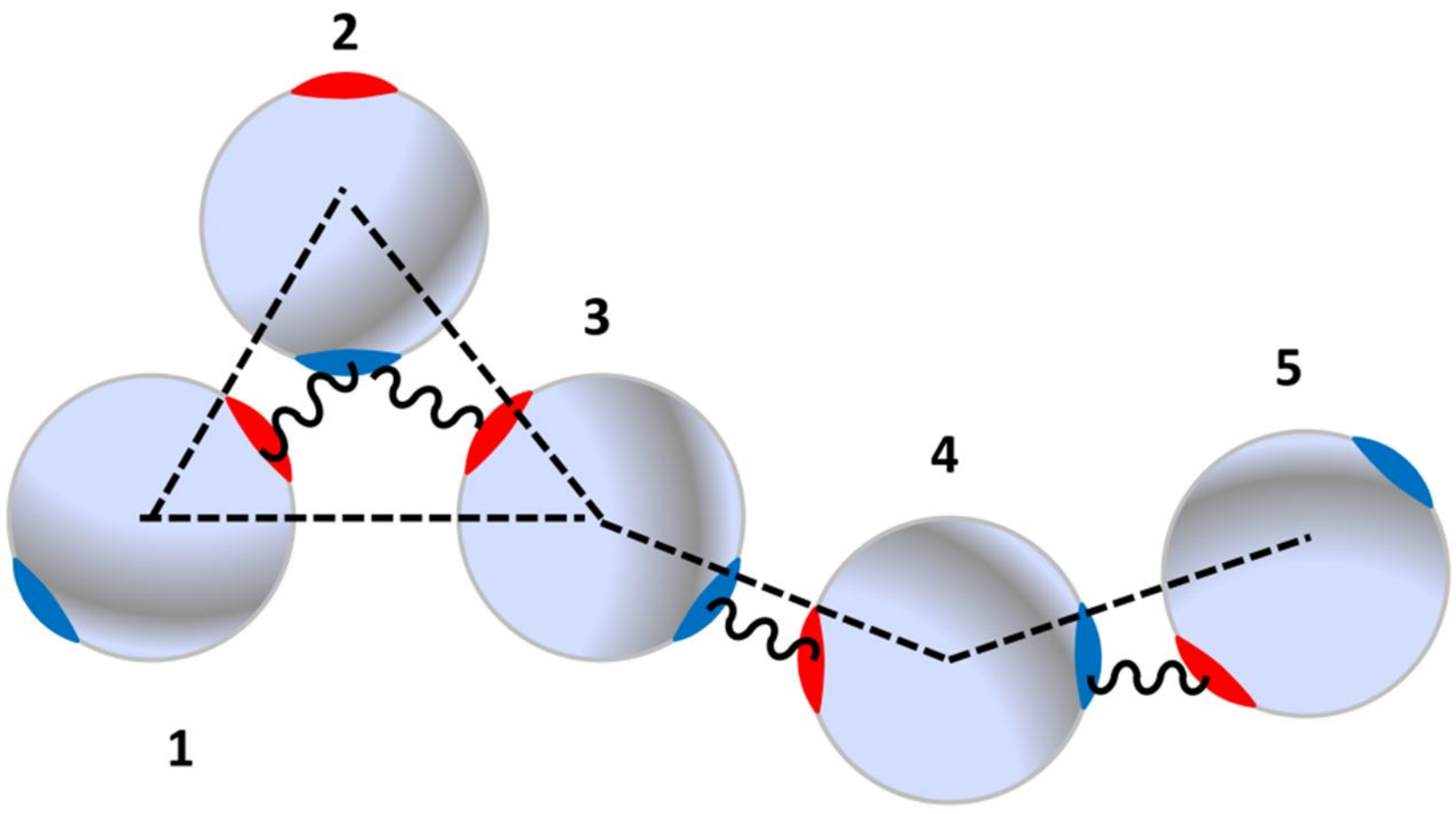

**Figure 1.4:** Representation of *s* – mer where wavy lines represent association bonds and dashed lines represent reference system *e* bonds

In the two density formalism for one site associating molecules, separate densities were assigned to molecules which where bonded and molecules which were not bonded. For multiple association sites this choice would result in the loss of information on site – site level interactions. For this reason, Wertheim expresses the total density as the sum over densities of individual bonding states of the molecules

$$\rho(1) = \sum_{\alpha \subset \Gamma} \rho_\alpha(1) \tag{1.31}$$

where $\rho_\alpha(1)$ is the density of molecules bonded at the set of sites $\alpha$. To aid in the reduction to irreducible graphs Wertheim defines the density parameters

$$\sigma_\gamma(1) = \sum_{\alpha \subset \gamma} \rho_\alpha(1) \tag{1.32}$$

Two important cases of Eq. (1.32) are $\sigma_o = \rho_o$ and $\sigma_\Gamma = \rho$. Using these density parameters, Wertheim transforms the theory from a fugacity expansion to an expansion in $\sigma_\gamma$ through the use of topological reduction, ultimately arriving at the following exact free energy

$$\frac{A}{k_B T} = \int \left( \rho(1) \ln\left( \rho_o(1) \Lambda^3 \right) + Q(1) \right) d(1) - c^{(o)} \tag{1.33}$$

The graph sum in Eq. (1.33) is now defined as

$$c^{(o)} = \left\{ \begin{array}{l} \text{sum of all irreducible graphs consisting of } s\text{-mer graphs (including monomer} \\ \text{hyperpoints) and } f_R \text{ bonds. Points which are bonded at a set of sites } \alpha \text{ carry a} \\ \text{factor } \sigma_{\Gamma-\alpha}(1) \end{array} \right\} \tag{1.34}$$

The term $Q(1)$ is given by

$$Q(1) = -\rho(1) + \sum_{\substack{\alpha \subset \Gamma \\ \alpha \neq \varnothing}} \sigma_{\Gamma-\alpha}(1) c_\alpha(1) \tag{1.35}$$

with

$$c_\alpha(1) = \frac{\delta c^{(o)}}{\delta \sigma_{\Gamma-\alpha}(1)} \tag{1.36}$$

The densities are related to the $c_\alpha(1)$ by the relation

$$\rho_\alpha(1) = \rho_o(1) \sum_{P(\alpha) = \{\gamma\}} \prod_\gamma c_\gamma(1) \tag{1.37}$$

Where $P(\alpha)=\{\gamma\}$ is the partition of $\alpha$ into subsets indexed by $\{\gamma\}$. For instance, the density $\rho_{AB}(1)$ is given by $\rho_{AB}(1)=\rho_o(1)(c_{AB}(1)+c_A(1)c_B(1))$.

The reference system Helmholtz free energy is given as

$$\frac{A_{ref}}{k_B T}=\int\left(\rho(1)\ln\left(\rho(1)\Lambda^3\right)-\rho(1)\right)d(1)-c_{ref}^{(o)} \tag{1.38}$$

The reference system $c_{ref}^{(o)}$ contains all of the graphs in Eq. (1.34) which are devoid of association bonds. Subtracting Eq. (1.38) from (1.33) we obtain

$$\frac{A-A_{ref}}{k_B T}=\int\left(\rho(1)\ln\left(\frac{\rho_o(1)}{\rho(1)}\right)+Q(1)+\rho(1)\right)d(1)-\Delta c^{(o)} \tag{1.39}$$

The association graph sum $\Delta c^{(o)}=c^{(o)}-c_{ref}^{(o)}$ contains all the graphs in Eq. (1.34) which contain association bonds. Figure 1.5 gives examples of graphs in the sum $\Delta c^{(o)}$ for the two site *AB* case discussed in Fig. 1.4.

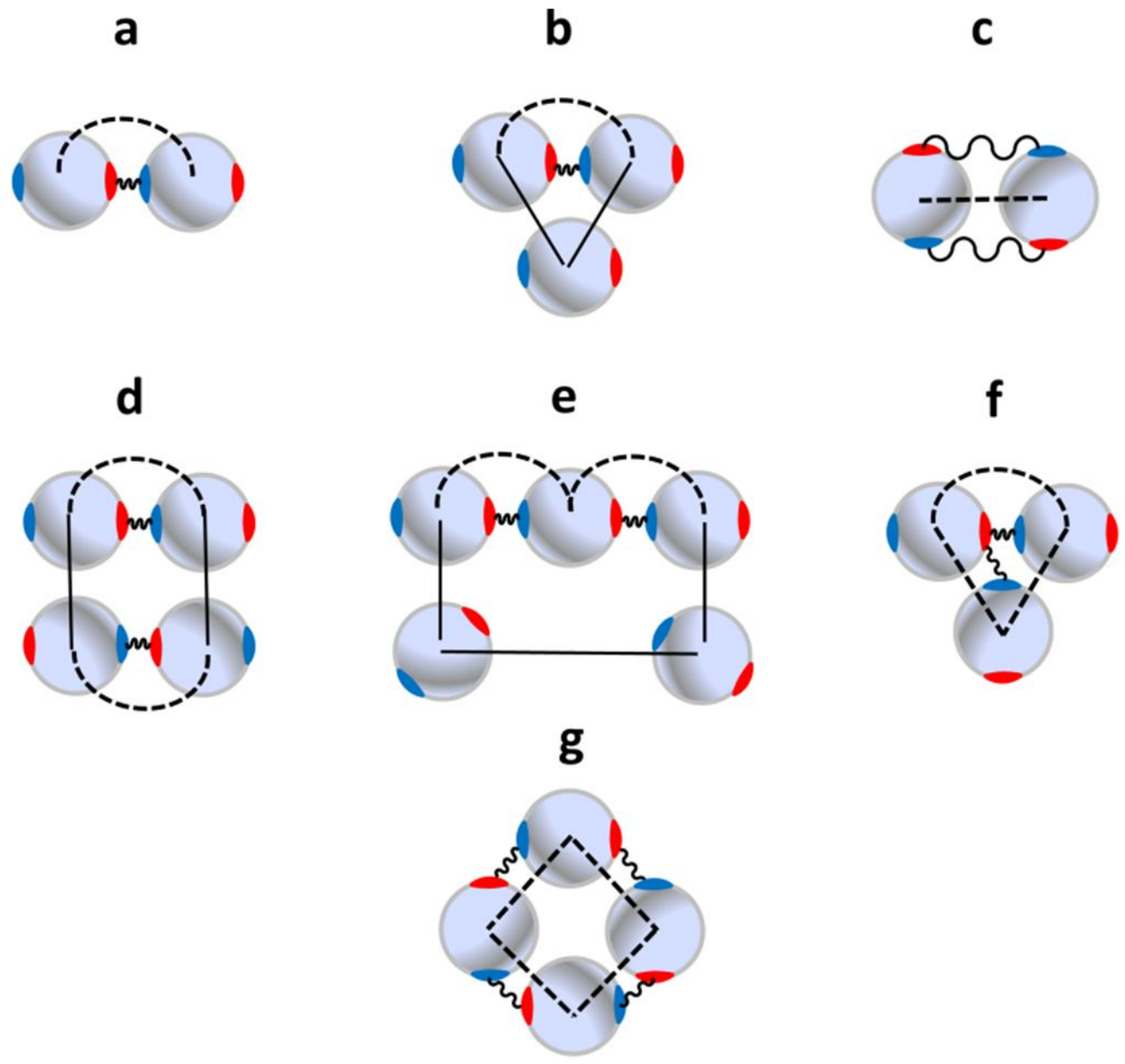

**Figure 1.5:** Examples of graphs in $\Delta c^{(o)}$ for the two site *AB* case. Wavy lines represent $f_{AB}$ bonds, dashed lines represent $e_{ref}$ bonds and solid lines represent $f_{ref}$ bonds

Equation (1.39) provides a very general starting point for the statistical mechanics of associating fluids and is exact so long as the system is pairwise additive, and the intermolecular potential can be separated into a reference and association portion as in Eq. (1.1). The challenge is to approximate the graph sum $\Delta c^{(o)}$ (here we assume the properties of the reference fluid are known).

A simple and widely used approximation of this sum forms the foundation of Wertheim's thermodynamic perturbation theory (TPT)[36, 37]. To start we decompose $\Delta c^{(o)}$ as

$$\Delta c^{(o)} = \sum_{k=1}^{\infty} \Delta \hat{c}_k^{(o)} \tag{1.40}$$

where $\Delta \hat{c}_k^{(o)}$ is the contribution for graphs which contain *k* associated clusters . For example, graph **d** in Fig. 1.5 belongs to the sum $\Delta \hat{c}_2^{(o)}$ while the remaining graphs all belong to $\Delta \hat{c}_1^{(o)}$. TPT can be defined as the neglect of all $\Delta \hat{c}_k^{(o)}$ for $k > 1$ giving

$$\Delta c^{(o)} \approx \Delta \hat{c}_1^{(o)} \tag{1.41}$$

The approximation Eq. (1.41) accounts for interactions within the associated cluster and between the clusters and the reference fluid, but not the interactions between associated clusters. This approximation is accurate for cases in which the pair correlation function between associating molecules is similar to that of the reference fluid.

Equation (1.40) accounts for multiply bonded association sites (graph **f** in Fig. 1.5 belongs to this class), cycles of association bonds (graph **g** in Fig. 1.5), multiple bonds between two molecules (graph **c** in Fig. 1.5) and chains (trees) of association bonds (graphs **a**, **b** and **e** in Fig. 1.5). For the time being we will assume the intermolecular potential and placement of association sites is such that contributions from cycle formation, multiply bonded association sites and multiple bonds between molecules can all be ignored; leaving only contributions for the formation of chain and tree like clusters. With these restrictions $\Delta \hat{c}_1^{(o)}$ can now be written as

$$\Delta \hat{c}_1^{(o)} = \Delta c_{TPT1}^{(o)} + \Delta c_{TPT2}^{(o)} + \Delta c_{TPT3}^{(o)} + \cdots \tag{1.42}$$

where $\Delta c_{TPT1}^{(o)}$ is the first order contribution which contains all contributions for association between a pair of molecules (graph **a** and **b** in Fig. 1.5 are examples), $\Delta c_{TPT2}^{(o)}$ is the second order contribution which contains information about the simultaneous association of three molecules (graph **e** in Fig. 1.5 belongs to this class) etc…For the case of molecules with a single association site $\Delta \hat{c}_1^{(o)} = \Delta c_{TPT1}^{(o)}$. If it can be assumed that steric hindrance between association sites is small, the sum in Eq. (1.42) can be truncated at first order (TPT1) giving the simple result[36]

$$\Delta c^{(o)} \approx \Delta \hat{c}_1^{(o)} \approx \Delta c_{TPT1}^{(o)} = \frac{1}{2} \sum_{A \in \Gamma} \sum_{B \in \Gamma} \int \sigma_{\Gamma - A}(1) g_{ref}(12) f_{AB}(12) \sigma_{\Gamma - B}(2) d(1) d(2) \tag{1.43}$$

It should be noted that while $\Delta c_{TPT1}^{(o)}$ only accounts for interactions between pairs of molecules, all possible trees of associated clusters can be reproduced. However, in first order, the only steric interactions are between nearest neighbors in the cluster giving a theory which is independent of bond angles.

Equation (1.43) allows the free energy Eq. (1.39) to be written in the following simple form[41]

$$\frac{A - A_{ref}}{k_B T} = \sum_{A \in \Gamma} \int \rho(1) \left( \ln X_A(1) - \frac{X_A(1)}{2} + \frac{1}{2} \right) d(1) \tag{1.44}$$

The fractions of molecules not bonded at site $A$, $X_A$ are solved for self consistently as

$$X_A(1) = \frac{\sigma_{\Gamma-A}(1)}{\rho(1)} = \frac{1}{1 + \int \rho(1) g_{ref}(12) \sum_{B \in \Gamma} f_{AB}(12) X_B(2) d(2)} \qquad (1.45)$$

This simple TPT1 result has proven to be a very powerful tool in the theoretical description of associating fluids. Equations (1.44) – (1.45) have been shown to be accurate for bulk fluids composed of hard spheres with one[42], two[42, 43] and four association sites.[44] The theory has also been shown to be accurate for associating Lennard – Jones spheres with one, two[45] and four[46] association sites and associating molecules with a square well reference fluid[47]. In addition to spherical molecules, TPT1 has been shown to be accurate for associating chains of tangentially bonded hard,[48] Lennard – Jones[49] and square well[47] spheres. When applying TPT1 to associating chains, a chain reference fluid must be used, which is also obtained using TPT1in the complete association limit.[41] In addition to model systems, TPT1 has been widely applied in both academia and industry as an engineering equation of state for hydrogen bonding fluids[50] in the form of the statistical associating fluid theory equation of state (SAFT).[51-54]

TPT1 has also found wide application for inhomogeneous associating fluids in the form of classical density functional theory (DFT). A small subset of these applications include the study of confined associating hard spheres,[40, 44, 55] associating chains near repulsive surfaces,[56] phase diagram of supramolecular polymers,[57] associating polymer brushes,[58] mixtures of alcohol and water near a hydrophobic surface,[59] n – alkane / water interface,[60] and the surface tension of associating chain molecules[47, 61] and water[61].

Though widely applied, TPT1 is far from perfect with a number of deficiencies resulting from the simplifying approximations employed. These approximations are summarized below

1) Single chain approximation – Neglects all graphs with more than one associated cluster. This is TPT, which assumes the structure of the fluid is similar to that of the reference fluid. The single chain approximation will fail, for instance, for fluids with a nematic phase.[62]

2) Singly bondable association sites – Assumes each association site saturates after sharing in a single association bond. This approximation is not valid for patchy colloids with large patch sizes.[7]

3) No multiple bonding of molecules – Assumes that any two molecules can share at most one association bond. Carboxylic acids[63] and water[64] are known to violate this condition.

4) No cycles of association bonds – Only chains and trees of association bonds are accounted for. Cycles are irreducible and cannot be reproduced in TPT1. It is well known that hydrogen fluoride exhibits significant ring formation.[65]

5) No steric hindrance between association sites – All contributions to the irreducible graph sum with more than a single association bond were neglected. For this reason, association at one site is independent of association at all other sites. Most polyfunctional associating molecules will exhibit some degree of steric hindrance between association sites.

6) Association is independent of bond angles – There is no information in TPT1 on location of association sites. This is intimately related to approximations 3 – 5 above.

7) In Wertheim’s multi – density formalism, pairwise additivity of the pair potential was assumed. Most polyfunctional hydrogen bonding molecules exhibit some degree of bond cooperativity (non – pairwise additivity).[66] Hydrogen bond cooperativity is particularly important for hydrogen fluoride.[65]

## 1.6: What about Andersen's approach for multiple sites?

While we have focused on Wertheim's multiple density approach, Andersen also developed an exact cluster expansion for associating fluids with multiple association sites.[32] Specifically considering the two site case depicted in Figs. 1.4 – 1.5, with the restriction that there are only *AB* attractions, the pair correlation function was decomposed as

$$g(12) = \tilde{F}_{AB}(12) + \tilde{F}_{BA}(12) + F_c(12) + F_{oo}(12) \tag{1.46}$$

where $\tilde{F}_{AB}$ are the renormalized Mayer functions which represent the sum of all graphs in $g(12)$ where the two root points are connected by a $F_{AB}$ bond, $F_c$ is the sum of diagrams in which the root points are connected by a chain of two or more association bonds, and $F_{oo}$ is the sum of all diagrams in which the root points are not connected by paths of association bonds.

Now, following a similar path of approximation as that used in the development of TPT1, we neglect all diagrams in (1.46) in which the root points are connected by a path of two or more association bonds. That is, we assume $F_c(12) = 0$. Further, in determining $N_{HB}$ the diagrams in $F_{oo}$ contribute little as compared to the renormalized Mayer functions $\tilde{F}_{AB}$, so they are neglected. This is consistent with Wertheim's treatment where $N_{HB} = 0$ when $\phi_{AB} = 0$. Then, assuming the two association sites are independent (no steric hindrance), Andersen approximately summed the series $\tilde{F}_{AB}$ as

$$\tilde{F}_{AB}(12) = F_{AB}(12) Y_p(12) \frac{1 + 2\Delta_{AB} - \sqrt{1 + 4\Delta_{AB}}}{2\Delta_{AB}^2} \tag{1.47}$$

where $\Delta_{AB}$ is given by Eq. (1.13) and $N_{HB}$ is obtained as

$$N_{HB} = \frac{\rho}{\Omega} \int \left( \tilde{F}_{AB}(12) + \tilde{F}_{BA}(12) \right) d(2) \tag{1.48}$$

Using the relation $X_A = 1 - N_{HB}/2$ and employing the approximation Eq. (1.19) we obtain the following result for the fraction of molecules not bonded at site $A$ (due to symmetry $X_A = X_B$)

$$X_A = \frac{-1 + \sqrt{1 + 4\Delta_{AB}}}{2\Delta_{AB}} \tag{1.49}$$

where $\Delta_{AB}$ is now given by $\Omega\Delta_{AB} = \rho \int y_{ref}(12) F_{AB}(12) d(2)$. Equation (1.49) is precisely the result obtained using TPT1 (Eq. 1.45) for this two site case in the limit of a homogeneous fluid.

Andersen's cluster expansion has also been applied to the case of a water like fluid with 4 association sites by Dahl and Andersen.[67, 68] After a formidable graphical analysis they arrive at a result which, in lowest order, has been shown to give identical numerical predictions[69] to TPT1.

While identical, or nearly identical, results can be obtained from both Wertheim's and Andersen's cluster expansions at lowest order in perturbation, the multi – density formalism of Wertheim is much simpler to navigate than the cluster expansion of Andersen. In a quest to relax the basic assumptions of TPT1 listed in section 1.5, Wertheim's multi – density formalism provides a much simpler framework to work in.

## 1.7: Scope of this work

Wertheim's first order perturbation theory has proven extremely useful in the theoretical description of associating fluids. However, as discussed in 1.6, the accuracy of TPT1 can deteriorate when certain assumptions are not satisfied. In this work we systematically address, to some degree, assumptions 2 – 7 above. Assumption 1 defines what is meant by perturbation theory and will not be addressed here. To move beyond the single chain approximation, the more complicated framework of integral equation theory[70-72] may be employed.

In chapter 2, TPT is extended to account for the possibility that the association site can bond more than once. Specifically we consider the one site case and allow the association site to bond a maximum of twice (the site is doubly bondable). In contrast to the singly bondable case discussed above, where only dimers can form, when the sites are doubly bondable extended associated chains may form as well as triatomic rings. This theory is most applicable to the study of patchy colloids with patch sizes such that patches are doubly (or singly) bondable. The theory is shown to be accurate in comparison to simulation data. In addition, the theory correctly predicts that three patch rings are the dominant type of associated cluster.

In chapters 3 – 4, TPT is extended to model mixtures of colloids with spherically symmetric attractions (*s* colloids) and patchy colloids (*p* colloids). Mixtures of this type have been recently synthesized by researchers.[8] The development of a theory to model mixtures of this type is challenging due to the fact that the patches of the *p* colloids are attracted to the *s* colloids. Since the *s* colloids are spherically symmetric, they can bond to multiple *p* colloids going well beyond the single bonding condition.

In chapters 5 – 6 our attention is restricted to the case of singly bondable association sites; however, in these chapters the angle between the centers of the association sites (we loosely call this the bond angle) is treated as an independent variable. In chapter 5 we develop a new theory for two site associating spheres which is valid over the full bond angle range. For the case of small bond angles, it is shown that steric hindrance, ring formation and double bonding must be accounted for. We treat each of these contributions as functions of bond angle to develop a complete theory. The theory is shown to be in excellent agreement with simulation results. This is the first time the effect of bond angle has been included in TPT. In chapter 6 we extend the theory developed in chapter 5 to the case that there are more than two association sites. The theory is shown to be in excellent agreement with simulation data for the case of associating spheres with three association sites. It is also shown that the phase diagram is strongly bond angle dependent.

In chapter 7 Wertheim's multi – density approach is extended to account for bond cooperativity in associating fluids. It is known[66] that when a multi – functional hydrogen bonding molecule forms a hydrogen bond, the polarization of the molecule is increased, thereby increasing the energetic benefit of additional hydrogen bonds. There is cooperativity and the system energy violates pairwise additivity. Since cooperativity is such a common feature of hydrogen bonding molecules, it seems prudent to extend Wertheim's multi – density formalism to account for these cooperative interactions. Unfortunately, Wertheim's theory is constructed around the assumption of pairwise additivty. That said, in chapter 7 we show how bond cooperativity can be incorporated by the development of a new TPT which treats bond cooperativity as a perturbation. The new theory is shown to be in excellent agreement with simulation data.

In chapter 8 a new density functional theory for associating fluids in spatially uniform external fields which act on orientation (orienting fields) is developed. A classic example of a fluid of this type would be hydrogen bonding dipolar molecules (hydrogen fluoride, water etc…) in an electric field. Wertheim's multi – density approach has never been applied to systems of this type. We derive some exact results for the orientational distribution function (ODF) of associating fluids in external fields using classical density functional theory in the canonical ensemble. We then approximate the exact ODF in TPT1, and compare the theory to simulations for the case of two site associating hard spheres in a linear orienting field.

Finally, in chapter 9 advances made throughout this dissertation are summarized. In addition, future extensions and applications of the developed theories are discussed.

# CHAPTER 2

# Doubly bondable association sites

One of the main assumptions in the development of Wertheim's first order perturbation theory (TPT1) is that association sites are singly bondable. Indeed, the entire multi – density formalism of Wertheim is constructed to enforce this condition. For the case of hydrogen bonding, the assumption of singly bondable sites is justified; however, for patchy colloids (see section 1.1 for a background on patchy colloids) it has been shown experimentally[7, 8] that the number of bonds per patch (association site) is dependent on the patch size. If TPT is to describe these types systems, the possibility of multiple bonds per association site must be accounted for. In this chapter we will restrict our attention to sites which can bond a maximum of two times. For this case the types of associated clusters which can form for the one site case are given in Fig. 2.1. For the case of singly bondable sites, only dimers can form. However, when sites can bond twice, additional contributions for chains and rings of doubly bonded sites must be accounted for.

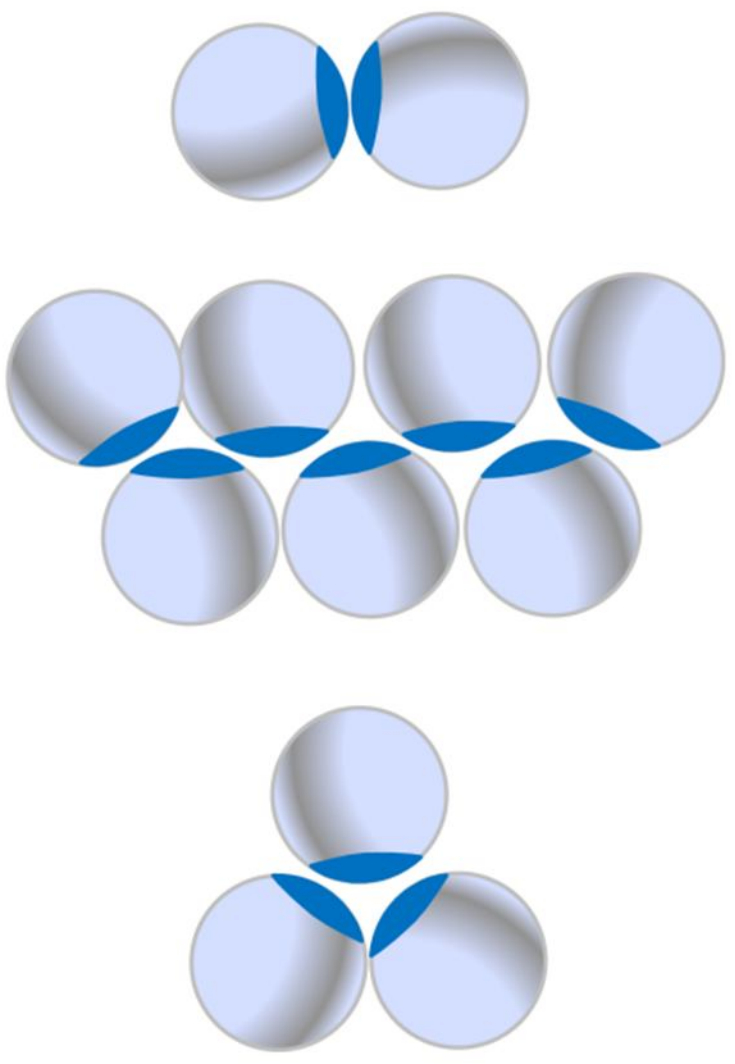

**Figure 2.1:** Associated clusters for patchy colloids with a single double bondable patch

For the case that only dimers can form, the TPT1 free energy given by Eq. (1.28) is highly accurate for the prediction of thermodynamic quantities. Recently, Kalyuzhnyi *et al*.[73, 74] developed a new resummed perturbation theory which explicitly included the effects of chains of doubly bonded sites, **b** in Fig. (2.1). The theory was developed in a formalism for associating fluids with spherically symmetric attractions developed by Kalyuzhnyi and Stell[75] . In their approach the possibility of ring formation was not accounted for. As will be shown, rings as given in **c** of Fig. 2.1 are the dominant type of associated cluster for this single patch system, so any accurate theory must properly account for ring formation. In this chapter we extend Wertheim's TPT to account for doubly bonded association sites. Both chain formation and ring formation are accounted for, giving a simple theory which is shown to be accurate in comparison to Monte Carlo simulation data. Before deriving the theory, the potential model is reviewed in the following section.

## 2.1: A primitive association model

Primitive models for association have provided a route to model hydrogen bonding fluids for many decades. One such potential first introduced by Bol[76], and later used by Chapman *et al.*[41, 42], considers association as a square well interaction which depends on the position and orientation of each molecule. Kern and Frenkel[77] later realized that this potential could describe the interaction between "patchy" colloids in a very precise way. For these conical sites the potential of interaction between two molecules / colloids is given by Eqns. (1.1) – (1.2) with the potential between any two association sites given by

$$\phi_{AB}(12) = \begin{cases} -\varepsilon_{AB}, & r_{12} \le r_c; \theta_{A1} \le \theta_c; \theta_{B2} \le \theta_c \\ 0 & otherwise \end{cases} \tag{2.1}$$

Where $\vec{r}_{12}$ is the vector of magnitude $r_{12}$ connecting the centers of two colloids, $r_c$ is the maximum separation between two colloids for which association can occur, $\theta_{A1}$ is the angle between $\vec{r}_{12}$ and the orientation vector passing through the center of patch *A* on colloid 1 and $\theta_c$ is the critical angle beyond which association cannot occur. Equation (2.1) states that if the colloids are close enough $r_{12} < r_c$, and both are oriented correctly $\theta_{A1} < \theta_c$ and $\theta_{B2} < \theta_c$, then an association bond is formed and the energy of the system is decreased by $\varepsilon_{AB}$. Figure 2.2 gives a depiction of the association potential for the case of a single association site. Furthermore, to isolate the effect of association, we will consider the reference fluid to be a fluid of hard spheres of diameter *d*.

$$\phi_{ref}(12) = \phi_{HS}(r_{12}) \tag{2.2}$$

Lastly, for this association potential, the association Mayer functions can be written as

$$f_{AB}(12) = O_{AB}(12) f_{AB} \tag{2.3}$$

where $O_{AB}(12) = -\phi_{AB}(12)/\varepsilon_{AB}$ is the overlap function and $f_{AB} = \exp(\varepsilon_{AB}/k_B T) - 1$.

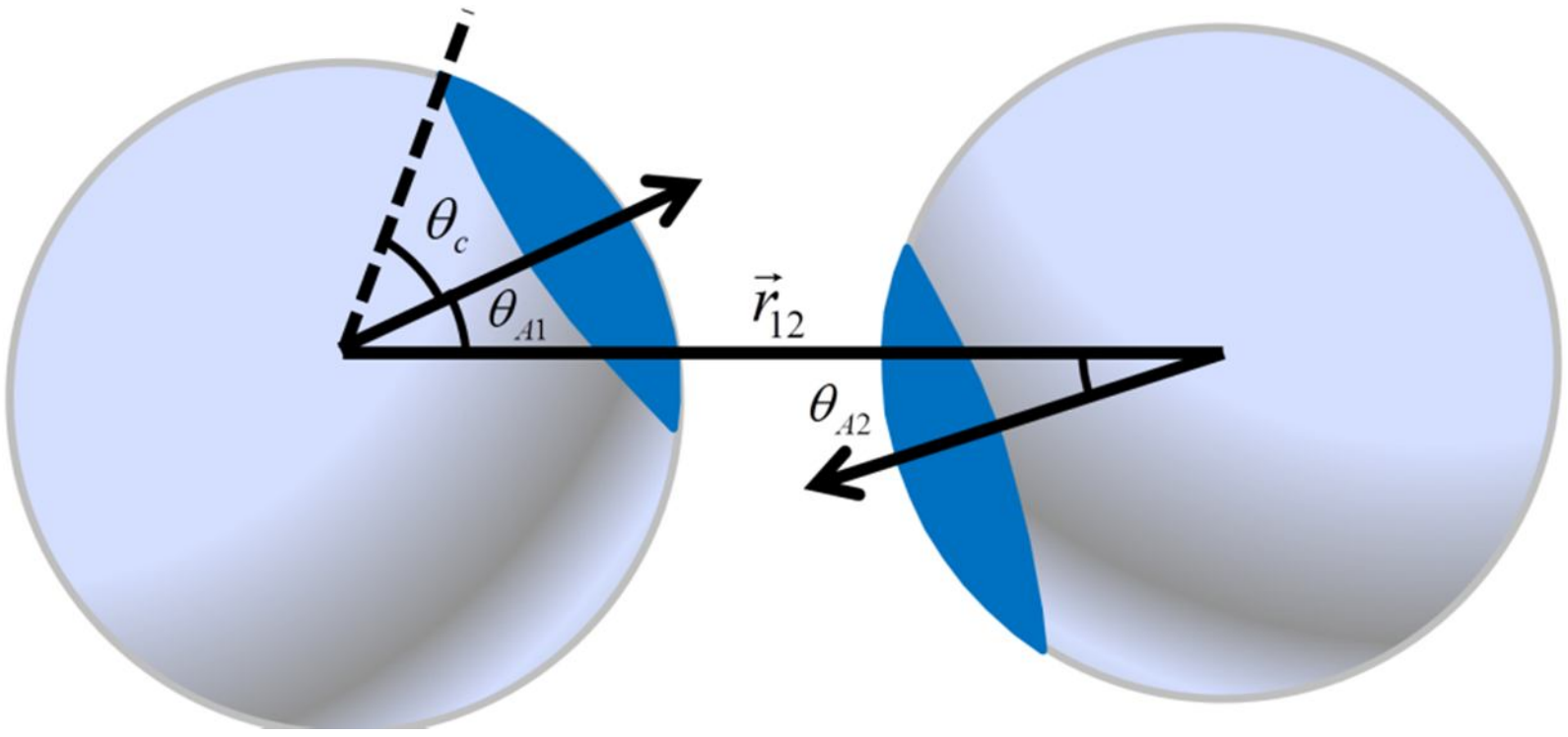

**Figure 2.2:** Association parameters for conical association sites

## 2.2: Theory

In this chapter we restrict our attention to the single site case depicted in Fig. 2.2. Assuming the absence of an external field the free energy Eq. (1.22) can be expressed as

$$\frac{A - A_{HS}}{V k_B T} = \rho \ln \frac{\rho_o}{\rho} - \rho_o + \rho - \frac{\Delta c^{(o)}}{V} \tag{2.4}$$

In accordance with Fig. 2.1 we separate $\Delta c^{(o)}$ into contributions for chain and ring formation

$$\Delta c^{(o)} \approx \Delta \hat{c}_1^{(o)} = \Delta c_{chain}^{(o)} + \Delta c_{ring}^{(o)} \tag{2.5}$$

The contribution $\Delta c_{chain}^{(o)}$ is further decomposed into contributions from chains of *n* bonds and *n* + 1 colloids

$$\Delta c_{chain}^{(o)} = \sum_{n=1}^{\infty} \Delta c_n^{chain} \tag{2.6}$$

Where, for instance, graph **e** in Fig. 1.3 belongs to the contribution $\Delta c_2^{chain}$ and graph **f** belongs to $\Delta c_{ring}^{(o)}$. Each of these contributions consists of an infinite series of graphs with a single associated cluster interacting with the reference fluid. These series can be summed as

$$\frac{\Delta c_n^{chain}}{V} = \frac{1}{2}\rho_o^{n+1} f_{AA}^n \int g_{HS}(1 \cdots n+1) \prod_{k=1}^{n} O_{AA}(k,k+1) \frac{d(k+1)}{\Omega} \tag{2.7}$$

and

$$\frac{\Delta c_{ring}^{(o)}}{V} = \frac{1}{6}\rho_o^3 f_{AA}^3 \int g_{HS}(123) O_{AA}(12) O_{AA}(23) O_{AA}(13) \frac{d(2)}{\Omega}\frac{d(3)}{\Omega} \tag{2.8}$$

Here $\Omega = 4\pi$ and the functions $g_{HS}(1 \cdots k)$ are the $k$ body correlation functions of the hard sphere reference system. Since little is known about the correlation functions $g_{HS}(1 \cdots k)$ for $k > 3$, we must approximate the higher order $g_{HS}(1 \cdots k)$ in superposition. For the current case, a particularly convenient approximation for the *chain contributions* will be the following

$$g_{HS}(1 \cdots k) = \prod_{j=1}^{k-1} g_{HS}(r_{j,j+1}) \prod_{i=1}^{k-2} e_{HS}(r_{i,i+2}) \tag{2.9}$$

The superposition given by Eq. (2.9) prevents overlap between nearest and next nearest neighbors in the chain and should be most accurate at low densities. We note that the probability that an isolated associated chain of $n$ + 1 colloids has a configuration $(1 \cdots n+1)$ is given by

$$P_{chain}^{(n)}(1 \cdots n+1) = \frac{\prod_{k=1}^{n} O_{AA}(k,k+1) e_{HS}(r_{k,k+1}) \prod_{i=1}^{n-1} e_{HS}(r_{i,i+2})}{Z_{chain}^{(n)}} \tag{2.10}$$

The probability in Eq. (2.10) accounts for steric interactions between nearest and next nearest neighbors in the chain, and the term $Z_{chain}^{(n)}$ is the chain partition function given by

$$Z_{chain}^{(n)} = \int \prod_{i=1}^{n-1} e_{HS}\left(r_{i,i+2}\right) \prod_{k=1}^{n} \mathrm{O}_{AA}(k,k+1) e_{HS}\left(r_{k,k+1}\right) d(k+1) \tag{2.11}$$

Combining Eqns. (2.7) and (2.9) – (2.11) we obtain

$$\frac{\Delta c_n^{chain}}{V} = \frac{1}{2}\rho_o^{n+1} f_{AA}^{n} \frac{Z_{ch}^{(n)}}{\Omega^n} \left\langle \prod_{j=1}^{n} y_{HS}\left(r_{j,j+1}\right) \right\rangle_{P_{chain}} \tag{2.12}$$

The cavity correlation function $y_{HS}\left(r_{j,j+1}\right) = g_{HS}\left(r_{j,j+1}\right) / e_{HS}\left(r_{j,j+1}\right)$. The brackets in Eq. (2.12) represent an average over the distribution function Eq. (2.10). To an excellent approximation this average can be evaluated as a product of individual averages over the bonding range

$$\left\langle \prod_{j=1}^{n} y_{HS}\left(r_{k,k+1}\right) \right\rangle_{P_{chain}} = \left\langle y_{HS}(r) \right\rangle_{br}^{n} \tag{2.13}$$

where

$$\left\langle y_{HS}(r) \right\rangle_{br} = \frac{4\pi \int_d^{r_c} y_{HS}(r) r^2 dr}{4\pi \int_d^{r_c} r^2 dr} = \frac{\xi}{v_b} \tag{2.14}$$

The constant $v_b$ is the volume of a spherical shell defined by the denominator of the second term in Eq. (2.14) and $\xi$ is defined by the numerator. For a hard sphere fluid with $r \geq d$ the cavity correlation function is equal to the pair correlation function $y_{HS}(r) = g_{HS}(r)$. Since the range of the integration in Eq. (2.14) is small, it is common practice to use a Taylor's series expansion of $g_{HS}(r)$ around the value at hard sphere contact $g_{HS}(d)$ such that

$$g_{HS}(r) = g_{HS}(d) + \left.\frac{\partial g_{HS}(r)}{\partial r}\right|_{r=d} (r-d) \tag{2.15}$$

However, for certain cases this approximation of $g_{HS}(r)$ can prove very inconvenient. As an alternative, we employ the fact that in the bonding range $\{d \le r \le r_c\}$ the following relation holds true to an excellent approximation[41]

$$r^p g_{HS}(r) = d^p g_{HS}(d) \tag{2.16}$$

Where $p$ is a density dependent quantity which we obtain by fitting Eq. (2.16) to the analytical solution for $g_{HS}(r)$ of Chang and Sandler[78]. The results can be represented by the simple polynomial $p = 17.87\eta^2 + 2.47\eta$, where $\eta = \pi\rho d^3/6$ is the packing fraction. Using Eq. (2.16) we evaluate the integral over the pair correlation function as

$$\xi = 4\pi\int_d^{r_c} g_{HS}(r) r^2 dr = 4\pi d^3 g_{HS}(d)\left(\frac{(r_c/d)^{3-p} - 1}{3-p}\right) \tag{2.17}$$

As has been shown[73], integrals of similar form to $Z_{chain}^{(n)}$ can be very accurately factored as

$$Z_{chain}^{(n)} = \left(Z_{chain}^{(1)}\right)^n \Phi_{chain}^{n-1} \tag{2.18}$$

where

$$\Phi_{chain} = \frac{Z_{chain}^{(2)}}{\left(Z_{chain}^{(1)}\right)^2} = \frac{1}{(v_b\kappa)^2}\int O_{AA}(12)O_{AA}(23)e_{HS}(r_{12})e_{HS}(r_{23})e_{HS}(r_{13})\frac{d(2)}{\Omega}\frac{d(3)}{\Omega} \tag{2.19}$$

When $\Phi_{chain} = 0$, multiple bonding of an association site (patch) is impossible. Combining the results above we obtain

$$\frac{\Delta c_n^{chain}}{V} = \frac{1}{2}\rho_o^{n+1}\left(f_{AA}\xi\kappa\right)^n \Phi_{chain}^{n-1} \tag{2.20}$$

The constant $\kappa$ represents the probability that two colloids are in mutual bonding orientations and is given by

$$\kappa = \frac{(1-\cos\theta_c)^2}{4} \tag{2.21}$$

Now using Eq. (2.20) to evaluate the infinite sum in Eq. (2.6) we obtain

$$\frac{\Delta c_{chain}^{(o)}}{V}=\frac{\frac{1}{2}\rho_o^2 f_{AA}\xi\kappa}{1-f_{AA}\xi\kappa\rho_o\Phi_{chain}} \tag{2.22}$$

When multiple bonding of a patch is impossible $\Phi_{chain}\to 0$, and we recover the TPT1 result Eq. (1.24) for the case $g_{oo}=g_{HS}$.

Now we turn our attention to the ring contribution $\Delta c_{ring}^{(o)}$ Eq. (2.8). For this case we approximate the triplet correlation function using Kirkwood superposition. Following a similar process to the one described above we obtain the result

$$\frac{\Delta c_{ring}^{(o)}}{V}=\frac{1}{6v_b}\left(\rho_o f_{AA}\xi\right)^3\kappa^2\Phi_{ring} \tag{2.23}$$

In Eq. (2.23) $\Phi_{ring}$ is given by

$$\Phi_{ring}=\frac{1}{\left(v_b\kappa\right)^2}\int O_{AA}(12)O_{AA}(23)O_{AA}(13)e_{HS}\left(r_{12}\right)e_{HS}\left(r_{23}\right)e_{HS}\left(r_{13}\right)\frac{d(2)}{\Omega}\frac{d(3)}{\Omega} \tag{2.24}$$

When multiple bonding of a site becomes impossible $\Phi_{ring}=0$, resulting in $\Delta c_{ring}^{(o)}=0$. Now that $\Delta c^{(o)}$ has been completely specified the free energy is minimized with respect to $\rho_o$ giving the following relation

$$\rho=\rho_o+\frac{\rho_o^2 f_{AA}\xi\kappa}{1-f_{AA}\xi\kappa\rho_o\Phi_{chain}}+\frac{1}{2}\rho_o^3\Phi_{chain}\left(\frac{f_{AA}\xi\kappa}{1-f_{AA}\xi\kappa\rho_o\Phi_{chain}}\right)^2+\frac{1}{2v_b}\left(\rho_o f_{AA}\xi\right)^3\kappa^2\Phi_{ring} \tag{2.25}$$

Equation (2.25) is simply conservation of mass. From Eq. (2.25) we identify the density of colloids boded twice in rings $\rho_2^{ring}$, bonded once $\rho_1^{chain}$ and bonded twice in a chain $\rho_2^{chain}$ as

$$\rho_2^{ring}=\frac{1}{2v_b}\left(\rho_o f_{AA}\xi\right)^3\kappa^2\Phi_{ring} \tag{2.26}$$

$$\rho_1^{chain} = \frac{\rho_o^2 f_{AA}\xi\kappa}{1 - f_{AA}\xi\kappa\rho_o\Phi_{chain}} \tag{2.27}$$

$$\rho_2^{chain} = \frac{1}{2}\rho_o^3\Phi_{chain}\left(\frac{f_{AA}\xi\kappa}{1 - f_{AA}\xi\kappa\rho_o\Phi_{chain}}\right)^2 \tag{2.28}$$

Using these density definitions the free energy can be simplified to

$$\frac{A - A_{HS}}{Vk_BT} = \rho\ln\frac{\rho_o}{\rho} + \rho - \rho_o - \frac{\rho_1^{chain}}{2} - \frac{\rho_2^{ring}}{3} \tag{2.29}$$

Equation (2.29) completes the theory for molecules / colloids with a single doubly bondable association site. To obtain the free energy, Eq. (2.25) is first evaluated for $\rho_o$ which allows the free energy to be calculated through Eq. (2.29).

All that remains to do is calculate the integrals $\Phi_{chain}$ (2.19) and $\Phi_{ring}$ (2.24). To evaluate these integrals we exploit the mean value theorem and employ Monte Carlo integration[79] to obtain

$$\Phi_{chain} = \left\{\begin{array}{l}\text{The probability that if the positions of two colloids are generated such} \\ \text{that they are correctly positioned to associate with a third colloid, that} \\ \text{there is no core overlap between the two generated colloids}\end{array}\right\} \tag{2.30}$$

$$\Phi_{ring} = \left\{\begin{array}{l}\text{The probability that if the positions and orientations of two colloids are} \\ \text{generated such that they are positioned and oriented correctly to bond to} \\ \text{a third colloid, that there is no core overlap between the two generated} \\ \text{colloids and that these two generated colloids are oriented and positioned} \\ \text{such that they share an association bond}\end{array}\right\} \tag{2.31}$$

Equations (2.30) and (2.31) are easily evaluated on a computer; the calculations are independent of temperature and density, they depend only on the potential parameters $r_c$ and $\theta_c$. Table 2.1 gives calculations for a critical radius of $r_c = 1.1d$.

**Table 2.1:** Numerical calculations of integrals $\Phi_{chain}$ and $\Phi_{ring}$ for a critical radius $r_c = 1.1d$

| $\theta_c$ | $\Phi_{chain}$ | $\Phi_{ring}$ |
|---|---|---|
| 27° | 0 | 0 |
| 28° | $2.89 \times 10^{-5}$ | 0 |
| 29° | $5.91 \times 10^{-4}$ | 0 |
| 30° | $2.82 \times 10^{-3}$ | 0 |
| 31° | $7.42 \times 10^{-3}$ | $5.10 \times 10^{-8}$ |
| 32° | $1.44 \times 10^{-2}$ | $2.41 \times 10^{-6}$ |
| 33° | $2.35 \times 10^{-2}$ | $1.79 \times 10^{-5}$ |
| 34° | $3.45 \times 10^{-2}$ | $6.20 \times 10^{-5}$ |
| 35° | $4.70 \times 10^{-2}$ | $1.51 \times 10^{-4}$ |
| 40° | 0.123 | $1.52 \times 10^{-3}$ |
| 45° | 0.207 | $4.37 \times 10^{-3}$ |
| 50° | 0.285 | $8.03 \times 10^{-3}$ |
| 55° | 0.355 | $1.18 \times 10^{-2}$ |

For each critical angle $\Phi_{ring} << \Phi_{chain}$; this shows that the entropic penalty for association into a triatomic ring is greater than association into a triatomic chain.

In the derivation of the theory it has been assumed that the association sites can bond a maximum of twice. This is not a valid assumption for large patch sizes. For instance, for $r_c = d$ it is possible for a patch to associate at most once for $0^{\circ} \leq \theta_c < 30^{\circ}$, twice for $30^{\circ} \leq \theta_c < 35.3^{\circ}$, 3 times for $35.3^{\circ} \leq \theta_c < 45^{\circ}$ and 4 times for $45^{\circ} \leq \theta_c < 58.3^{\circ}$.[73] For most cases it should be expected that the theory will be accurate for $\theta_c < 40^{\circ}$. In what follows we will validate the new theory by comparison to new Monte Carlo simulation data.

## 2.3: Numerical results

In this section we compare theoretical predictions to "exact" Monte Carlo simulations for the self – assembly and thermodynamics of single patch colloids which interact with the potential given in section 2.1. All simulations are performed in the canonical ensemble (constant number of colloids *N*, volume *V* and temperature *T*) with a critical radius $r_c = 1.1d$. Here, *N* = 256 are used for each simulation. Simulations were carried out using standard methodology.[79]

We begin by considering the fraction of colloids which are bonded $k$ times $X_k$ given in the top panel of Fig. 2.3. We consider two packing (volume) fractions η = 0.2 and 0.4 and plot the results against the reduced association energy $\varepsilon^* = \varepsilon_{AA} / k_B T$ . For all calculations in Fig. 2.3, the critical angle is set to $\theta_c = 35^{\circ}$ so the probability of a patch bonding more than twice is not significant. The symbols are simulation data and solid curves give the predictions of the current theory. For comparison, the dashed curves give theoretical predictions using first order perturbation theory TPT1. At each η, increasing $\varepsilon^*$ increases the degree of association. At low $\varepsilon^*$ the fraction $X_1$ is larger than $X_2$ due to the fact that the entropic penalty of positioning and orienting two colloids for association is small compared to the penalty paid to position and orient multiple colloids into larger clusters. As $\varepsilon^*$ increases, the energetic benefit of forming multiple bonds overcomes this entropic penalty until $X_2$ dominates at large $\varepsilon^*$. This energetic – entropic tug of war results in a maximum in $X_1$ whose location increases with decreasing packing fraction. In each case the current theory and simulation are in excellent agreement; TPT1 fails to accurately predict $X_o$ and $X_1$ for $\varepsilon^* > 6$, and predicts $X_2 = 0$ for all cases.

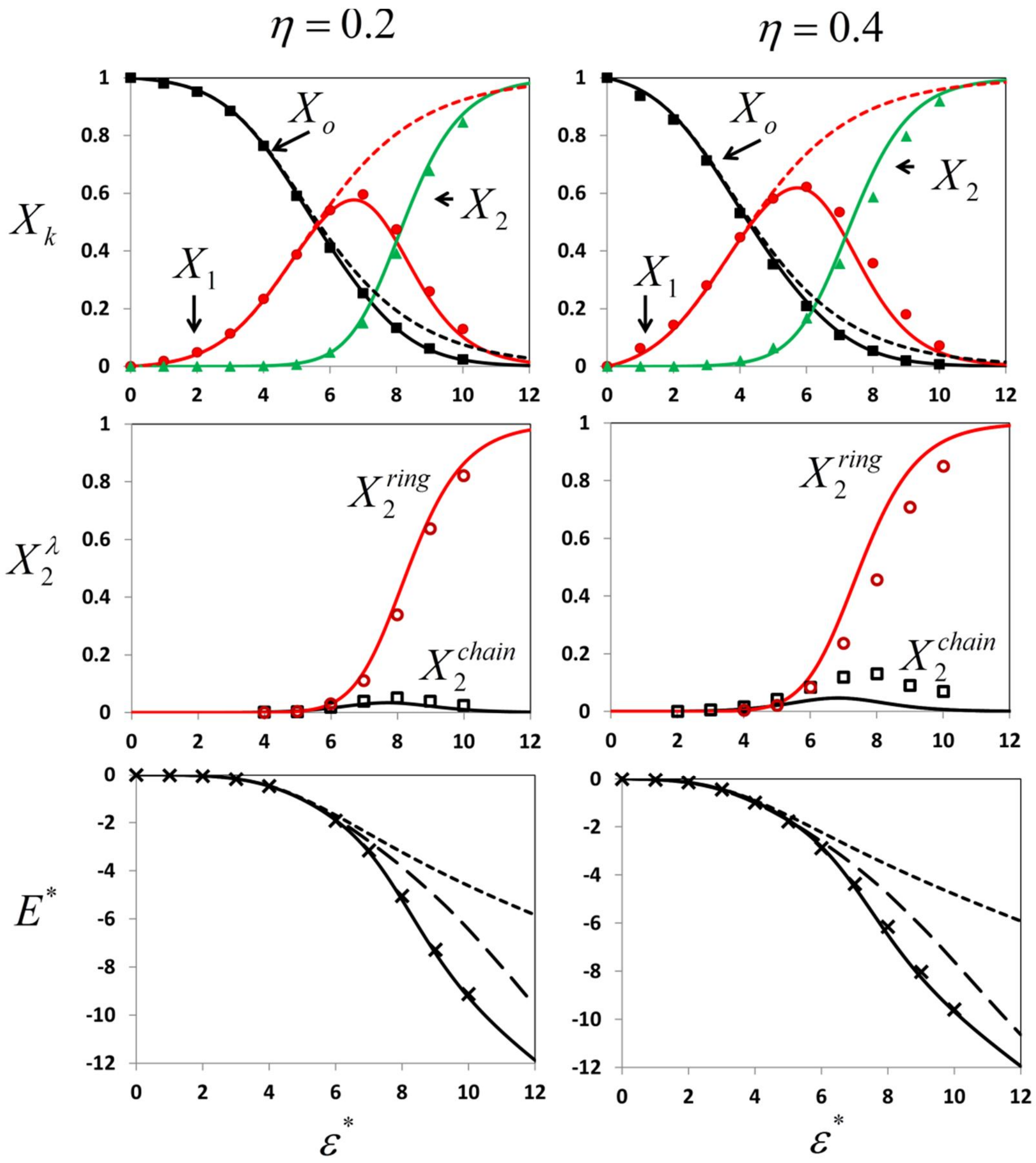

**Figure 2.3:** Left panel gives calculations for a packing fraction of 0.2 and the right for 0.4. The independent variable is reduced association energy. Critical angle is $\theta_c$ = 35° for all calculations. Top panel gives fractions bonded $k$ times. Symbols give simulation results, small dashed curves give TPT1 predictions and solid curves are predictions from the new theory. Center panel compares theory and simulation predictions for the fractions of colloids bonded twice in chains or rings. Bottom panel compares simulation and theory for the excess internal energy. Long dashed curve gives theoretical predictions using the method of Kalyuzhnyi et *al.*[73]

In the center panel of Fig. 2.3, theory and simulation predictions for the fraction of colloids bonded twice in rings $X_2^{ring} = \rho_2^{ring} / \rho$ and twice in chains $X_2^{chain} = \rho_2^{chain} / \rho$ are compared. At low $\varepsilon^*$ the fraction $X_2^{chain}$ dominates due to the fact that the entropic penalty of positioning and orienting three colloids to bond in a chain is less than that paid to form a triatomic ring; however, upon increasing $\varepsilon^*$ this entropic penalty is rapidly overtaken by the energetic benefit of forming the triatomic rings with higher density favoring ring formation. Overall the theory and simulation are in fair agreement. The theory is accurate for η = 0.2, but at η = 0.4 predicts values of $X_2^{ring}$ which are too large and predicts values of $X_2^{chain}$ which are too low; this is a result of the simple superposition approximations used for the correlation functions.

In the bottom panel of Fig. 2.3, simulation and theoretical predictions for the reduced excess internal energy per colloid $E^* = E_{AS} / Nk_B T$ are compared. Predictions using TPT1 (short dashed curve) and the approach of Kalyuzhnyi *et al.*[73] (long dashed curve) have also been included. As can be seen, TPT1significantly under predicts the magnitude of $E^*$ due to the fact that the possibility of two bonds per patch is not accounted for. The equation of state of Kalyuzhnyi *et al.*[73] (long dashed curve) is accurate for $\varepsilon^* < 5$, but underpredicts the magnitude of the energy at higher $\varepsilon^*$. The theory derived here (solid curve) is in excellent agreement with the simulation data over the full range of $\varepsilon^*$.

A key approximation made here is the neglect of all graphs which account for patches bonded more than twice. This is a rigorous graph cancelation for small $\theta_c$; however, for larger $\theta_c$ the neglect of these higher order graphs is an approximation whose accuracy will depend on $\varepsilon^*$, $\theta_c$ and η. In Fig. 2.4 we compare theoretical and simulation predictions of $E^*$ as a function of $\theta_c$ at a packing fraction of η = 0.3 and a moderate $\varepsilon^* = 5$; for comparison we have include predictions from TPT1. The current theory is in excellent agreement with simulation for $\theta_c \leq 40^\circ$, while for larger patch sizes the theory fails. The genesis of this failure is illustrated by the inset figure, which shows the rapid increase in the fraction of colloids bonded 3 times $X_3$ for $\theta_c > 40^\circ$. If the theory is to be extended to account for these higher order interactions additional graphs must be added to the graph sum $\Delta c^{(o)}$.

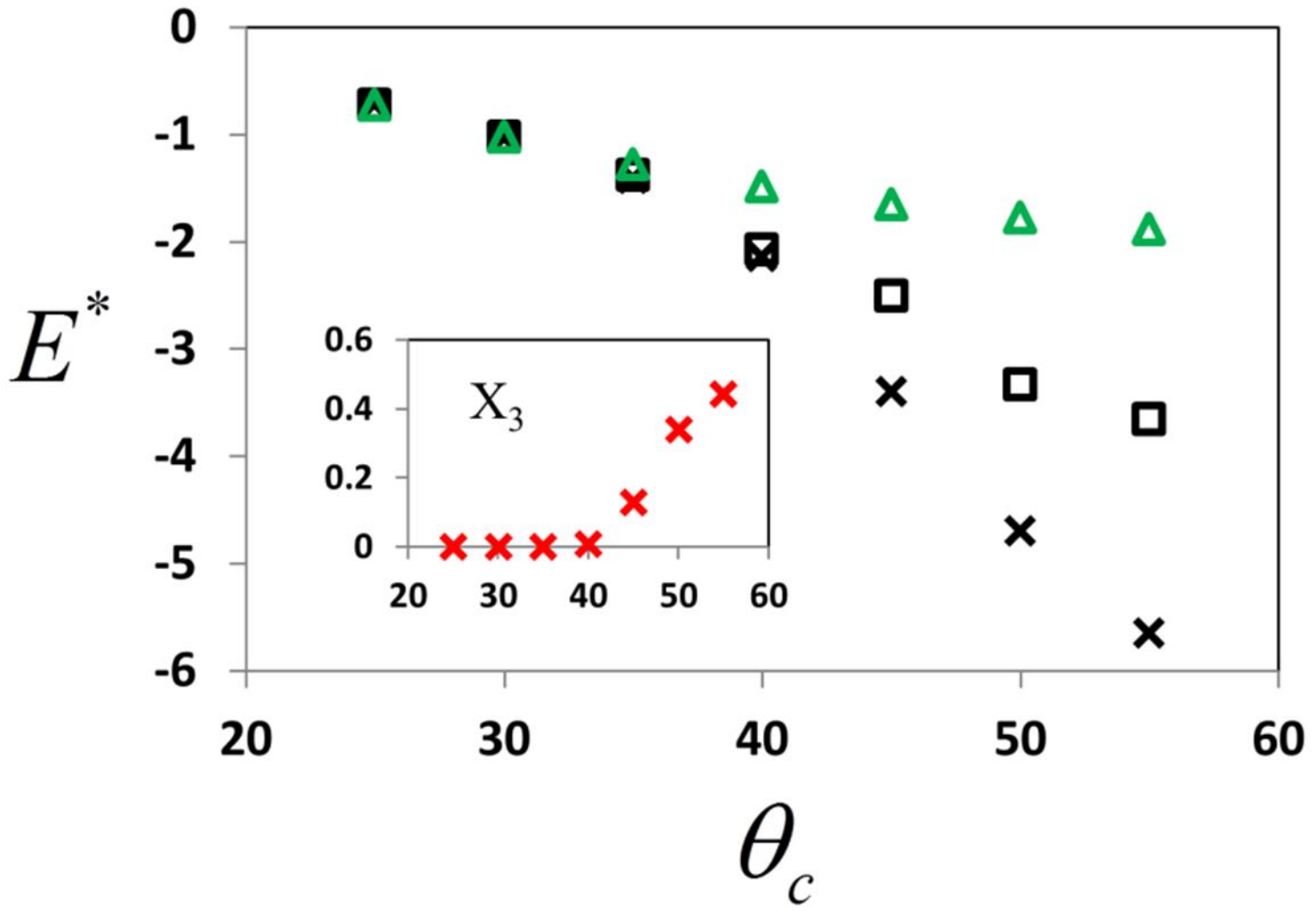

**Figure 2.4:** Comparison of simulation (crosses), new theory (squares) and TPT1 (triangles) predictions of the reduced excess internal energy versus critical angle at a packing fraction η = 0.3. Inset gives simulations for the fraction of colloids bonded 3 times versus critical angle. Association energy is set to $\varepsilon^* = 5$.

## 2.4: Summary and conclusions

In this chapter we extended TPT to account for the possibility that the association site can bond a maximum of twice. This case is particularly relevant as a primitive model for the self – assembly of patchy colloids. Both chain and ring formation were accounted for in a systematic way. The many body correlation functions which appeared in the graph sum were approximated in superposition such that the infinite sum of chain diagrams could be evaluated while still preserving the steric effects associated with doubly bonded association sites. The resulting theory was relatively simple and much more accurate than existing applicable theories. Only the single patch case was considered, however extension to the case of multiple patches should be possible. Wertheim’s two density formalism can account for higher order interactions (bonded more than twice), one simply needs to include the relevant graphs. However, proper accounting for all association probabilities when the association sites can bond more than 3 times would likely be prohibitively complex in a simple perturbation theory.

# CHAPTER 3

# Mixtures of single patch and spherically symmetric colloids

In Chapter 2 it was shown how TPT for one site associating fluids can be extended to account for association sites capable of forming two association bonds. In this chapter we turn our attention to a new type of binary mixture. In a recent paper[8], researchers synthesized mixtures of patchy *p* and spherically symmetric *s* colloids by binding DNA to the surfaces of the colloids. The *p* colloids had a single sticky patch terminated with type *A* single stranded DNA sticky ends and the *s* colloids were uniformly coated with DNA terminated with complementary type *B* single stranded DNA sticky ends. See Fig. 3.1 for an illustration. The DNA types were chosen such that there were *AB* attractions but no *AA* or *BB* attractions. That is, *s* colloids attract the anisotropic *p* colloids, but *s* colloids do not attract other *s* colloids and *p* colloids do not attract other *p* colloids. It was shown that this mixture would reversibly self - assemble into clusters where a single *s* colloid would be bonded to some number of *p* colloids into colloidal star molecules consisting of *n* arms (*n* patchy colloids). Of course *n* is not uniform and falls onto some distribution.

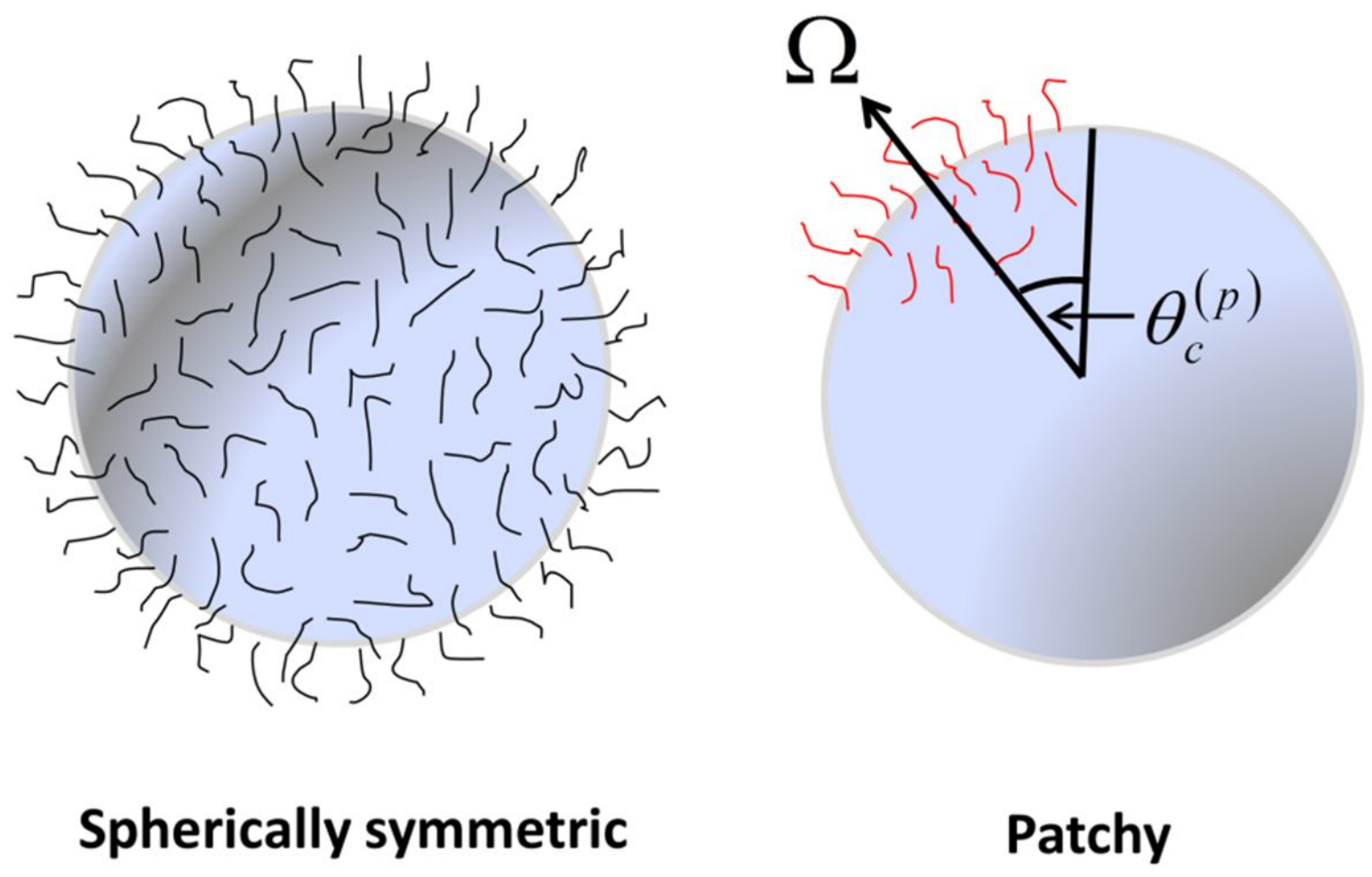

**Figure 3.1:** Diagram of spherically symmetric and patchy colloids. The orientation of the patchy colloid is defined by Ω and the size of the patch by the angle $\theta_c^{(p)}$

In this chapter we wish to derive a new theory to model this type of *s* and *p* colloid mixture. We will choose the patch size of the *p* colloid such that the single bonding condition holds; however, since the *s* colloid is a single uniform patch we must account for the fact that it can bond multiple times going well beyond the single bonding condition. Actually, the maximum number of bonds will be determined by the maximum number of *p* colloids which can be fit into the "bonding shell" of the *s* colloid. For equal sized *p* and *s* colloids this number must be a minimum of 12 which corresponds to the coordination number of hexagonal cubic closest packing of hard spheres. We will develop the one patch theory in Wertheim's two density formalism[33, 34] for one site associating fluids. To account for the fact that the *s* colloids can bond to multiple *p* colloids we must include the contribution for each type of associated cluster in $\Delta c^{(o)}$

. To quantitatively test the new theory we perform new Monte Carlo simulations to test the effect of temperature, density and composition on the internal energy, pressure, average solvation number of *s* colloids, *n* distribution of s colloids and fraction of *p* colloids bonded. The theory and simulation are found to be in excellent agreement.

## 3.1: Theory

In this section we derive the theory for a two component mixture of patchy colloids, denoted *p*, and spherically symmetric colloids, denoted *s*. We consider the case that both colloids have the same diameter *d*. The *p* colloid has a single attractive type *A* patch the size of which is determined by the critical angle $\theta_c^{(p)}$ which defines the solid angle of the patch as $2\pi\left(1-\cos\theta_c^{(p)}\right)$. The *s* colloid has a single large type *B* patch covering the entire surface of the colloid with a critical angle $\theta_c^{(s)} = 180^{\circ}$. We allow *AB* attractions, but no *AA* or *BB* attractions. This model draws inspiration from recent experiments where researchers synthesized mixtures of *s* and *p* colloids by binding DNA to the surfaces of the colloids.[8] A diagram of these colloids can be found in Fig. 3.1.

Since there are no attractions between *p* colloids, their potential of interaction is simply that of a hard sphere system $\phi^{(p,p)}(r_{12}) = \phi_{HS}(r_{12})$. Similarly, since there are no attractions between *s* colloids their potential of interaction is also $\phi^{(s,s)}(r_{12}) = \phi_{HS}(r_{12})$. The potential of interaction between *s* and *p* colloids contains a hard sphere contribution and an attractive association contribution $\phi^{(s,p)}(r_{12}) = \phi_{HS}(r_{12}) + \phi_{as}^{(s,p)}(12)$. Here we follow a similar approach to that presented in chapter 2.1 with the *p* colloids treated as a hard sphere with a conical square well association site and the *s* colloid with a spherically symmetric square well association site giving the association potential

$$\phi_{as}^{(s,p)}(12)=\begin{cases}-\varepsilon_{AB}, & r_{12}\le r_c \text{ and } \theta_{\mathrm{A}}\le\theta_c^{(p)}\\ 0 & \text{otherwise}\end{cases} \tag{3.1}$$

which states that if colloids 1 and 2 are within a distance $r_c$ of each other, and the *p* colloid is oriented such that the angle between the site orientation vector and the vector connecting the two segments $\theta_{\mathrm{A}}$ is less than the critical angle $\theta_c^{(p)}$, the two colloids are considered bonded and the energy of the system is decreased by a factor $\varepsilon_{AB}$. While simple in form, the parameters of this potential can be related back to the properties of the grafted single strand DNA. For instance, the critical radius $r_c$ will depend on the persistence length of the grafted DNA as well as solvent conditions, and the square well depth $\varepsilon_{AB}$ will depend on both DNA sequences and grafting densities among other things.[80]

The free energy of a mixture of associating molecules with one association site each (assuming a homogeneous fluid) is a simple generalization of Eq. (1.22)

$$\frac{A-A_{HS}}{Vk_BT}=\sum_k\left(\rho^{(k)}\ln\frac{\rho_o^{(k)}}{\rho^{(k)}}-\rho_o^{(k)}+\rho^{(k)}\right)-\Delta c^{(o)}/V \tag{3.2}$$

For the current case the *s* colloid is a single spherical association site which can clearly not be modeled in the single bonding condition. The maximum number of bonds is simply the maximum number of *p* colloids $n^{\max}$ which can pack in the *s* colloid's bonding shell $d\le r\le r_c$. To account for all association possibilities we will have to include contributions for each association possibility explicitly (one *s* colloid with one *p* colloid, two *p* colloids, three *p* colloids etc…). To accomplish this we decompose $\Delta c^{(o)}$ as

$$\Delta c^{(o)} \approx \Delta \hat{c}_1^{(o)} = \sum_{n=1}^{n^{\max}} \Delta c_n^{(o)} \tag{3.3}$$

where $\Delta c_n^{(o)}$ is the contribution for $n$ patchy colloids bonded to a $s$ colloid. Equation (3.3) is depicted in Fig. 3.2.

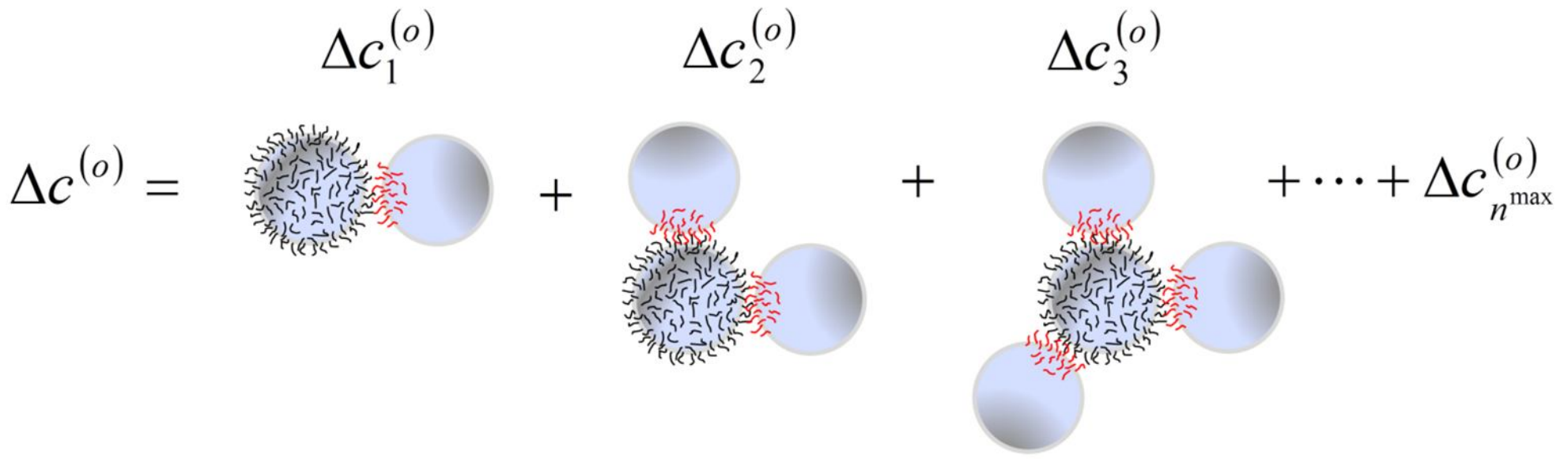

**Figure 3.2:** Diagram of contributions to the graph sum $\Delta c^{(o)}$

To evaluate $\Delta c_n^{(o)}$ in perturbation theory, we consider all graphs consisting of a single associated cluster with $n$ patchy colloids bonded to a $s$ colloid with $n$ association bonds. We also assume that the patch size on the $p$ colloid is such that double bonding between one $p$ colloid and multiple $s$ colloids cannot occur. For each $n$ this results in an infinite sum of graphs consisting of this associated cluster interacting with the hard sphere reference fluid which can be summed to yield

$$\Delta c_n^{(o)} = \frac{\rho_o^{(s)} \left(\rho_o^{(p)}\right)^n}{\Omega^{n+1} n!} \int d(1)...d(n+1) \prod_{k=2}^{n+1} \left(f_{as}^{(s,p)}(1,k)\right) g_{HS}(1...n+1) \tag{3.4}$$

where $f_{as}^{(s,p)}(12) = \exp\left(-\phi_{as}^{(s,p)}(12)/k_B T\right) - 1$ is the association Mayer function, $\Omega = 4\pi$ is the total number of orientations and $g_{HS}(1...n+1)$ is the $n+1$body hard sphere reference correlation function. In Eq. (3.4) the $s$ colloid is labeled (1).

To evaluate the reference system correlation function $g_{HS}(1...n+1)$ we first use the definition of the hard sphere cavity correlation function

$$g_{HS}(1...n+1) = y_{HS}(1...n+1) \prod_{\substack{all\ pairs \\ \{l,k\}}} e_{HS}(r_{lk}) \tag{3.5}$$

Where $e_{HS}(r) = \exp(-\phi_{HS}(r)/k_B T)$ are the reference system *e* bonds which serve to prevent hard sphere overlap in the cluster. We note from Fig. 3.2 that the associated clusters of *n* patchy colloids are simply star molecules with *n* arms of length 1. In Wertheim's perturbation theory, in the limit of infinitely strong association, the change in free energy in going from a mixture of hard spheres to a polyatomic star molecule consisting of *m* hard spheres bonded at contact can be written as[81, 82]

$$\frac{A_{star}}{N_m k_B T} = -\ln y_{HS}^{(m)} \tag{3.6}$$

where $y_{HS}^{(m)}$ is the *m* body cavity correlation function averaged over the states of the star molecule and $N_m$ is the number of star molecules. While the correlation function $y_{HS}^{(m)}$ is not known in general we can approximate it as follows. Recently Marshall and Chapman[83] obtained an approximate general branched chain solution to the free energy (for spheres bonded at contact) in second order perturbation theory[37]. For our current problem, the cavity function $y_{HS}(1...n+1)$ can be replaced by the average $y_{HS}^{(n+1)}$ (mean value theorem) which we obtain by solution of Eqn. (3.6) and the second order result of Marshall and Chapman[83]. The result is

$$y_{HS}(1...n+1) = y_{HS}^{(n+1)} \approx y_{HS}(d)^n \delta^{(n)} \tag{3.7}$$

where

$$\delta^{(n)} = \begin{cases} (1+4\lambda)^{\frac{n-3}{2}} \left( \dfrac{1+\sqrt{1+4\lambda}}{2} \right)^3 & for \quad n>1 \\ 1 \quad for \quad n=1 \end{cases} \tag{3.8}$$

The term $\lambda$ is quadratic in density and given as $\lambda = 0.2336\eta + 0.1067\eta^2$. The packing fraction is given by $\eta = \pi\left(\rho^{(s)} + \rho^{(p)}\right)d^3/6$.

With the association potential between *p* colloids and *s* colloids given by Eq. (3.1), Eq. (3.4) can now be simplified as

$$\Delta c_n^{(o)}/V = \frac{1}{n!}\rho_o^{(s)}\Delta^n \delta^{(n)}\Xi^{(n)} \tag{3.9}$$

The term $\Delta$ is given by

$$\Delta = f_{as}^{(s,p)}\rho_o^{(p)} P y_{HS}(d) \tag{3.10}$$

where $f_{as}^{(s,p)} = \exp(\varepsilon_{AB}/k_B T) - 1$ is the magnitude of the association Mayer function, $P = \sqrt{\kappa} = \left(1 - \cos\theta_c^{(p)}\right)/2$ is the fractional patch coverage of the *p* colloid. In Eq. (3.9) the integral $\Xi^{(n)}$ is given by

$$\Xi^{(n)} = \prod_{k=2}^{n+1}\left(\int_0^{2\pi}\int_{-1}^{1}\int_d^{r_c} d\phi_{1,k}\, d\cos\beta_{1,k}\, dr_{1,k}\, r_{1,k}^2\right)\prod_{\substack{all\ pairs \\ \{l,k\}}} e_{HS}(r_{lk}) \tag{3.11}$$

Equation (3.11) is simply a single cluster partition function for a cluster of 1 spherical colloid and *n* patchy colloids. Here $\phi_{1,k}$, $\beta_{1,k}$ and $r_{1,k}$ define the azimuthal angle, polar angle and radial distance of *p* colloid *k* in a spherical coordinate system centered on the *s* colloid 1. We discuss the evaluation of Eq. (3.11) in detail in the section 3.2.

Now that $\Delta c^{(o)}$ has been completely specified we can minimize the free energy with respect to monomer densities to obtain the following mass action equations

$$\rho^{(s)} = \rho_o^{(s)} + \rho_o^{(s)}\sum_{n=1}^{n^{\max}}\frac{1}{n!}\Delta^n \delta^{(n)}\Xi^{(n)} \tag{3.12}$$

$$\rho^{(p)} = \rho_o^{(p)} + \rho_o^{(s)} \sum_{n=1}^{n^{\max}} \frac{n}{n!} \Delta^n \delta^{(n)} \Xi^{(n)} \tag{3.13}$$

It is clear that the second term on the right hand side of these two equations is the density of bonded colloids $\rho_b^{(k)}$. For the *s* colloids it is convenient to introduce the densities $\rho_n^{(s)}$, which represent the density of *s*pherically symmetric colloids which are bonded to *n* patchy colloids. By conservation these densities must satisfy the relation

$$\rho^{(s)} = \sum_{n=0}^{n^{\max}} \rho_n^{(s)} \tag{3.14}$$

Comparing Eqns. (3.12) and (3.14) we can deduce the relation for $n > 0$

$$\rho_n^{(s)} = \frac{\rho_o^{(s)}}{n!} \Delta^n \delta^{(n)} \Xi^{(n)} \qquad n > 0 \tag{3.15}$$

Introducing the fractions $X_n^{(k)} = \rho_n^{(k)} / \rho^{(k)}$ we can write the average cluster size $\bar{n}$ (average number of *p* colloids bonded to an *s* colloid) as

$$\bar{n} = \sum_{n=0}^{n^{\max}} n X_n^{(s)} \tag{3.16}$$

Equations (3.12) and (3.13) are now simplified as

$$\frac{1}{X_o^{(s)}} = 1 + \sum_{n=1}^{n^{\max}} \frac{1}{n!} \Delta^n \delta^{(n)} \Xi^{(n)} \tag{3.17}$$

and

$$X_o^{(p)} = 1 - \frac{x^{(s)}}{\left(1 - x^{(s)}\right)} \frac{\sum_{n=1}^{n^{\max}} \frac{n}{n!} \Delta^n \delta^{(n)} \Xi^{(n)}}{1 + \sum_{n=1}^{n^{\max}} \frac{1}{n!} \Delta^n \delta^{(n)} \Xi^{(n)}} \tag{3.18}$$

where $x^{(s)}$ is the mole fraction of *s* colloids. To obtain the monomer fractions, Eq. (3.18) is first solved numerically for $X_o^{(p)}$ and then Eq. (3.17) can be used to evaluate $X_o^{(s)}$.

Combining these results we simplify the free energy to the following form

$$\frac{A - A_{HS}}{Nk_BT} = x^{(s)} \ln X_o^{(s)} + \left(1 - x^{(s)}\right)\left(\ln X_o^{(p)} - X_o^{(p)} + 1\right) \tag{3.19}$$

where *N* is the total number of colloids. Equations (3.17) – (3.19) give the fundamental equations for the theory of single patch colloids interacting with spherically symmetric colloids. In the following section we discuss the evaluation of the cluster partition functions $\Xi^{(n)}$

## 3.2: Evaluation of cluster partition functions

In this section we evaluate the cluster partition functions $\Xi^{(n)}$ given by Eq. (3.11). For $n = 1$ we solve the integral analytically to obtain

$$\Xi^{(1)} = \frac{4}{3}\pi\left(r_c^3 - d^3\right) = v_b \tag{3.20}$$

where $v_b$ is the bonding volume. We can also obtain an analytical solution for $n = 2$

$$\Xi^{(2)} = \frac{v_b^2}{2} + \pi^2\left(r_c^3 - d^2 r_c\right)^2 \tag{3.21}$$

The remaining $\Xi^{(n)}$ are evaluated numerically using Monte Carlo integration[79] as

$$\Xi^{(n)} = v_b^n P^{(n)} \tag{3.22}$$

The term $P^{(n)}$ is the probability that if we randomly generate *n* patchy colloids in the bonding shell of the *s* colloid that there is no hard sphere overlap. For the case $r_c = 1.1d$, Eq. (3.11) was evaluated using this Monte Carlo integration routine for cluster sizes $1 \le n \le 9$. However, for $n \ge 9$ this method becomes very inefficient due to the low probability of generating this many hard spheres in the bonding shell of the *s* colloid and there being no overlap. A much more efficient method to evaluate $P^{(n)}$ in this case is the following

$$P^{(n)} = P_{insert}^{(n)} P^{(n-1)} \tag{3.23}$$

Equation (3.23) states that the probability that *n* randomly generated *p* colloids will not have any hard sphere overlap, is simply the probability that *n* – 1 randomly generated *p* colloids do not overlap multiplied by an insertion probability $P_{insert}^{(n)}$. This insertion probability is simply the probability that a randomly generated *p* colloid in the bonding shell of the *s* colloid, with *n* – 1 non-overlapping *p* colloids already in place, will not overlap with any of the existing *n* – 1 patchy colloids. Mathematically the insertion probability is given by

$$P_{insert}^{(n)} = \left\langle \prod_{j=1}^{n-1} e_{HS}\left(r_{j,n}\right) \right\rangle_{n-1} \tag{3.24}$$

where $\langle \ \rangle_{n-1}$ represents an ensemble average in a cluster of *n* – 1 non-overlapping *p* colloids in the bonding shell of a *s* colloid. This is similar to Widom test particle insertions.[84] Solving Eq. (3.23) recursively we obtain

$$P^{(n)} = \prod_{k=1}^{n} P_{insert}^{(k)} \tag{3.25}$$

Equation (3.24) was evaluated using standard Monte Carlo simulation techniques.[79] In total $10^8$ – $10^9$ configurations were generated with $10^8$ – $10^9$ insertions used to evaluate the average in Eq. (3.24). Numerical calculations for the probabilities $P_{insert}^{(n)}$ and $P^{(n)}$ can be found in Table 3.1 for the case $r_c = 1.1d$. As expected, increasing *n* results in a decrease in both $P_{insert}^{(n)}$ and $P^{(n)}$. For this case, the maximum number of *p* colloids for which we obtained a non-zero insertion probability was found to be $n^{\max} = 13$. This had a much lower probability than the case $n$ = 12, which corresponds to hexagonal cubic closest packing if the colloids were restricted to bond at contact.

**Table 3.1:** Insertion probabilities $P_{insert}^{(n)}$ and generation probabilities $P^{(n)}$ calculated for $r_c = 1.1d$

| $n$ | $P_{insert}^{(n)}$ | $P^{(n)}$ |
|---|---|---|
| 1 | 1 | 1 |
| 2 | 0.774 | 0.774 |
| 3 | 0.573 | 0.444 |
| 4 | 0.401 | 0.178 |
| 5 | 0.261 | 0.0463 |
| 6 | 0.153 | $7.11\times10^{-3}$ |
| 7 | 0.0790 | $5.61\times10^{-4}$ |
| 8 | 0.0340 | $1.91\times10^{-5}$ |
| 9 | 0.0111 | $2.12\times10^{-7}$ |
| 10 | $6.97\times10^{-3}$ | $1.48\times10^{-9}$ |
| 11 | $3.37\times10^{-3}$ | $4.98\times10^{-12}$ |
| 12 | $2.23\times10^{-4}$ | $1.11\times10^{-15}$ |
| 13 | $\sim10^{-9}$ | $\sim10^{-24}$ |

## 3.3: Simulation methodology

As a test of the theory, we perform new Monte Carlo simulations (not to be confused with the simulations discussed in section 3.2) for the case of a mixture of $s$ and $p$ type colloids which are hard spheres with an additional attractive potential given by Eq. (3.1). Unless otherwise stated, we use the potential parameters $r_c = 1.1d$ and $\theta_c^{(p)} = 27^\circ$ such that only single bonding of a $p$ colloid will occur. Constant $NVT$ (number of colloids, volume, temperature) simulations were performed using standard methodology.[79] Each $NVT$ simulation was allowed to equilibrate for $10^8$ – $10^9$ trial moves and averages where taken for another $10^8$ – $10^9$ trial moves. A trial move consists of an attempted relocation of a $s$ colloid or an attempted relocation and reorientation of a $p$ colloid. For each simulation we used a total of $N$ = 864 colloids. Constant $NPT$ (number of colloids, pressure, temperature) simulations were performed in the same manner as the $NVT$ simulations with the addition of an attempted volume change each $N$ trial moves.

The simulations equilibrated relatively quickly due to the fact that the only attractions were between $p$ and $s$ colloids, meaning there were no extended associated clusters. A simulation was considered to be equilibrated once the fraction of accepted moves, average of the internal energy and averages of the bonding fractions stabilized as the simulation progressed. Even when the $s$ colloids were bonded to a large number of $p$ colloids, reasonable displacement parameters could be used due to the critical radius being $r_c = 1.1d$ and the fact that reorientation of the $s$ colloids was not required.

## 3.4: Numerical results

Now we use the theory derived in sections 3.1 and 3.2 to study the self – assembly of mixtures of $s$ and $p$ colloids. At each point we will also compare to Monte Carlo simulation results to validate the new theory. We begin with a discussion of the dependence on $s$ colloid mole fraction $x^{(s)}$ when association energy (inverse temperature) $\varepsilon^* = 1/T^* = \varepsilon_{AB} / k_B T$ and density $\rho^* = \rho d^3$ are both held constant. Comparison of theory and simulation predictions of the excess internal energy $E^* = E_{AS} / N k_B T$ , average number of bonds per $s$ type colloid $\bar{n}$ and fraction of patchy colloids bonded $X_1^{(p)} = \rho_b^{(p)} / \rho^{(p)}$ can be found in Fig. 3.3, for both low $\rho^* = 0.2$ (packing fraction of $\eta \sim 0.105$) and high $\rho^* = 0.7$ ($\eta \sim 0.366$) density cases. Here the association energy is set to $\varepsilon^* = 7$.

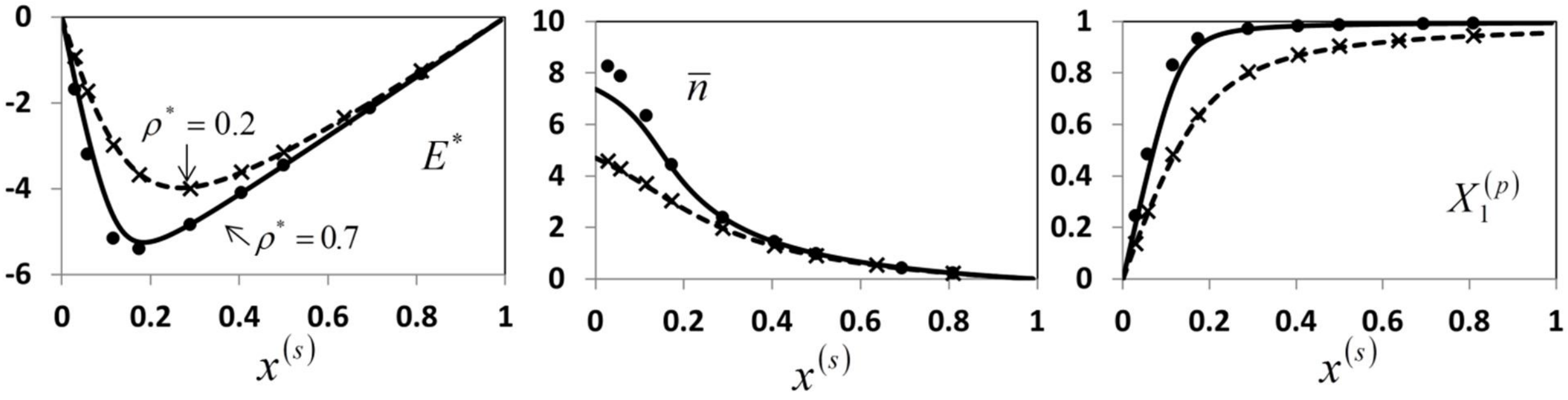

**Figure 3.3:** Excess internal energy $E^*$ (left), average number of bonds per *s* colloid $\bar{n}$ (middle) and fraction of patchy colloids bonded (right) versus mole fraction of *s* type colloids $x^{(s)}$ at an association energy $\varepsilon^* = 7$ and densities $\rho^* = 0.2$(dashed curve - theory, crosses - simulation) and $\rho^* = 0.7$ (solid curve – theory, circles - simulation)

For each case, $\bar{n}$ increases with decreasing $x^{(s)}$ reaching a maximum when $x^{(s)} \to 0$. The reasoning behind this is simple, when $x^{(s)}$ is small there is an abundance of *p* colloids available to "solvate" the *s* colloids. As $x^{(s)}$ is increased, $\bar{n}$ decreases because there are less *p* colloids available for association due to a decreased fraction of *p* colloids and competition with other *s* colloids. Increasing density increases $\bar{n}$ as expected. The theory is in excellent agreement with simulation for $\rho^* = 0.2$ over the full range of $x^{(s)}$ and is in good agreement for the higher density case $\rho^* = 0.7$ over much of the $x^{(s)}$ range; however, at this higher density the accuracy of the theory decreases somewhat for $x^{(s)} \to 0$. It is in this limit that $\bar{n}$ is a maximum and the high order contributions $\Delta c_7^{(o)}$, $\Delta c_8^{(o)}$ etc… come into play. To evaluate these graphs we approximated the many body correlation functions $y_{HS}(1....n+1)$ by the approximation given by Eq. (3.7). We expect this to be most accurate at low densities and less accurate at higher densities. For this reason the accuracy of the theory decreases for $x^{(s)} \to 0$ and high density. That said, the overall agreement between theory and simulation is very good.

The fraction $X_1^{(p)}$ shows the opposite $x^{(s)}$ dependence as compared to $\bar{n}$. $X_1^{(p)}$ is a maximum for $x^{(s)} \rightarrow 1$ when there are an abundance of *s* colloids available for association and a minimum for $x^{(s)} \rightarrow 0$ when there are few *s* colloids available for association. The $x^{(s)}$ dependence of $E^*$ is more interesting. For small $x^{(s)}$, increasing the fraction of *s* colloids increases association due to the fact that the system is *s* colloid limited. This results in a more negative $E^*$. For large $x^{(s)}$, increasing the fraction of *s* colloids decreases association due to the fact that the system is now *p* colloid limited. This results in a less negative $E^*$. Near $x^{(s)} \sim 0.2$ association is a maximum resulting in a clear minimum in $E^*$ at both densities. We see that the minimum in $E^*$ shifts to lower $x^{(s)}$ as density is increased. This is due to the increase in $\bar{n}$ as density is increased. Theory and simulation are in excellent agreement.

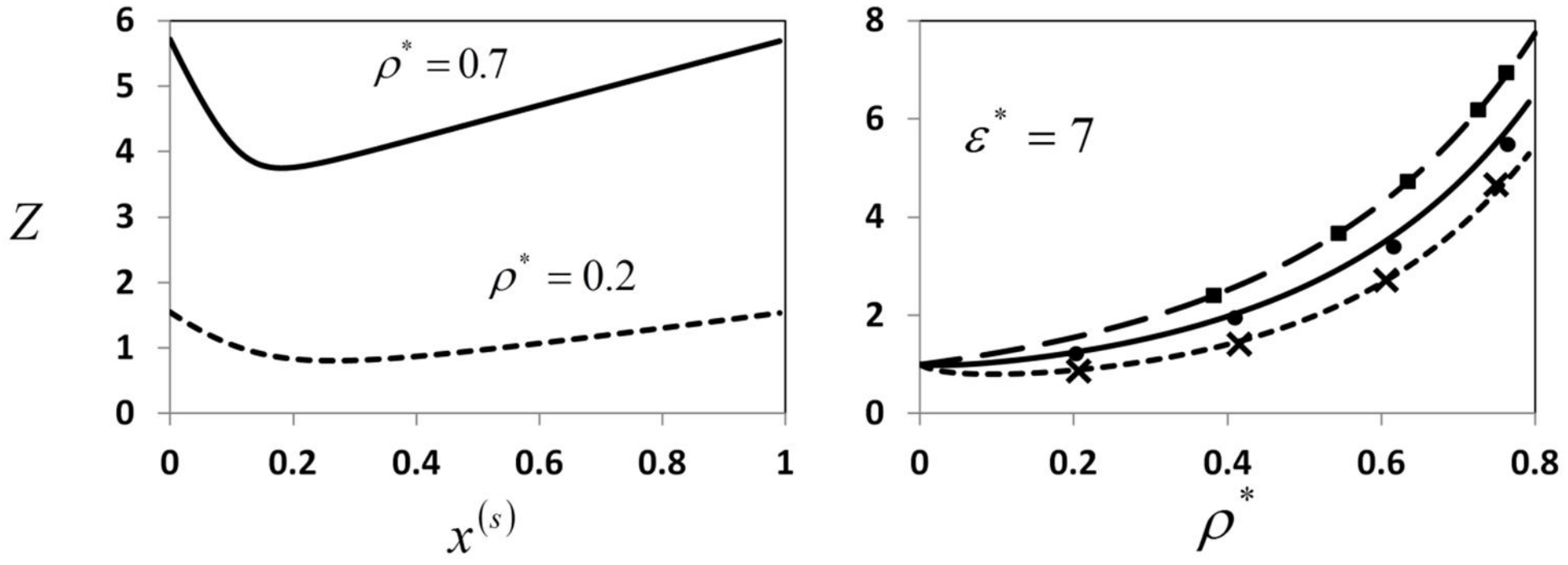

**Figure 3.4:** Left: Compressibility factor Z versus mole fraction of *s* type colloids $x^{(s)}$ at an association energy $\varepsilon^* = 7$. Right: Z versus density $\rho^*$ at $\varepsilon^* = 7$ for $x^{(s)} = 0$ (long dashed curve - theory, squares – simulation), $x^{(s)} = 0.0579$ (solid curve - theory, circles - simulation) and $x^{(s)} = 0.174$ (short dashed curve - theory, crosses simulation)

The left panel of Fig. 3.4 shows calculations for the compressibility factor *Z* at the same conditions shown in Fig. 3.3. At $x^{(s)} = 0$ there is no association in the system and *Z* is that of a hard sphere fluid. Increasing $x^{(s)}$, when $x^{(s)}$ is small, results in a decrease in *Z* as *s* and *p* type colloids associate into larger clusters. The opposite is true for large $x^{(s)}$ where increasing $x^{(s)}$ decreases association and results in an increase in *Z*. Like the internal energy $E^*$, *Z* shows a distinct minimum due to these competing effects near $x^{(s)} \sim 0.2$ when association is maximized in the system. In the right panel of Fig. 3.4 we compare theory to *NPT* simulations at an association energy $\varepsilon^* = 7$ and mole fractions $x^{(s)} = 0$, 0.0579 and 0.174. The agreement between theory and simulation is good, although the theory slightly overpredicts *Z* at high density for the case $x^{(s)} = 0.0579$. This is due to the underprediction of $\overline{n}$ in this regime, Fig. 3.3.

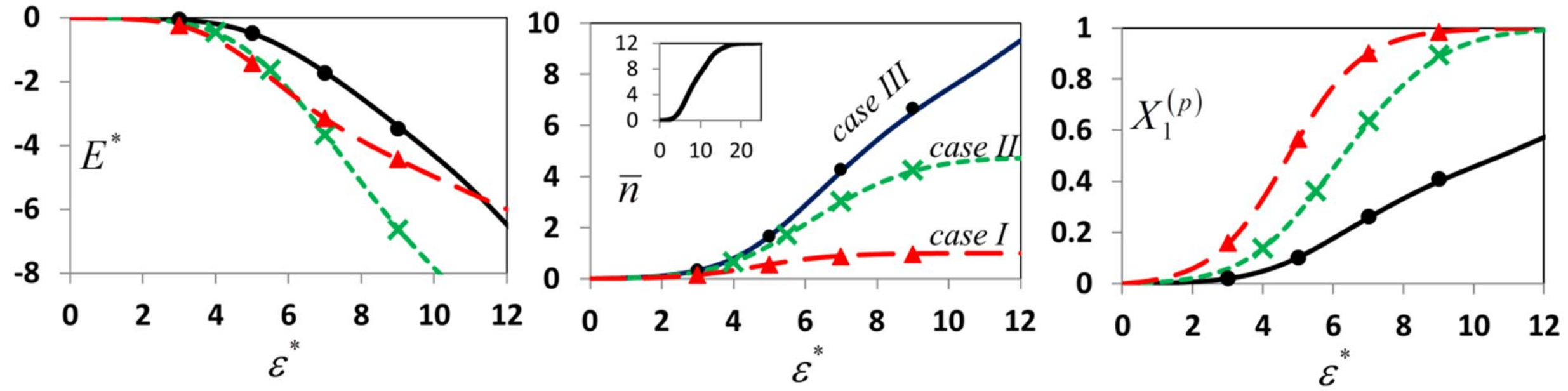

**Figure 3.5:** Excess internal energy $E^*$ (left), average number of bonds per *s* colloid $\overline{n}$ (middle) and fraction of *p* colloids bonded (right) versus association energy $\varepsilon^*$ at a density $\rho^*$ = 0.2 and $x^{(s)}$ = 0.0579 (solid curve - theory, circles - simulation), $x^{(s)}$ = 0.174 (short dashed curve - theory, crosses - simulation) and $x^{(s)}$ = 0.5 (long dashed curve - theory, triangles - simulation). Inset in middle panel gives $\overline{n}$ versus $\varepsilon^*$ for case III and large $\varepsilon^*$

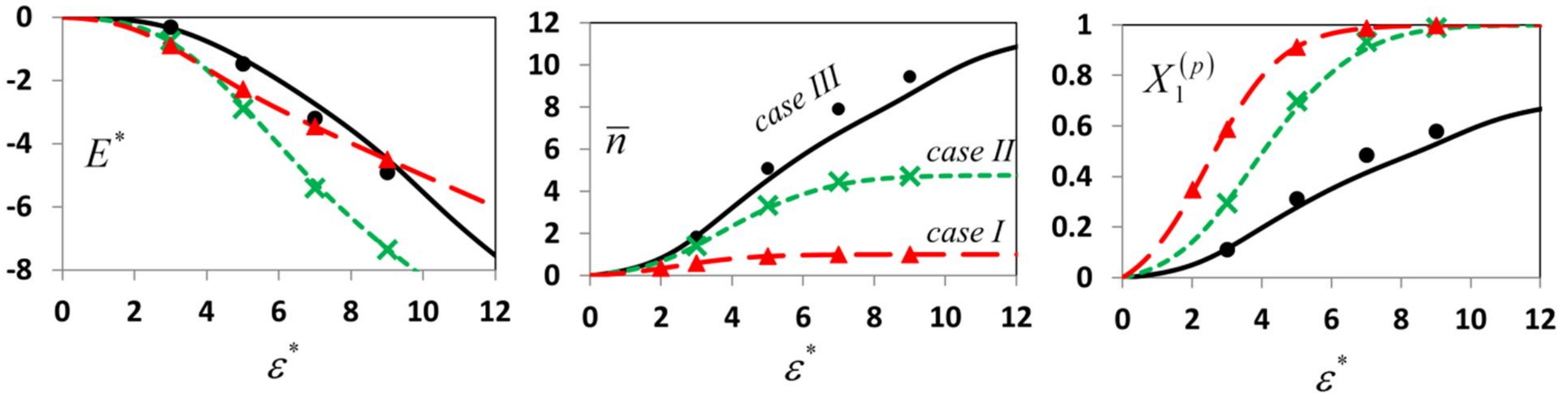

**Figure 3.6:** Same as Fig. 3.5 with $\rho^*$ = 0.7

Now we will specifically consider the effect of association energy (inverse temperature) at constant density and composition. These results can be found in Figs. 3.5 and 3.6 for $\rho^* = 0.2$ and $\rho^* = 0.7$ respectively. At each density we perform calculations for $x^{(s)} = 0.5$ (case I), 0.174 (case II) and 0.0579 (case III). We begin our discussion of these figures with case I. For case I there is an equal number of *s* and *p* colloids, so we should expect the relation $\bar{n} = X_1^{(p)}$ to hold exactly for each $\varepsilon^*$; this is observed in both theory and simulation. When there is an abundance of *s* colloids, as in case I, the entropic penalty of association is quite low. For this reason we see a rapid increase in $X_1^{(p)}$ even at low $\varepsilon^*$. This results in case I having the lowest $E^*$ for the given $x^{(s)}$ when $\varepsilon^*$ is low. Of course, for this case, when $\varepsilon^*$ is high $\bar{n}$ must reach a limiting value of 1 due to simple stoichiometry. Decreasing the mole fraction to $x^{(s)} = 0.174$ (case II) the fraction $X_1^{(p)}$ now increases more slowly with $\varepsilon^*$ due to the fact that there are less *s* colloids available for association as compared to case I. For large $\varepsilon^*$, $\bar{n}$ reaches a limiting value of 4.76 which is simply the ratio of *p* colloids to *s* colloids. Of the three cases, case II has the most negative $E^*$ for strong association energies. In case III the ratio of *p* to *s* colloids is 814/50 = 16.28 which is greater than $n^{\max}$. This means that there is no stoichiometric limit to $\bar{n}$. The inset in the middle panel of Fig. 3.5 shows $\bar{n}$ levels out at $\bar{n} = 12$ in strongly associating systems; although, as $\varepsilon^*$ is increased further contributions from n = 13 will be become significant. Also, $X_1^{(p)}$ does not approach 1 as in cases I and II. Comparing the low and high density cases (Figs. 3.5 and 3.6) we see that the results are qualitatively similar with association being enhanced for $\rho^* = 0.7$. The theory and simulation are in good agreement; however, for case III at $\rho^* = 0.7$ there is some error for the same reasons discussed previously.

Until now, for convenience, we have only considered the average number of bonds per *s* colloid (solvation number) $\bar{n}$. However, the mole fraction of *s* colloids in each cluster of *n* patchy colloids $X_n^{(s)}$ is easily calculated through Eq. (3.15). We show these fractions for the cases I – III, discussed above, in Fig. 3.7 at a density of $\rho^* = 0.2$ and association energies $\varepsilon^* = 6$ and 12.

Simulations were performed for $\varepsilon^* = 6$ and are represented by the star symbols. First we will focus on the right panel, case III. As discussed above, for this case there are enough *p* colloids for the *s* colloids to become fully bonded. At an association energy of $\varepsilon^* = 6$, the average number of bonds per *s* colloid is $\bar{n} = 2.89$, and the distribution of $X_n^{(s)}$ is more or less Gaussian with significant contributions ranging between *n* = 0 and *n* = 6. Increasing the association energy to $\varepsilon^* = 12$ the shape of the distribution is similar to the $\varepsilon^* = 6$ case with $\bar{n}$ shifted to 9.32 with non-negligible contributions as high as *n* = 12. For case II, with a mole fraction $x^{(s)} = 0.174$, the $\bar{n}$'s shift to lower $n, \bar{n} = 2.14$ for $\varepsilon^* = 6$ and $\bar{n}$ = 4.72 for $\varepsilon^*$ = 12, due to the fact that the *p* colloids are now limiting with a maximum possible average solvation number of 4.76. Even with the low $\bar{n} = 2.14$ for $\varepsilon^* = 6$ there are still significant contributions for clusters with *n* = 5 patchy colloids. Lastly, for case I, there are an equal number of *p* and *s* colloids rendering a maximum average solvation number of 1. For this case the distributions are asymmetric with the majority of *s* colloids being in clusters $0 \le n \le 3$ with clusters of larger *n* contributing to a lesser degree. Theory and simulation are in excellent agreement for each case.

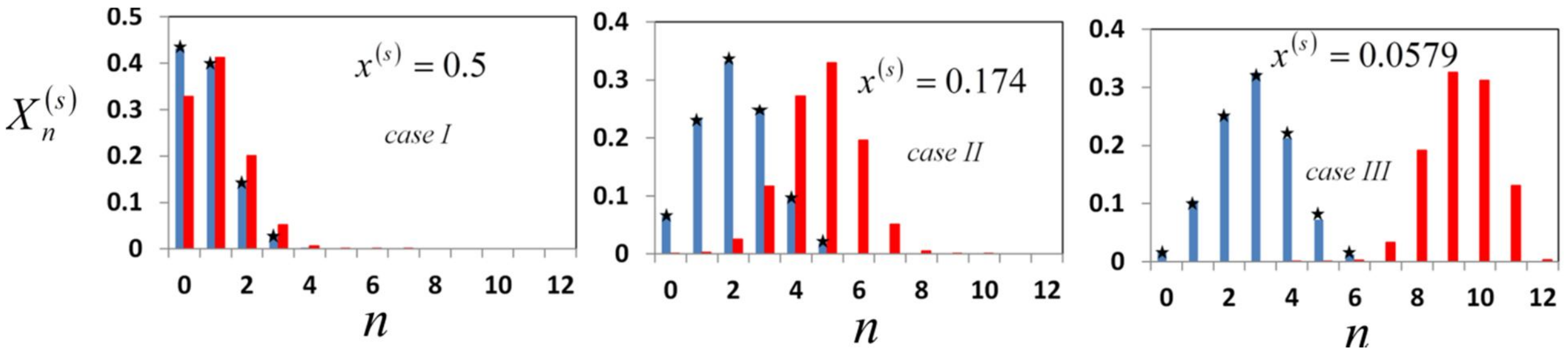

**Figure 3.7:** Fractions of s colloids bonded *n* times versus *n* at a density of $\rho^*$ = 0.2 and association energies $\varepsilon^*$ = 6 (blue bars – theory, stars – simulation) and $\varepsilon^*$ = 12(red bars – theory)

Lastly, we will consider the effect of fractional patch coverage $P$ of the patchy colloid on $\bar{n}$. As shown in Chapter 2, one has to account for the possibility of multiple bonds per patch once patch size increases beyond a certain point, about $\theta_c^{(p)} \sim 30^\circ$ for $r_c = 1.1d$.[73, 8573, 8570,82] For the current case, in which the patchy colloids do not self-attract, the possibility of a patchy colloid with large patch size forming a double bond will vanish for small enough $x^{(s)}$. Simply put, if the *s* type colloids are very dilute the probability of a *p* colloid simultaneously interacting with more than one *s* colloid (regardless of patch size) is vanishing. In this regime the theory derived in this work should be accurate for the full range of patch sizes. We validate this in Fig. 3.8 which shows the $P$ dependence of $\bar{n}$ for the case $\varepsilon^* = 8$ and $\rho^* = 0.2$ at a spherical colloid mole fraction of $x^{(s)} = 0.00579$. As can be seen, $\bar{n}$ increases logarithmically with $P$. The theory and simulation are in excellent agreement.

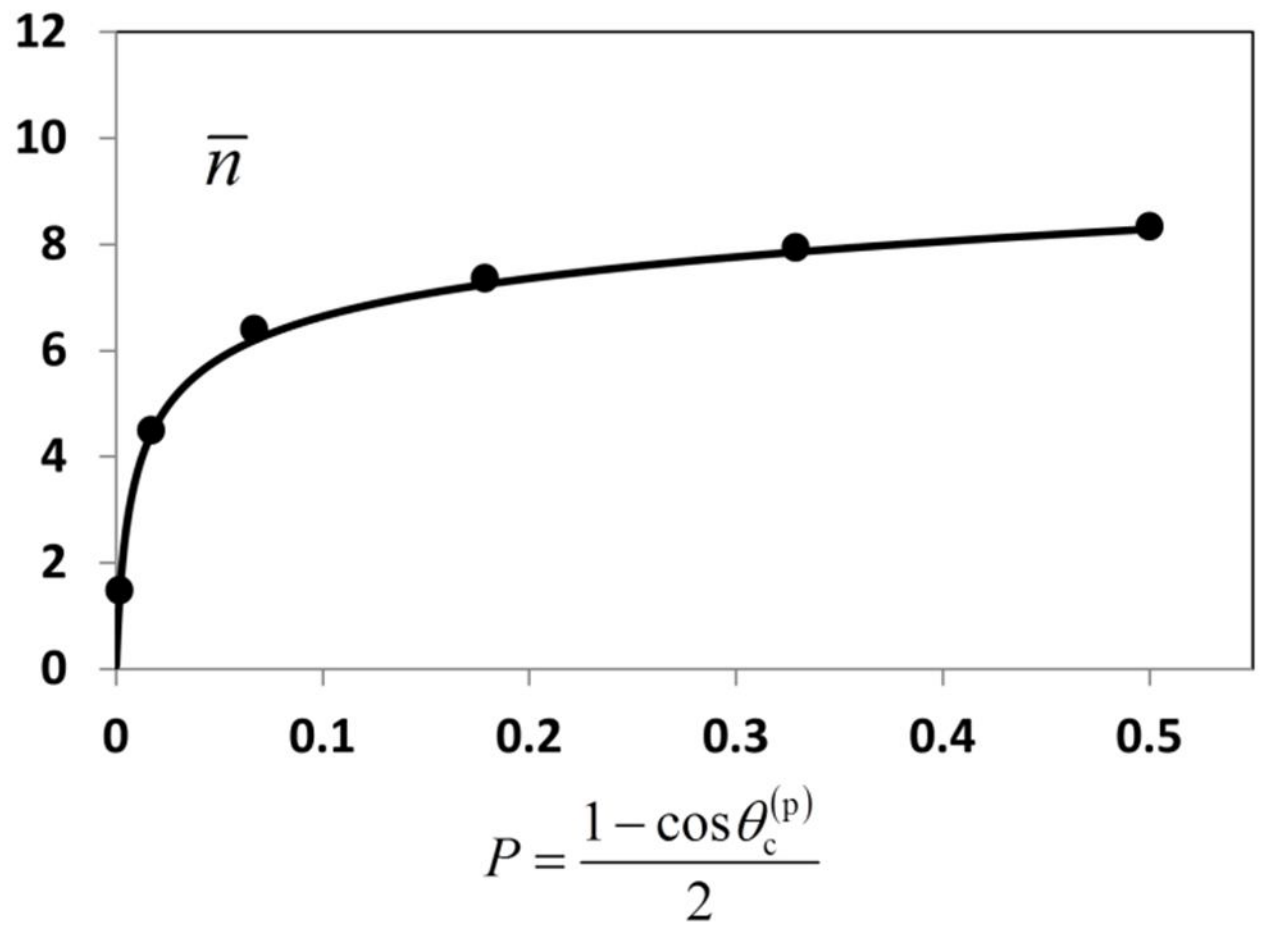

**Figure 3.8:** Average number of bonds per *s* colloid $\bar{n}$ versus fractional patch coverage $P$ at a density $\rho^* = 0.2$ and mole fraction $x^{(s)} = 0.00579$. Curve gives theory predictions and symbols are simulation results

This logarithmic dependence can be explained by the following simple model. We can write the change in Helmholtz free energy due to forming a single bond as

$$\Delta A^b = \Delta E^b - T\left(\Delta S^b_{config} + \Delta S^b_{orient}\right) \tag{3.26}$$

where $\Delta E^b = -\varepsilon_{AB}$ is the change in internal energy, $\Delta S^b_{config}$ is the change in configurational entropy and $\Delta S^b_{orient} \sim \ln P$ is the change in orientational entropy[8] due to bond formation. Since $\Delta E^b$ and $\Delta S^b_{config}$ are both independent of $P$ we can say

$$\frac{\partial \Delta A^b}{\partial P} \sim -\frac{1}{P} \tag{3.27}$$

Equation (3.27) states that a small change in patch size results in a large decrease in the free energy for small $P$, while for larger $P$ the decrease in free energy upon increasing patch size is less. This is due to the fact that the penalty in decreased orientational entropy due to association is much less for large patches than for small patch sizes. This is the genesis of the logarithmic dependence observed in Fig. 3.8.

## 3.5: Summary and conclusions

We have derived a simple perturbation theory to model mixtures of patchy *p* and spherically symmetric *s* colloids. In the current derivation we have assumed that the *p* colloids have a single patch which can engage in a single bond to a *s* colloid only. The *s* colloids consist of a single spherically symmetric site which can bond to as many *p* colloids as can physically fit in the *s* colloids bonding shell with no overlap. We have enforced the restriction that the spherically symmetric colloid cannot bond to other spherically symmetric colloids. Inspiration for this model was drawn from the recent work of Feng et *al.*[8] who synthesized mixtures of spherical and patchy single stranded DNA coated nanoparticles which have the same type of restrictions as discussed above. The new theory was extensively tested against new Monte Carlo simulation results and was found to be accurate. We extend this theory to the case that the patchy colloid can have multiple patches in chapter 4.

# CHAPTER 4

# Mixtures of multi - patch and spherically symmetric colloids

In Chapter 3 a new TPT was developed to model mixtures of single patch colloids and colloids with spherically symmetric attractions. The theory was shown to be highly accurate, but is limited by the fact that the patchy colloids can only have a single patch. What about the many patch case? For instance, consider the case where the *p* colloids have two patches (an *A* patch and *B* patch located on opposite poles of the colloid) and the *s* colloids are only attracted to the type *A* patch on the *p* colloid. As before, there are *AB* attractions but no *AA* or *BB* attractions. See Fig. 4.1 for an illustration of the types of associated clusters which may form.The simple addition of this extra patch on the *p* colloid significantly increases the complexity and richness of the behavior of the *s* and *p* colloid mixture, as compared to the one patch case. Now there will be a competition between the *p* colloids associating into free chains of colloids and the *p* colloids bonding to the *s* colloids to form colloidal star molecules. Unlike the one patch case, now the arm number distribution of colloidal star molecules as well as arm length distribution will vary with temperature, composition and density. Since there are no *AA* or *BB* attractions, no two *s* colloids can be connected by a path of association bonds (two arms on separate star molecules cannot form an association bond).

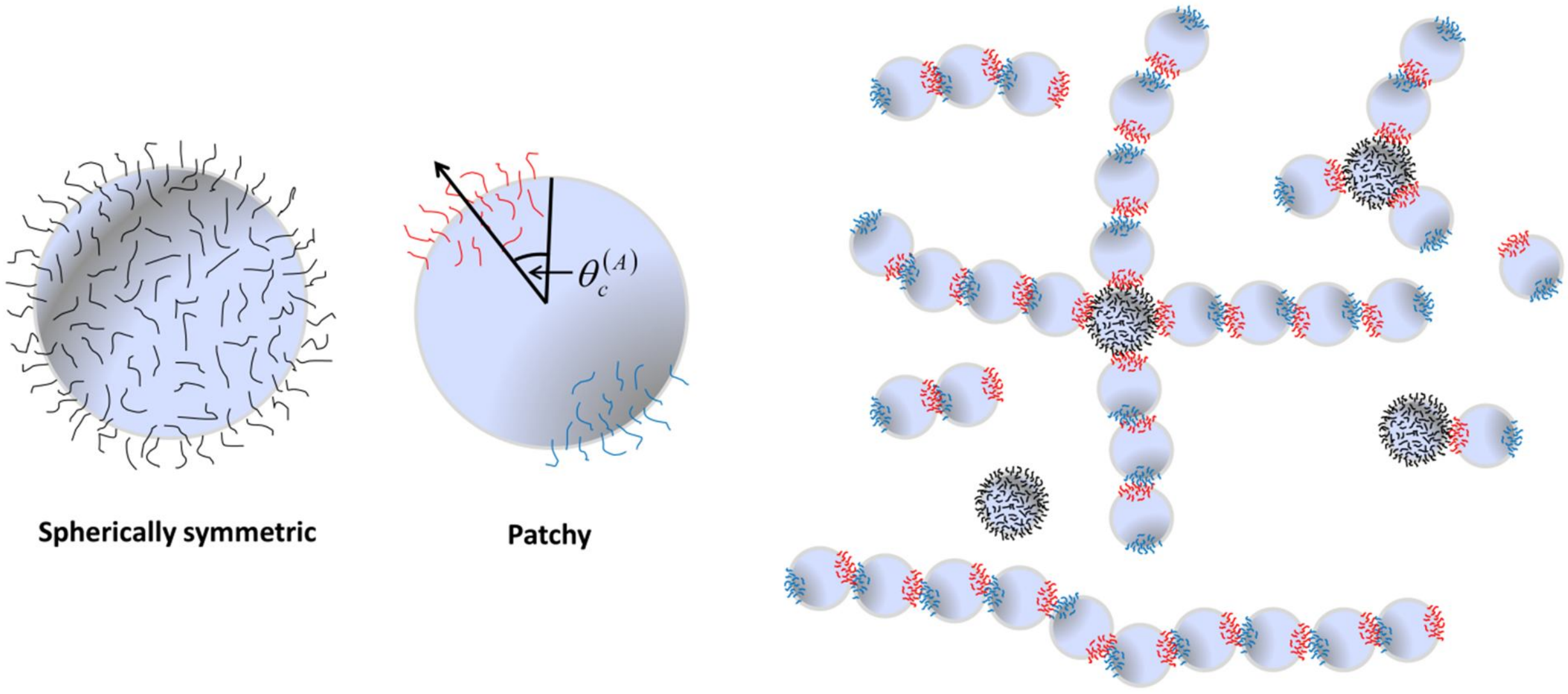

**Figure 4.1:** Types of associated clusters possible for mixture of two patch colloids and spherically symmetric colloids

These *s* and *p* colloid mixtures represent a new class of materials with an enormous potential. In this chapter we develop a simple perturbation theory to model mixtures of this type. We will develop the theory in Wertheim's multi – density formalism for multi-site associating fluids presented in section 1.5. [35] The anisotropic attractions between *p* colloids will be treated in standard first order perturbation theory (TPT1)[37, 38, 42] while the attractions between *p* and *s* colloids are treated in a modification of the results of chapter 3. The *p* colloids are allowed to have any number of patches and throughout the chapter we consider a hard sphere reference system. To include attractions between *s* colloids one simply needs to use an appropriate reference system.

Once developed, we apply the theory to the two patch case discussed above. We show that this *s* and *p* colloid mixture self – assembles into a mixture of free colloidal chains consisting of only *p* colloids and colloidal star molecules which consist of an *s* colloid articulation segment and *p* colloid arms. We show that the fraction of chains which are star arms, average number of star arms per *s* colloid, average length of star arms, fraction of *p* colloids bonded *k* times and resulting thermodynamics can be manipulated by varying temperature, density, composition and relative attractions. As a quantitative test, we compare theoretical predictions to new Monte Carlo simulation results. The theory is shown to be accurate.

## 4.1: General theory

In this section we derive the theory for a two component mixture of patchy *p* and spherically symmetric *s* colloids of diameter $d$ . We choose the case of equal diameters for notational simplicity, extension of the general approach to mixtures of different diameters is straight forward. The *p* colloids have a set $\Gamma^{(p)} = \{A, B, C...\}$ of short range attractive patches whose sizes are determined by the critical angles $\theta_c^{(L)}$ which define the solid angle of patch $L$ as $2\pi\left(1-\cos\theta_c^{(L)}\right)$. The *s* colloids are thought of as a colloid with a single large patch of critical angle $\theta_c^{(s)} = 180^{\circ}$ . A diagram of these types of colloids can be found in Fig. 4.1 for the case that attractions are mediated by grafted single stranded DNA with sticky ends, and the *p* colloid has two patches. The derivation given in this section allows for an arbitrary number of patches. This is a generalization of the one patch case studied theoretically in chapter 3 and experimentally by Feng *et al*.[8]

The potential of interaction between two *p* colloids is given by the sum of a hard sphere potential $\phi_{HS}(r_{12})$ and orientation dependant attractive patchy potential

$$\phi^{(p,p)}(12) = \phi_{HS}(r_{12}) + \sum_{A\in\Gamma^{(p)}} \sum_{B\in\Gamma^{(p)}} \phi_{AB}^{(p,p)}(12) \tag{4.1}$$

The potential $\phi_{AB}^{(p,p)}$ is that of conical sites, as described in chapter 2.1

$$\phi_{AB}^{(p,p)}(12) = \begin{cases} -\varepsilon_{AB}^{(p,p)}, & r_{12} \le r_c \textit{ and } \theta_{\mathrm{A}} \le \theta_c^{(A)} \textit{ and } \theta_{\mathrm{B}} \le \theta_c^{(B)} \\ 0 & \textit{otherwise} \end{cases} \tag{4.2}$$

which states that if colloids 1 and 2 are within a distance $r_c$ of each other and each colloid is oriented such that the angles between the site orientation vectors and the vector connecting the two segments, $\theta_{\mathrm{A}}$ for colloid 1 and $\theta_{\mathrm{B}}$ for colloid 2, are both less than the critical angle, $\theta_c^{(A)}$ for *A* and $\theta_c^{(B)}$ for *B*, the two sites are considered bonded and the energy of the system is decreased by a factor $\varepsilon_{AB}^{(p,p)}$.

The interaction between *p* and spherically symmetric *s* colloids is similarly defined as

$$\phi^{(s,p)}(12) = \phi_{HS}(r_{12}) + \sum_{A\in\Gamma^{(p)}} \phi_{A}^{(s,p)}(12) \tag{4.3}$$

Where the attractive potential is given by

$$\phi_{A}^{(s,p)}(12) = \begin{cases} -\varepsilon_{A}^{(s,p)}, & r_{12} \le r_c \textit{ and } \theta_{\mathrm{A}} \le \theta_c^{(A)} \\ 0 & \textit{otherwise} \end{cases} \tag{4.4}$$

which states that if *s* colloid 1 and *p* colloid 2 are within a distance $r_c$ of each other, and the *p* colloid is oriented such that the angle between the site *A* orientation vector and the vector connecting the two segments $\theta_{\mathrm{A}}$ is less than the critical angle $\theta_c^{(A)}$, the two colloids are considered bonded and the energy of the system is decreased by a factor $\varepsilon_A^{(s,p)}$. Lastly, the *s* colloids are assumed to interact with hard sphere repulsions only, that is

$$\phi^{(s,s)}(12) = \phi_{HS}(r_{12}) \tag{4.5}$$

We develop the theory in the multi - density formalism of Wertheim[35-3735-3735-3735-37] where each bonding state of a colloid is assigned a number density. The density of species $k = \{p, s\}$ bonded at the set of patches $\alpha$ is given by $\rho_\alpha^{(k)}$. The extension of the fundamental equations presented in section 1.5 to mixtures is straightforward. For the case of mixtures the Helmholtz free energy is now given by

$$\frac{A - A_{HS}}{Vk_BT} = \sum_k \left( \rho^{(k)} \ln\left( \frac{\rho_o^{(k)}}{\rho^{(k)}} \right) + Q^{(k)} + \rho^{(k)} \right) - \Delta c^{(o)} / V \tag{4.6}$$

where we are summing over all species *k*. All other relationships remain unchanged, except for the insertion of species labels. For instance, for the case of mixtures, Eq. (1.35) is now

$$Q^{(k)} = -\rho^{(k)} + \sum_{\substack{\gamma \subset \Gamma^{(k)} \\ \gamma \neq \varnothing}} c_\gamma^{(k)} \sigma_{\Gamma^{(k)} - \gamma}^{(k)} \tag{4.7}$$

The graph sum $\Delta c^{(o)}$ is decomposed as

$$\Delta c^{(o)} \approx \Delta \hat{c}_1^{(o)} = \Delta c_{pp}^{(o)} + \Delta c_{sp}^{(o)} \tag{4.8}$$

Where $\Delta c_{pp}^{(o)}$ accounts for the attractions between *p* colloids and $\Delta c_{sp}^{(o)}$ accounts for the attraction between *p* and *s* colloids. We treat the interaction between *p* colloids in first order perturbation theory (TPT1) giving $\Delta c_{pp}^{(o)}$ as

$$\Delta c_{pp}^{(o)} / V = \frac{1}{2} \sum_{L \in \Gamma^{(p)}} \sum_{M \in \Gamma^{(p)}} \sigma_{\Gamma^{(p)} - L}^{(p)} \xi \kappa_{LM} f_{LM}^{(p,p)} \sigma_{\Gamma^{(p)} - M}^{(p)} \tag{4.9}$$

where $\kappa_{LM}=\left(1-\cos\theta_c^{(L)}\right)\left(1-\cos\theta_c^{(M)}\right)/4$ is the probability that two colloids are oriented such that patch *L* on colloid 1 can bond to patch *M* on colloid 2 and

$$\xi \approx 4\pi d^2 \left(r_c - d\right) g_{HS}\left(d\right) \tag{4.10}$$

In Eq. (4.10) we have assumed $p$ = 2 in Eq. (2.17). [41] This results in a simpler chemical potential and is accurate for this first order contribution to the graph sum.

The contribution $\Delta c_{sp}^{(o)}$ cannot be obtained in TPT1 due to the fact that we must move beyond Wertheim's single bonding condition[35] which restricts each patch to bonding only once. The *s* colloids exhibit spherical symmetry meaning they can bond multiple times. The maximum number of bonds is simply the maximum number of *p* colloids $n^{\max}$ which can pack in the *s* colloids bonding shell. We can rewrite $\Delta c_{sp}^{(o)}$ as

$$\Delta c_{sp}^{(o)} = \sum_{n=1}^{n^{\max}} \Delta c_n \tag{4.11}$$

where $\Delta c_n$ is the contribution for *n* patchy colloids bonded to a single *s* colloid. We approximate $\Delta c_n$ in a generalization of Wertheim's single chain approximation[36, 37] and consider all graphs consisting of a single associated cluster with *n* patchy colloids bonded to a *s* colloid with *n* association bonds. We then simplify the results as described in section 3.1 to obtain

$$\Delta c_n^{(o)} / V = \frac{1}{n!} \rho_o^{(s)} \Delta^n \delta^{(n)} \Xi^{(n)} \tag{4.12}$$

where

$$\Delta = y_{HS}\left(d\right) \sum_{L\in\Gamma^{(p)}} \sigma_{\Gamma^{(p)}-L}^{(p)} f_L^{(s,p)} \sqrt{\kappa_{LL}} \tag{4.13}$$

Now that $\Delta c^{(o)}$ has been fully specified the densities of the various bonding states can be calculated through the relation[35]

$$\frac{\rho_{\gamma}^{(k)}}{\rho_{o}^{(k)}} = \sum_{P(\gamma)=\{\tau\}} \prod_{\tau} c_{\tau}^{(k)} \tag{4.14}$$

where $P(\gamma)$ is the partition of the set $\gamma$ into non-empty subsets. Since the spherically symmetric colloid is a single patch, Wertheim's theory only assigns two densities; the density of *s* colloids not bonded $\sigma_o^{(s)} = \rho_o^{(s)}$ and the density of *s* colloids which are bonded $\rho_b^{(s)}$. The density $\rho_b^{(s)}$ is obtained as

$$\rho_b^{(s)} = \rho_o^{(s)} \frac{\partial \Delta c^{(o)} / V}{\partial \rho_o^{(s)}} = \sum_{n=1}^{n^{\max}} \rho_n^{(s)} \tag{4.15}$$

where $\rho_n^{(s)}$ is the density of *s* colloids bonded *n* times which we identify as

$$\rho_n^{(s)} = \frac{\Delta c_n}{V} \qquad \textit{for } n > 0 \tag{4.16}$$

Turning our attention to the *p* colloids we note from Eqns. (4.8) and (1.36) that $c_{\gamma}^{(p)} = 0$ for $n(\gamma) > 1$ which results in the following rule

$$\frac{\rho_{\gamma}^{(p)}}{\rho_{o}^{(p)}} = \prod_{A \in \gamma} \frac{\rho_{A}^{(p)}}{\rho_{o}^{(p)}} \tag{4.17}$$

Equation (4.17) leads to the following relation for the fraction of colloids *not bonded* at patch *A* $X_A^{(p)} = \sigma_{\Gamma^{(p)}-A}^{(p)} / \rho^{(p)}$ as

$$X_A^{(p)} = \frac{1}{1 + c_A^{(p)}} \tag{4.18}$$

as well as the relation for the monomer fraction

$$X_o^{(p)} = \prod_{A \in \Gamma^{(p)}} X_A^{(p)} \tag{4.19}$$

We obtain $c_A^{(p)}$ as

$$c_A^{(p)} = \sum_{M\in\Gamma^{(p)}} \xi\kappa_{AM} f_{AM}^{(p,p)} \rho^{(p)} X_M^{(p)} + \sum_{n=1}^{n^{\max}} \frac{\rho_o^{(s)} n}{n!} \sqrt{\kappa_{AA}}\, y_{HS}(d) f_A^{(s,p)} \Delta^{n-1} \Xi^{(n)} \delta^{(n)} \tag{4.20}$$

Since $c_\gamma^{(p)} = 0$ for $n(\gamma) > 1$ we obtain from Eq. (4.7)

$$\frac{Q^{(p)}}{\rho^{(p)}} = -1 + \sum_{A\in\Gamma^{(p)}} c_A^{(p)} X_A^{(p)} \tag{4.21}$$

which when combined with Eq. (4.18) gives

$$\frac{Q^{(p)}}{\rho^{(p)}} = -1 + \sum_{A\in\Gamma^{(p)}} \left(1 - X_A^{(p)}\right) \tag{4.22}$$

For the spherically symmetric *s* colloids Eq. (4.7) simply gives

$$\frac{Q^{(s)}}{\rho^{(s)}} = -X_o^{(s)} \tag{4.23}$$

We can now write $\Delta c_{pp}^{(o)} / V$ as

$$\frac{\Delta c_{pp}^{(o)}}{V} = \frac{1}{2} \sum_{A\in\Gamma^{(p)}} \rho^{(p)} X_A^{(p)} c_A^{(p)} - \sum_{n=1}^{n^{\max}} \frac{n}{2} \rho^{(s)} X_n^{(s)} \tag{4.24}$$

The term $X_n^{(s)} = \rho_n^{(s)} / \rho^{(s)}$ is the fraction of spherically symmetric colloids bonded *n* times obtained from Eq. (4.16). Combining these results we obtain the free energy

$$\frac{A - A_{HS}}{Nk_BT} = x^{(s)} \left( \ln X_o^{(s)} - X_o^{(s)} + 1 + \sum_{n=1}^{n^{\max}} \left( \frac{n}{2} - 1 \right) X_n^{(s)} \right) + \left(1 - x^{(s)}\right) \sum_{A\in\Gamma^{(p)}} \left( \ln X_A^{(p)} - \frac{X_A^{(p)}}{2} + \frac{1}{2} \right) \tag{4.25}$$

where $N$ is the total number of colloids in the system and $x^{(s)}$ is the mole fraction of *s* colloids. Using the relation

$$\sum_{n=0}^{n^{\max}} X_n^{(s)} = 1 \tag{4.26}$$

and the definition of the average number of bonds per *s* colloid

$$\bar{n} = \sum_{n=0}^{n^{\max}} n X_n^{(s)} \tag{4.27}$$

Eq. (4.25) can be further simplified as

$$\frac{A - A_{HS}}{N k_B T} = x^{(s)}\left(\ln X_o^{(s)} + \frac{\bar{n}}{2}\right) + \left(1 - x^{(s)}\right)\sum_{A \in \Gamma^{(p)}}\left(\ln X_A^{(p)} - \frac{X_A^{(p)}}{2} + \frac{1}{2}\right) \tag{4.28}$$

Where the fractions $X_A^{(p)}$ are obtained by solving Eqns. (4.18) in conjunction with the relation

$$X_o^{(s)} = \frac{1}{1 + \sum_{n=1}^{n^{\max}} \frac{1}{n!} \Delta^n \, \Xi^{(n)} \delta^{(n)}} \tag{4.29}$$

which was obtained using Eq. (4.26).

## 4.2: Application to two patch colloids

Here we consider the case discussed in the introduction of this chapter, where the *p* colloid has an *A* type and *B* type patch where $\varepsilon_{AA}^{(p,p)} = \varepsilon_{BB}^{(p,p)} = 0$. For attractions between the *p* and *s* colloid we set

$$\varepsilon_{A}^{(s,p)} = C \varepsilon_{AB}^{(p,p)} \qquad \qquad \varepsilon_{B}^{(s,p)} = 0 \tag{4.30}$$

where the constant $C$ is defined by Eq. (4.30). The restrictions $\varepsilon_{B}^{(s,p)} = \varepsilon_{BB}^{(p,p)} = \varepsilon_{AA}^{(p,p)} = 0$ will suppress the formation of a network, meaning a branch emanating from an *s* colloid cannot terminate on another *s* colloid. This situation is depicted in Fig. 4.1. This system represents a mixture of colloids which exhibits the reversible and temperature dependent self – assembly into colloidal star molecules and free chains, where the *s* colloids are the articulation segments for the star molecules and the *p* colloids make up the arms of the star molecules and the segments of the free chains.

From the results of the previous section we obtain

$$c_A^{(p)} = \xi \kappa_{AB} f_{AB}^{(p,p)} \rho^{(p)} X_B^{(p)} + \frac{\rho^{(s)}}{\rho^{(p)}} \frac{\overline{n}}{X_A^{(p)}} \tag{4.31}$$

and

$$c_B^{(p)} = \xi \kappa_{AB} f_{AB}^{(p,p)} \rho^{(p)} X_A^{(p)} \tag{4.32}$$

From which we obtain the fractions not bonded as

$$X_B^{(p)} = \frac{1}{1 + \xi \kappa_{AB} f_{AB}^{(p,p)} \rho^{(p)} X_A^{(p)}} \tag{4.33}$$

and

$$X_A^{(p)} = \frac{1}{1 + \xi \kappa_{AB} f_{AB}^{(p,p)} \rho^{(p)} X_B^{(p)} + \frac{\rho^{(s)}}{\rho^{(p)}} \frac{\overline{n}}{X_A^{(p)}}} = X_B^{(p)} - \frac{\rho^{(s)}}{\rho^{(p)}} \overline{n} \tag{4.34}$$

To compare to simulations we will use the fraction of *p* colloids bonded *k* times $X_k^{(p)}$ which we obtain from Eqns. (4.17) and (4.19) as

$$X_o^{(p)} = X_A^{(p)} X_B^{(p)}$$

$$X_1^{(p)} = X_o^{(p)} \left( \frac{1}{X_A^{(p)}} + \frac{1}{X_B^{(p)}} - 2 \right) \tag{4.35}$$

$$X_2^{(p)} = X_o^{(p)} \left( 1 - \frac{1}{X_A^{(p)}} \right) \left( 1 - \frac{1}{X_B^{(p)}} \right)$$

We will place the $p$ colloids into two classes. The first class consists of $p$ colloids which are part of a chain which emanates from an $s$ colloid, we will call these star $p$ colloids with a colloid density $\rho_{arm}^{(p)}$. The second class consist of colloids which are not in a bonded network which includes an $s$ colloid, we will call these free $p$ colloids with a colloid density $\rho_{free}^{(p)}$ (note the monomer density $\rho_{o}^{(p)}$ is included in $\rho_{free}^{(p)}$). Unfortunately, the densities $\rho_{arm}^{(p)}$ and $\rho_{free}^{(p)}$ are not directly accessible with the current approach; however, the densities of free chains (including monomers) $\rho_{free}^{(chain)} = \sigma_{B}^{(p)}$ and star arms $\rho_{arm}^{(star)} = \rho^{(s)}\bar{n}$ are known. A quantity which will provide insight into the competition between self – assembled star molecules and free chains is the fraction

$$\Psi = \frac{\rho^{(s)}\bar{n}}{\rho^{(s)}\bar{n} + \sigma_{B}^{(p)}} \tag{4.36}$$

Where $\Psi$ is the fraction of chains (free chains including monomers + star arms) which are star arms. The last quantity we would like to determine is the average length of the star arms $L_{arm}$. To determine this length (in a strict way) the densities $\rho_{arm}^{(p)}$ and $\rho_{free}^{(p)}$ would have to be known. They are not. As an alternative, an approximation of $L_{arm}$ can be constructed as follows. The total density of chains *not including* free monomers is $\rho_{A}^{(p)}$. For the star arms, the chains terminate on one side with a $p$ colloid (this colloid is bonded at both patches $A$ and $B$) bonded to an $s$ colloid and the other side with a $p$ colloid only bonded at patch $A$. To estimate $L_{arm}$ we will assume that the average number of double bonded colloids in a chain (not including free monomers in this average) is equal for free chains (not including monomers) and star arms. With this we can approximate $L_{arm}$ as follows

$$L_{arm} \approx \frac{\rho_{AB}^{(p)} + \rho_{A}^{(p)}}{\rho_{A}^{(p)}} = \frac{1}{X_{B}^{(p)}} \quad (4.37)$$

Equation (4.37) completes our theoretical analysis. In the following section we compare the new theory to Monte Carlo simulation results.

## 4.3: Numerical results

In this section we compare theory and simulation results for the case that the *p* colloid has two patches as discussed in sections 4.1 – 4.2 and illustrated in Fig. 4.1. Simulations were performed in the same manner as discussed in section 3.2 (with $\theta_c^{(p)} = 27^\circ$ and $r_c = 1.1d$ )

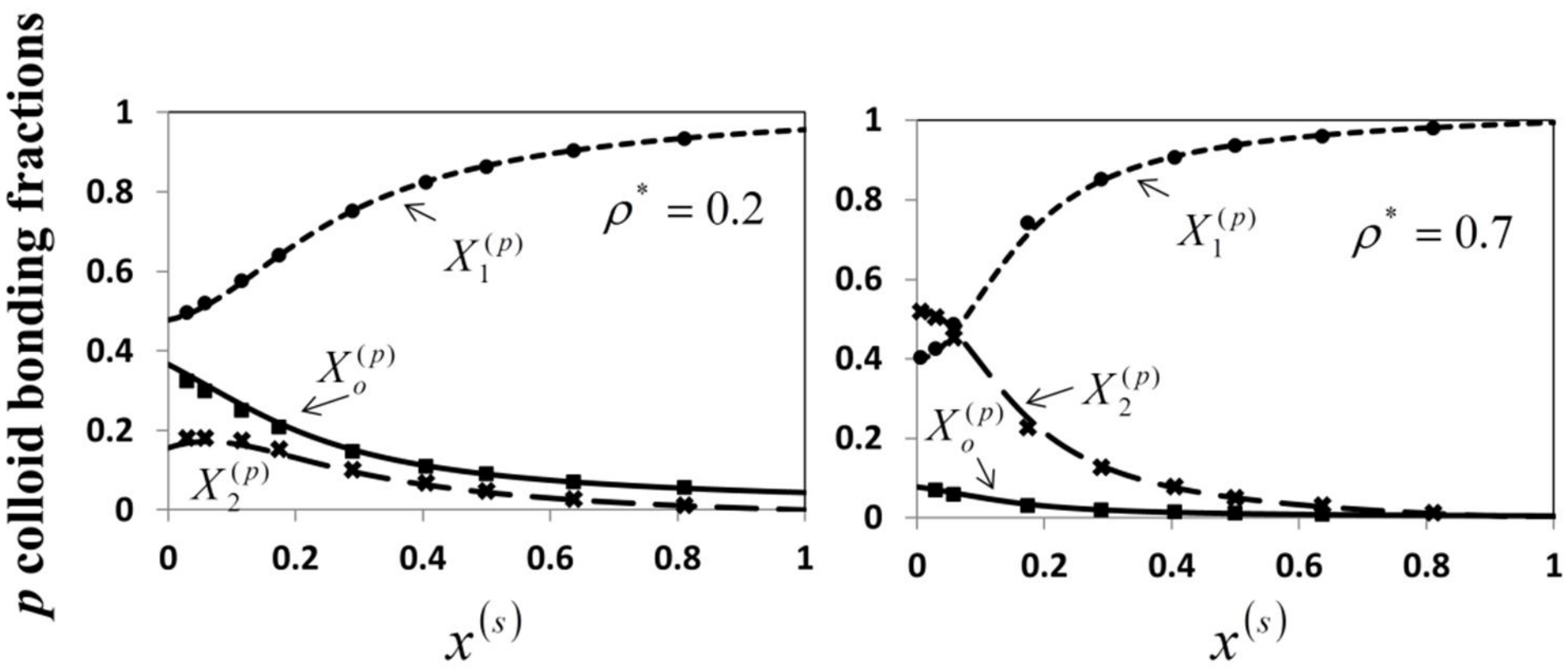

**Figure 4.2:** Fractions of *p* colloids bonded *k* times versus mole fraction of *s* colloids $x^{(s)}$ at densities $\rho^*$ = 0.2(left) and $\rho^*$ = 0.7 (right) and an association energy $\varepsilon^*$ = 7 with *C* = 1. Curves give theoretical predictions and symbols give simulation results

We begin by comparing theory and simulation for the case that the association energy between patchy colloids is set to $\varepsilon^* = \varepsilon_{AB}^{(p,p)} / k_B T$ = 7 with the association energy ratio set to *C* = 1. We consider both low density $\rho^* = \rho d^3 = 0.2$ and high density $\rho^* = 0.7$ cases. Figure 4.2 gives the fraction of *p* colloids bonded *k* times versus mole fraction of *s* colloids $x^{(s)}$. For $x^{(s)} \rightarrow 0$ the fluid is a pure component fluid of *p* colloids with $X_1^{(p)}$ being the dominant fraction for the low density case and $X_2^{(p)}$ being dominant for the high density case. Introducing *s* type colloids into the system, increasing $x^{(s)}$, results in a decrease in $X_2^{(p)}$ and $X_o^{(p)}$ with an increase in $X_1^{(p)}$. The decrease in $X_2^{(p)}$ and increase in $X_1^{(p)}$ is a result of the fact that longer free chains are being sacrificed to form star arms. Since $C = 1$ there is no energetic difference between a bond between two *p* colloids (*pp* bond) and a bond between *s* and *p* colloids (*sp* bond); however, the penalty in decreased orientational entropy for forming a *pp* bond is double that which is paid for an *sp* bond. Theory and simulation are in excellent agreement.

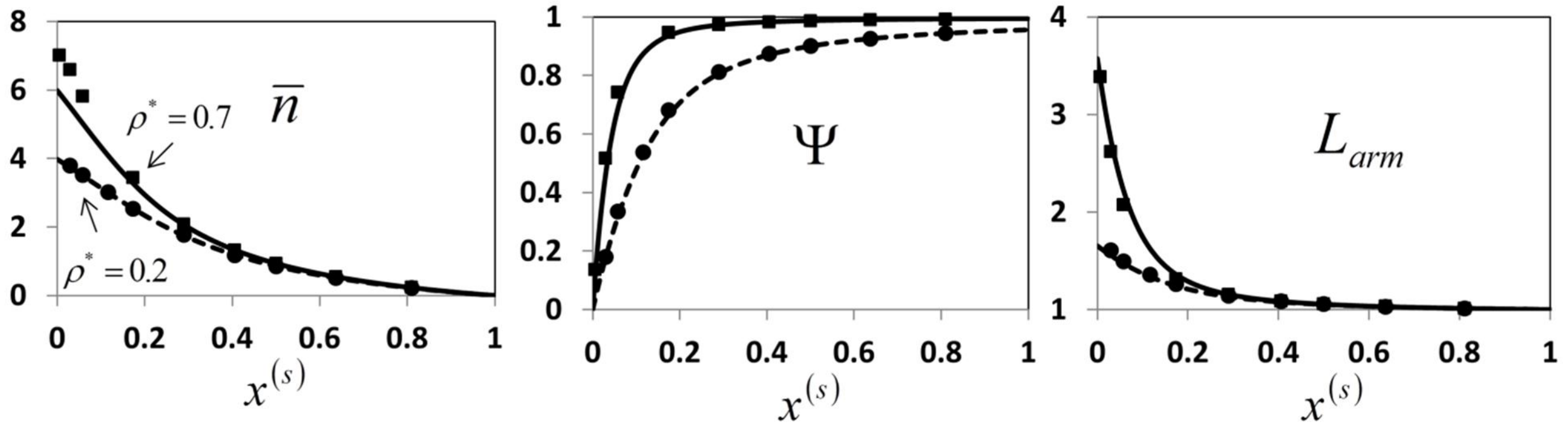

**Figure 4.3:** Average number of bonds per s colloid $\bar{n}$ (left), fraction of chains which are star arms $\Psi$ (middle) and average length of star arms $L_{arm}$ (right) versus mole fraction *s* colloids $x^{(s)}$ at $\rho^* = 0.2$ (dashed curve – theory , circles – simulation) and $\rho^* = 0.7$ (solid curve – theory , squares – simulation). Association energy is $\varepsilon^* = 7$ with $C = 1$

Figure 4.3 shows calculations for the average number of arms (bonds) per *s* colloid $\bar{n}$, fraction of chains which are star arms $\Psi$ and average length of star arms $L_{arm}$. For $x^{(s)} \rightarrow 0$ the *s* colloids are dilute. Since there are an abundance of *p* colloids available to “solvate” the *s* colloids, it is in this realm where $\bar{n}$ is a maximum and $\Psi$ is a minimum. It is also in this region where $L_{arm}$ is maximum. In fact, for $x^{(s)} \rightarrow 0$, $L_{arm} = L_{free}$ where $L_{free}$ is the average length of free chains. Increasing $x^{(s)}$ results in a decrease in $\bar{n}$ and $L_{arm}$ as there is now less *p* colloids to solvate the *s* colloids and the introduction of *s* colloids breaks longer chains of *p* colloids. As expected, increasing $x^{(s)}$ results in an increase in $\Psi$ as there are now more *s* colloids to seed star arms. As $x^{(s)} \rightarrow 1$, there are now few *p* colloids to solvate the *s* colloids and the probability of forming *pp* bonds becomes very small. In this limit $\bar{n} \to 0$, $\Psi \to 1$ and $L_{arm} \to 1$. According to these results, if one desired to create a small number of colloidal star molecules with many long arms, this is best achieved for small $x^{(s)}$. On the other hand, if one wished to create a larger number of stars with a few short arms, this would be best achieved for larger $x^{(s)}$. Comparing the cases for low and high density we see that increasing density increases $\bar{n}$, $\Psi$ and $L_{arm}$. Overall the theory and simulation are in excellent agreement; however, the theory predicts $\bar{n}$ to be too small for $x^{(s)} \to 0$ at high density. This is the result of the approximation of the *n* + 1 body cavity correlation functions given by Eq. (3.7) .

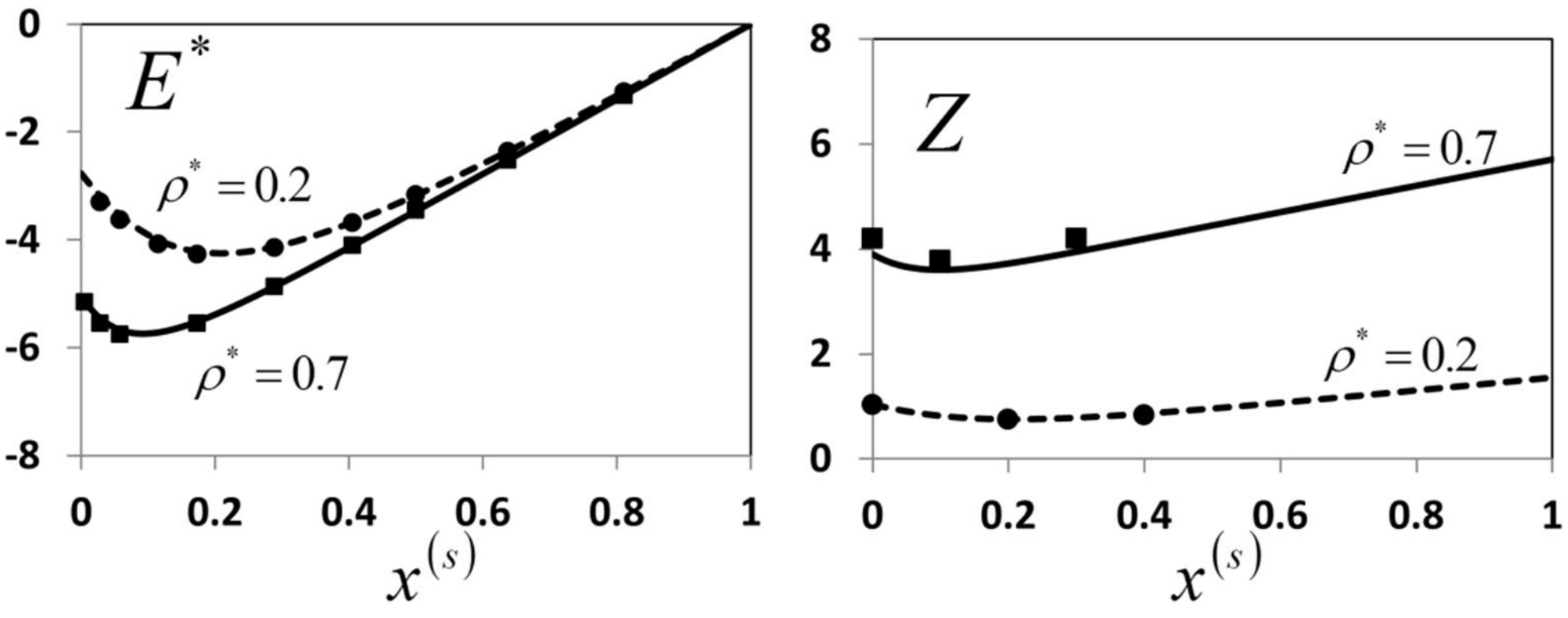

**Figure 4.4:** Same as Fig. 4.3 except the excess internal energy $E^*$ is the dependent variable (left) and compressibility factor $Z$ (right)

Figure 4.4 shows the excess internal energy $E^* = E^{AS} / Nk_BT$ and compressibility factor $Z = P / \rho k_B T$ for these same conditions. Initially, for $x^{(s)} \to 0$, increasing $x^{(s)}$ results in a decrease in both $E^*$ and Z as association in the system is increased. On the other extreme, $x^{(s)} \to 1$, *p* colloids are limiting and increasing $x^{(s)}$ decreases association in the system. This results in an increase in both $E^*$ and Z. Association is maximized at the location that $E^*$ and Z show distinct minimums, about $x^{(s)} \approx 0.1$ for $\rho^* = 0.7$ and $x^{(s)} \approx 0.21$ for $\rho^* = 0.2$. Theory and simulation are in excellent agreement for $E^*$. Note, the energy in the simulations is calculated as $E = -\langle N_b \rangle \varepsilon_{AB}$; where $\langle N_b \rangle$ is the ensemble average number of bonds in the system. Hence, when $\langle N_b \rangle$ (association) is a maximum, $E$ is a minimum. As can be seen, the theory accurately predicts the location of both the minimum in *E* and *Z*. The *NPT* simulations for *Z* were performed in an iterative manner; for a given $x^{(s)}$ the pressure would be optimized to reproduce the desired density. To provide a further quantitative test for the theoretical predictions of *Z* we performed *NPT* simulation for various *s* colloid mole fractions $x^{(s)}$ as a function of $\rho^*$. These results can be found in Fig. 4.5. As can be seen, theory and simulation are in good agreement.

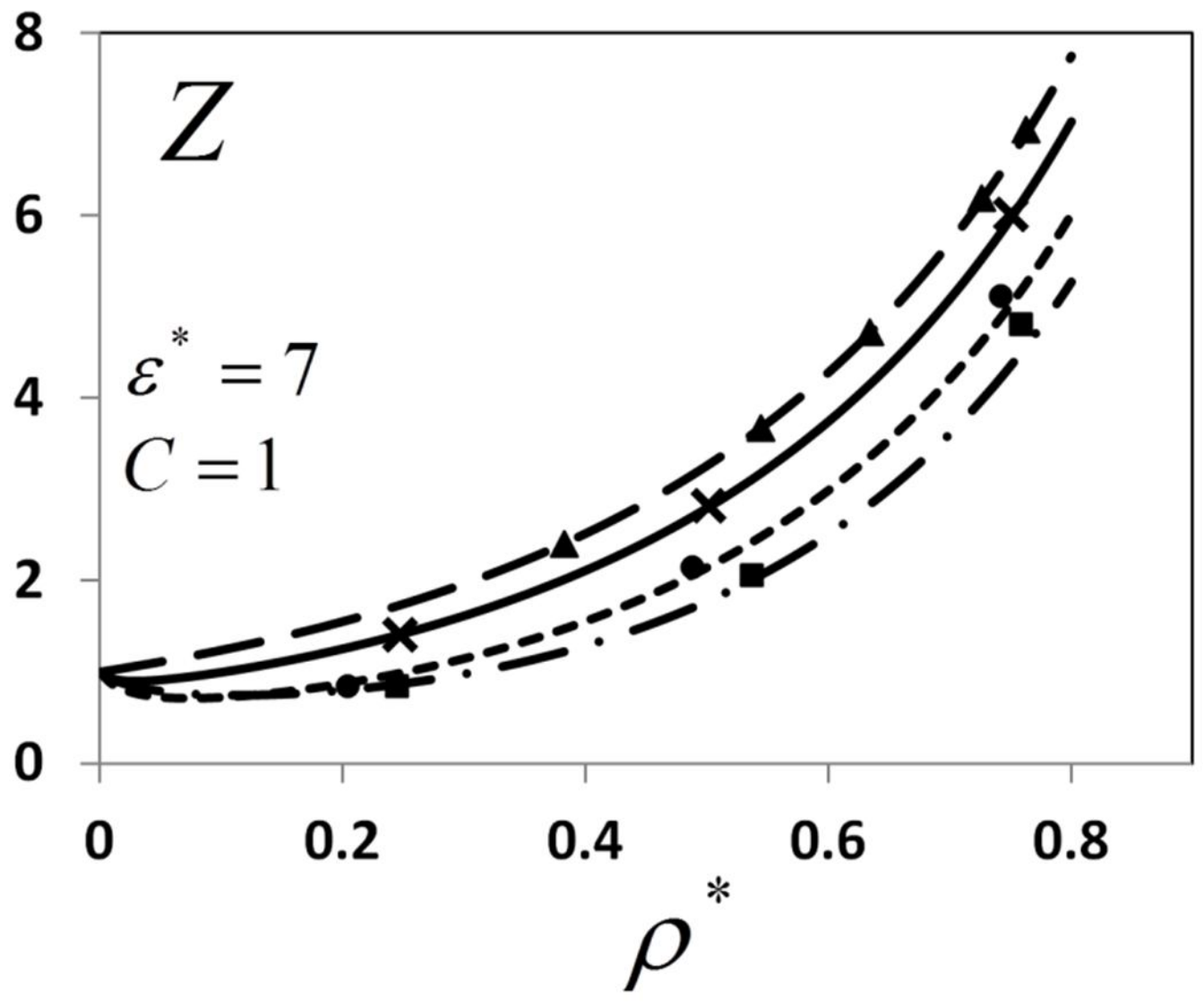

**Figure 4.5:** Compressibility factor Z versus density $\rho^*$ at an association energy $\varepsilon^*$ = 7 with *C* = 1. Curves give theoretical predictions and symbols give *NPT* simulation results for $x^{(s)}$ = 1(long dashed curve – theory, triangles – simulation), $x^{(s)}$ = 0.752 (solid curve – theory, crosses – simulation), $x^{(s)}$ = 0.405 (short dashed curve – theory, circles – simulation) and $x^{(s)}$ = 0.116 (dashed dotted curve – theory, squares – simulation)

Now we will analyze the effect of association energy $\varepsilon^*$ on fluid properties when composition remains constant. Specifically we will consider the case where $x^{(s)} = 50/864$ =

0.05787; again, we will keep the energy of a *pp* bond equal to a *ps* bond $\varepsilon_{AB}^{(p,p)} = \varepsilon_{A}^{(s,p)}$. We choose this mole fraction to set up a competition between star and chain formation. We perform calculations for both low $\rho^* = 0.2$ and high $\rho^* = 0.7$ density cases.

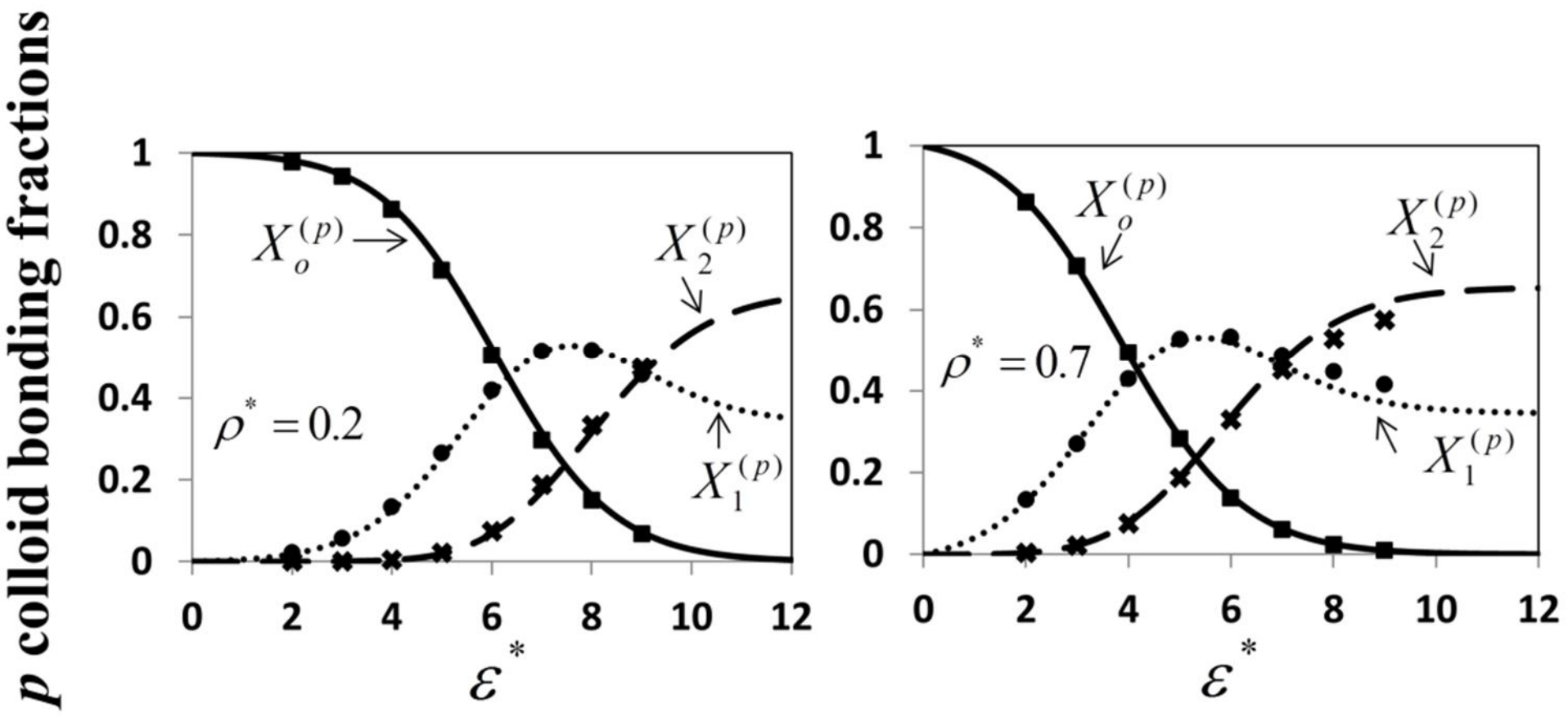

**Figure 4.6:** Fractions of *p* colloids bonded *k* times versus association energy $\varepsilon^*$ at densities $\rho^* = 0.2$ (left) and $\rho^* = 0.7$ (right) and an *s* colloid mole fraction $x^{(s)} = 0.05787$ with $C = 1$. Curves give theoretical predictions and symbols give simulation results

Figure 4.6 shows the fractions of *p* colloids bonded *k* times versus $\varepsilon^*$ for this case. For small $\varepsilon^*$, the monomer fraction $X_o^{(p)}$ is the dominant contribution, due to the fact that the entropic penalty of bond formation outweighs the energetic benefit for this low $\varepsilon^*$. Increasing $\varepsilon^*$ results in an increase in $X_1^{(p)}$ as the energetic benefit of forming a single bond outweighs the entropic penalty. At around $\varepsilon^* = 5$ for $\rho^* = 0.2$ and $\varepsilon^* = 3$ for $\rho^* = 0.7$ the fraction $X_2^{(p)}$ begins to have a significant increase with increasing $\varepsilon^*$. This increase in $X_2^{(p)}$ results in a maximum in $X_1^{(p)}$ near $\varepsilon^* = 7.5$ for $\rho^* = 0.2$ and $\varepsilon^* = 5.5$ for $\rho^* = 0.7$. Eventually the energetic benefit of being fully bonded outweighs the entropic penalty and $X_2^{(p)}$ becomes the dominant fraction. Overall theory and simulation are in excellent agreement, although for the density $\rho^* = 0.7$, the theory does slightly overpredict $X_2^{(p)}$ and underpredict $X_1^{(p)}$ for large $\varepsilon^*$.

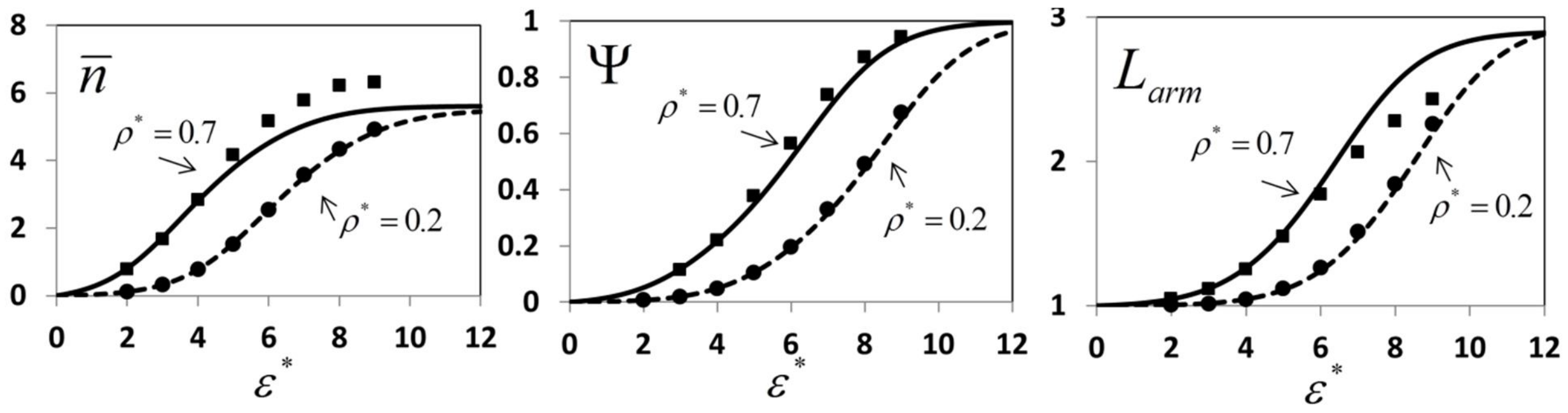

**Figure 4.7:** Average number of arms (bonds) per *s* colloid $\bar{n}$ (left), fraction of chains which are star arms Ψ (middle) and average length of star arms $L_{arm}$ (right) versus association energy $\varepsilon^*$ at $\rho^*$ = 0.2 (dashed curve – theory , circles – simulation) and $\rho^*$ = 0.7 (solid curve – theory , squares – simulation). Mole fraction of *s* colloids is $x^{(s)}$ = 0.05787 with $C$ = 1

Figure 4.7 shows calculations for the average number of arms (bonds) per *s* colloid $\bar{n}$, fraction of chains which are star arms $\Psi$ and average length of star arms $L_{arm}$; while Fig. 4.8 gives the results for the excess internal energy $E^*$. Increasing $\varepsilon^*$, or equivalently decreasing *T*,

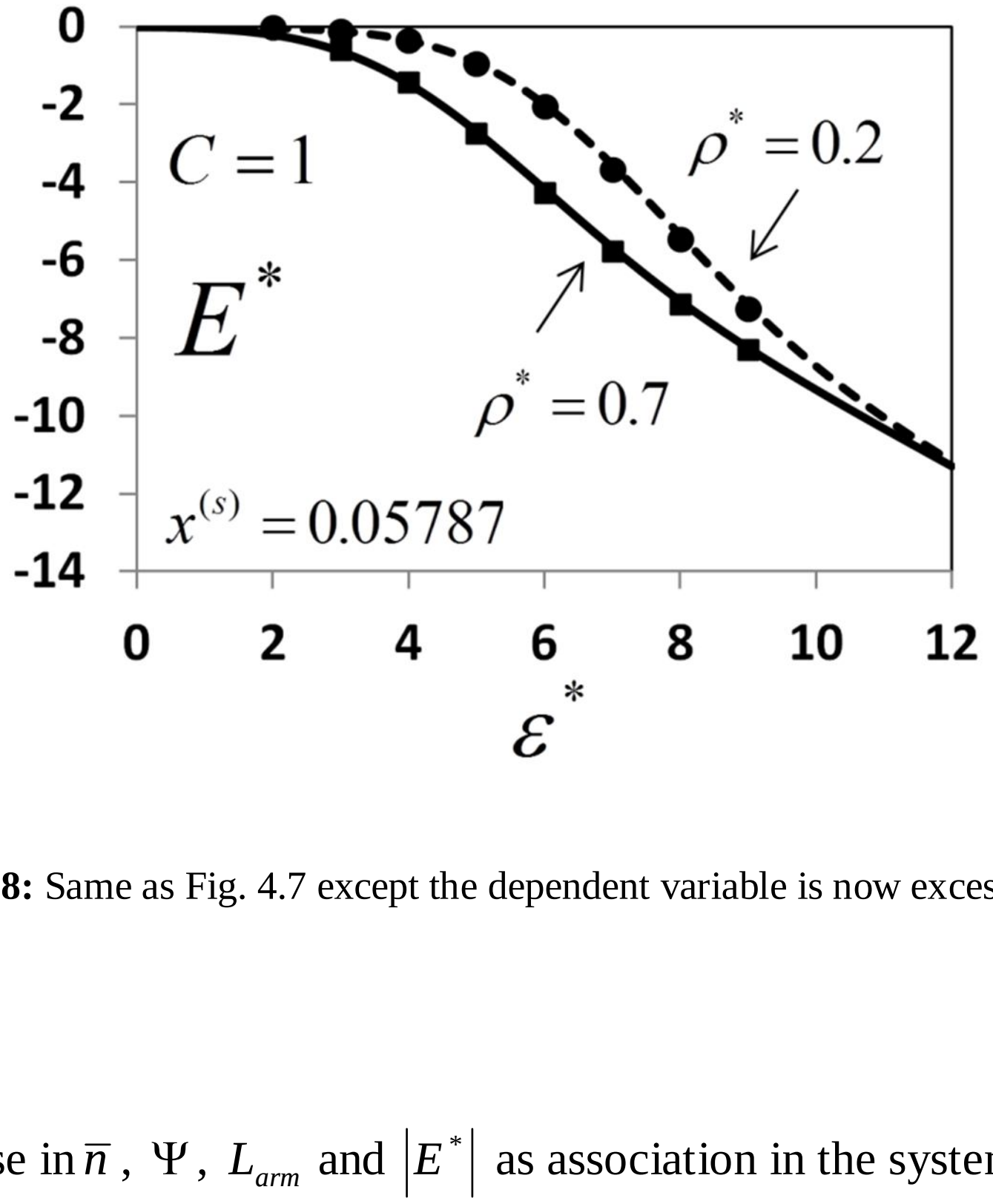

**Figure 4.8:** Same as Fig. 4.7 except the dependent variable is now excess internal energy

results in an increase in $\bar{n}$, $\Psi$, $L_{arm}$ and $|E^*|$ as association in the system increases. Overall theory and simulation are in excellent agreement for each quantity at low density. For the high

density case theory and simulation are in good agreement, although the theory underpredicts $\bar{n}$ and overpredicts $L_{arm}$ for large $\varepsilon^*$.

Now we consider the specific effect of the ratio *C*. For $C < 1$ the *pp* attractions are stronger than *ps* attractions, while for $C > 1$the opposite is true. Figure 4.9 shows calculations for the average number of arms (bonds) per *s* colloid $\bar{n}$, fraction of chains which are star arms $\Psi$ and average length of star arms $L_{arm}$ versus $\varepsilon^*$ at a density of $\rho^* = 0.2$ and composition $x^{(s)} = 0.05$ for ratios *C* = 0.6, 0.8, 1, 1.2.

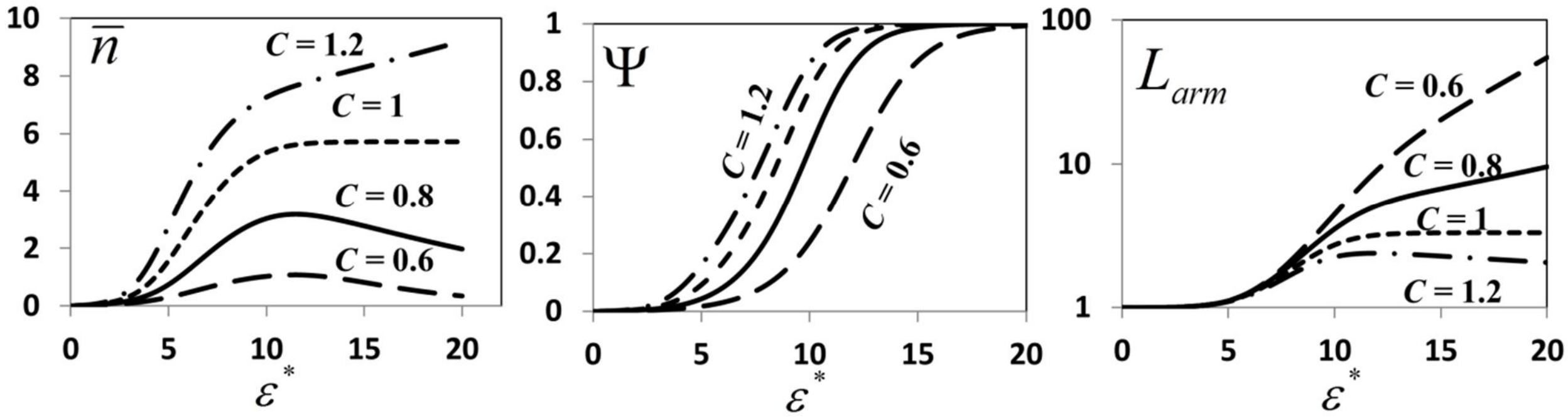

**Figure 4.9:** Average number of arms per *s* colloid $\bar{n}$ (left), fraction of chains which are star arms Ψ(middle) and average length of star arms $L_{arm}$ (right) versus association energy $\varepsilon^*$. Density is $\rho^*$ = 0.2, with an *s* colloid mole fraction $x^{(s)}$ = 0.05 for various values of the association energy ratio *C*

First we focus on the case C = 1.2. For this case the *sp* attractions are greater than *pp* attractions. Since *sp* attractions are favored, $\bar{n}$ is large, $\bar{n} \sim 7.3$ at $\varepsilon^*$ = 10, and increases steadily as $\varepsilon^*$ is increased. For the studied cases, $\bar{n}$ is largest at C = 1.2. In contrast, the average star arm length $L_{arm}$ is smallest for this case. This is a direct result of the large $\bar{n}$. Since there are a large number of arms and a finite number of *p* colloids, the arms must be shortest for this case. Initially $L_{arm}$ increases with $\varepsilon^*$ as association increases in the system, goes through a maximum and then decreases as *pp* bonds are traded for *ps* bonds. We also note that $\Psi$ is largest for this case.

Now considering the case $C$ = 0.6, where *pp* attractions are greater than *ps* attractions, we see that $\bar{n}$ initially increases with $\varepsilon^*$, goes through a maximum as *sp* bonds are traded for *pp* bonds and then decreases as $\varepsilon^* \to \infty$. It is for this $C$ that $\bar{n}$ is smallest, $\bar{n} \sim 1$ at $\varepsilon^* \sim 10$, and $L_{arm}$ is largest, $L_{arm} \sim 4.5$ at $\varepsilon^* \sim 10$. Increasing $\varepsilon^*$ further to $\varepsilon^* \sim 20$ we find $\bar{n} \sim 0.34$ and $L_{arm} \sim 550$, meaning the *s* colloids exist primarily as monomers and chain ends for long chains of *p* colloids. For the case $C$ = 0.8 we see similar behavior to the case $C$ = 0.6, although less pronounced. Finally considering the case $C$ = 1, we see that there are no maximums and both $L_{arm}$ and $\bar{n}$ reach a low temperature limiting value around $\varepsilon^* \sim 12$. From these results it is clear that the attraction ratio $C$ can be tuned to achieve a range of colloidal star molecules.

## 4.4: Summary and conclusions

We have developed a new theory to model binary mixtures of multi – patch *p* colloids and colloids with spherically symmetric attractions (*s* colloids). We developed the theory in Wertheim's [35, 3635, 36] multi – density formalism for associating fluids using modifications of the graphs developed in section 3.1 for the case of a mixture of single patch colloids and *s* colloids. We applied the theory to the case of a mixture of bi-functional *p* colloids, consisting of an *A* patch and *B* patch located on opposite poles of the colloid, and *s* colloids which only attract the *A* patch of the *p* colloid. There were only *AB* attractions between the patches on the *p* colloid, and there were no attractions between *s* colloids. This system was shown to self – assemble into a mixture of free chains and colloidal star molecules. The average arm length of star molecules, ratio of free chains to star arms and average number of arms per colloidal star can be manipulated by varying density, temperature, composition and the ratio of association energies. The theory was shown to be accurate in comparison to Monte Carlo simulation data. In the development of the theory we assumed there were no attractions between *s* colloids which allowed us to write the theory as a perturbation to a hard sphere reference fluid. To include attraction between the *s* colloids one would need to use an appropriate reference system for the *s* colloids. Besides the definition of the reference system, the general results presented here would still be valid.

# CHAPTER 5

# Bond angle dependence in two site associating fluids

Chapters 2 – 4 focused on the development of new perturbation theories which went beyond the single bonding condition allowing multiple bonds per association site. To account for multiple bonds per association site, steric effects had to be incorporated which accounted for the fact that when one molecule bonds at an association site, the bond volume available for a second molecule to bond at this same association site is decreased. Another type of steric hindrance which may arise, and has been neglected thus far, is the steric hindrance between association sites. When the association sites are widely separated, as in Fig. 1.4, steric hindrance between sites is negligible; when association sites are in close proximity to each other steric hindrance between association sites is significant. We will define the angle between the center of two association sites $A$ and $B$ to be $\alpha_{AB}$ which we loosely call the bond angle. Figure 5.1 gives examples of association into linear chains for both cases of large (case **a**), and small (case **b**) $\alpha_{AB}$. For case **a** there is no steric hindrance between sites, while for case **b** there will be significant steric hindrance which results in a decrease in the number of available associated states of the triatomic cluster. In other words, case **a** has a higher entropy than case **b**.

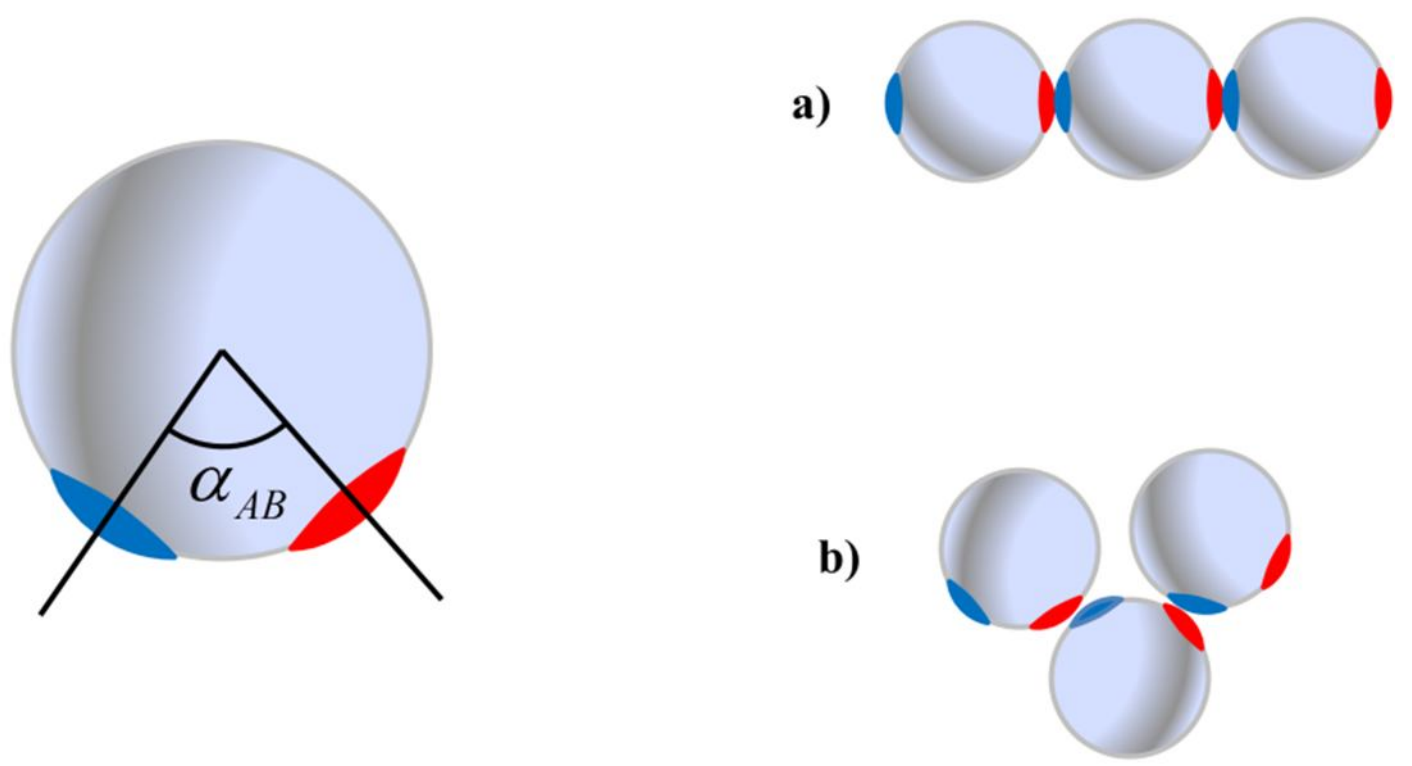

**Figure 5.1:** Examples of linear triatomic clusters for two angles $\alpha_{AB}$

As discussed in chapter 1, first order perturbation theory treats each association site independently meaning the theory is independent of $\alpha_{AB}$ . To account for steric effects between association sites Wertheim carried out a resummed perturbation theory,[37] which in first order requires only the pair correlation function of the hard sphere reference system and a blockage parameter which accounts for the decrease in bonding volume of one patch caused by association of the other. This approach was applied to model flexible chains of molecules and has never been applied to study the effects of bond angle on association.

In addition to steric effects, when bond angles are small as in case **b**, ring formation can become significant. Where now, as opposed to chapter 2, we are assuming association sites are singly bondable and rings are composed of cycles of *AB* bonds as depicted in Fig. 5.2. It was Sear and Jackson (SJ)[86] who were the first to introduce contributions for association into rings. In this approach the associated rings were treated ideally such that non-adjacent neighbors along the ring can overlap. The probability that a chain of colloids was in a valid ring state was approximated by the expression of Treolar[87] for the distribution of the end to end vector in a polymer chain. In this approach any dependence on $\alpha_{AB}$ is neglected when in reality $\alpha_{AB}$ plays a dominant role in determining if association into rings will occur. A recent study using lattice simulations has shown that ring formation is strongly dependent on $\alpha_{AB}$.[88] For instance, it is impossible to form 4-mer rings (and satisfy the one bond per site condition) when $\alpha_{AB} = 180^{\circ}$; however, decreasing $\alpha_{AB}$ to $90^{\circ}$ this type of ring would be possible. Tavares et *al.*[89] explored the possibility of ring formation in 2 patch colloid fluids with $\alpha_{AB} = 180^{\circ}$ by extending the approach of SJ[86] and found that to achieve appreciable ring formation the parameters of the interaction potential had to be chosen such that the one bond per patch condition would be violated. To correct for this in the simulations they used a model which restricts bonding to at most one bond per patch.[90]

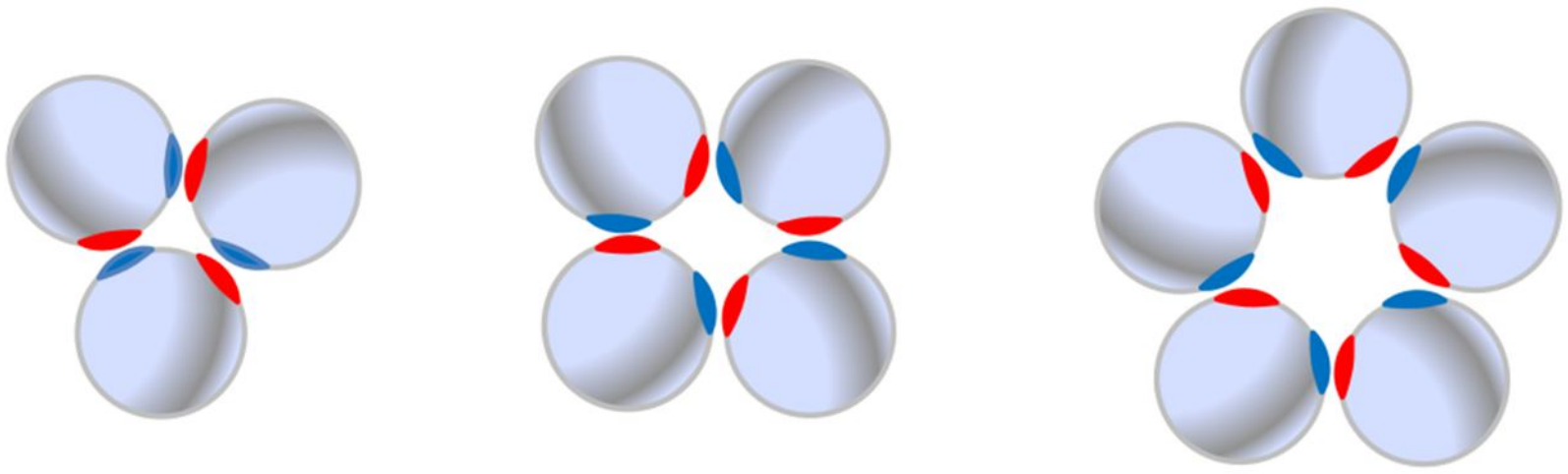

**Figure 5.2:** Examples of associated rings

Lastly, when $\alpha_{AB}$ is small enough, double bonded dimers may form. This interaction is depicted in graph **c** of Fig. 1.5. This contribution is neglected in TPT1. The formation of double bonded dimers was a problem initially tackled by SJ[91] who included the additional contribution for the double bonded dimer. The theory includes a geometric quantity which accounts for the probability that two colloids will be oriented such that double bonding can occur. This quantity was never explicitly evaluated and was treated as a parameter allowing only qualitative comparisons to be made. The ability of molecules / colloids to double bond will be strongly dependent on the angle $\alpha_{AB}$.

In this chapter we extend Wertheim's perturbation theory to model these two site fluids with small (or large) angles $\alpha_{AB}$. Our goal is to derive a single theory which will be accurate over the full range of $\alpha_{AB}$. To accomplish this we will combine and extend the resummed perturbation theory of Wertheim[37], theory for double bonded dimers of SJ[91] and a modified version of the approach of SJ[86] for ring formation. We explicitly include the dependence on bond angle $\alpha_{AB}$ in each contribution of the theory and evaluate all required geometric integrals rigorously. To test the new theory we perform new Monte Carlo simulations to determine the effect of $\alpha_{AB}$ on the fractions of colloids associated into chains, rings, double bonded dimers as well as the effect of $\alpha_{AB}$ on internal energy and pressure. The theory is found to be in excellent agreement with simulation.

## 5.1: Theory

In this section the theory will be developed for colloids / molecules of diameter $d$ with an $A$ patch and a $B$ patch with the centers of the patches having a bond angle $\alpha_{AB}$ in relation to each other as illustrated in Fig. 5.1. We will refer to the associating units as colloids; however, as in the rest of this dissertation, the results can also be applied as a primitive model of molecules. Again we consider the primitive conical site model discussed in section 2.1. We restrict association such that there are $AB$ attractions, but no $AA$ or $BB$ attractions. Also, we restrict $\theta_c$ such that each association site is singly bondable. For this case the interaction potential between two colloids is given as

$$\phi(12) = \phi_{HS}(r_{12}) + \phi_{AB}(12) + \phi_{BA}(12) \tag{5.1}$$

To ensure that each patch can only bond once we choose $r_c = 1.1\sigma$ and $\theta_c = 27^\circ$.

For this two site case, in the absence of an external field, Eq. (1.39) can be simplified to

$$\frac{A - A^{HS}}{Vk_B T} = \rho \ln\left(\frac{\rho_o}{\rho}\right) - \sigma_A - \sigma_B + \frac{\sigma_A \sigma_B}{\rho_o} + \rho - \frac{\Delta c^{(o)}}{V} \tag{5.2}$$

For colloids with small bond angles we will have to account for chain formation $\Delta c_{ch}^{(o)}$, association into double bonded dimers $\Delta c_d^{(o)}$ and lastly rings of associated colloids $\Delta c_{ring}^{(o)}$ giving the graph sum

$$\Delta c^{(o)} \approx \Delta \hat{c}_1^{(o)} = \Delta c_{ch}^{(o)} + \Delta c_d^{(o)} + \Delta c_{ring}^{(o)} \tag{5.3}$$

For the chain contribution we use the first order resummed perturbation theory RTPT1 of Wertheim[37] to account for the fact that at small bond angles, association at one patch can block association at the other

$$\frac{\Delta c_{ch}^{(o)}}{V} = \frac{\sigma_A \sigma_B \kappa f_{AB} \xi}{1 + (1 - \Psi)\kappa f_{AB} \rho_o \xi} \tag{5.4}$$

where $f_{AB} = \exp(\varepsilon_{AB} / k_B T) - 1$ is the magnitude of the association Mayer function, $\kappa$ is the probability that two monomers are oriented such that a certain patch on one colloid can bond to a certain patch on the other and is given by Eq. (2.21), and $\xi$ is given by Eq. (2.17). The last term to consider in $\Delta c_{ch}^{(o)}$ is the blockage integral $\Psi$ which accounts for the fact that as the bond angle is decreased bonding at one patch will decrease the available bond volume of the other patch due to steric hindrance. Wertheim developed RTPT1 in the context of spheres which bond at contact with 2 glue spots *A* and *B*. In Wertheim's treatment once spheres formed a bond they were stuck and would not rattle around in the bond volume defined by Eqn. (2.1). In the current work $\alpha_{AB}$ defines the angle between patch centers which we call the bond angle; for a given $\alpha_{AB}$ the actual angle $\alpha'_{AB}$ between the first and third spheres in an associated linear triatomic cluster can vary in the range $\alpha_{AB} - 2\theta_c \le \alpha'_{AB} \le \alpha_{AB} + 2\theta_c$. In Wertheim's analysis of RTPT1 this was not the case; there was no rattling in the bond volume meaning $\alpha'_{AB} = \alpha_{AB}$. To account for bond flexibility Wertheim introduced normalized bond angle distribution functions $\zeta(\alpha_{AB})$. For hard spheres with glue spot bonding Wertheim found $\Psi = 1 - L = \int_{\pi/3}^{\pi} \zeta(\alpha_{AB}) \sin(\alpha_{AB}) d\alpha_{AB}$; which is simply the fraction of $\alpha_{AB}$ states which will not result in hard sphere overlap when both glue spots are bonded. Our case here is somewhat different since we do not have glue spot bonding. We set $\alpha_{AB}$, $r_c$ and $\theta_c$, not $\zeta(\alpha_{AB})$; however the interpretation of $\Psi$ is similar. In our case $\Psi$ is the ratio of the number of states where three colloids associate to form a linear triatomic cluster in which there is no hard sphere overlap between the unbounded pair, to the number of states if there were no steric interference and the patches were independent. For the model considered here this fraction $\Psi$ is approximated by the integral

$$\Psi = \frac{\int_0^{2\pi}\int_0^{\theta_c}\int_d^{r_c}\int_0^{2\pi}\int_0^{\theta_c}\int_d^{r_c} dr_{12} r_{12}^2 d\theta_{12} \sin\theta_{12} d\phi_{12} dr_{13} r_{13}^2 d\theta_{13} \sin\theta_{13} d\phi_{13} e_{HS}(r_{23})}{\left(\int_0^{2\pi}\int_0^{\theta_c}\int_d^{r_c} dr_{12} r_{12}^2 d\theta_{12} \sin\theta_{12} d\phi_{12}\right)\left(\int_0^{2\pi}\int_0^{\theta_c}\int_d^{r_c} dr_{13} r_{13}^2 d\theta_{13} \sin\theta_{13} d\phi_{13}\right)} \tag{5.5}$$

Where in Eq. (5.5) colloid 2 and 3 are both bonded to colloid 1, $\theta_{12}$ is the polar angle that the vector $\vec{r}_{12} = \vec{r}_2 - \vec{r}_1$ makes in a coordinate system centered on colloid 1 whose *z* axis lies on the site vector $\vec{r}_A$ of colloid 1 and $\phi_{12}$ is the corresponding azimuthal angle. The angles $\theta_{13}$ and $\phi_{13}$ are similarly defined with respect to the site vector $\vec{r}_B$ of colloid 1. The $e_{HS}$ prevents hard sphere overlap between the colloids in the associated cluster and is given by $e_{HS}(r) = H(r-d)$ where $H$ is the Heaviside step function. For the case of total blockage of one patch by the other the reference system $e_{HS}(r_{23}) = 0$ for all configurations of the cluster resulting in $\Psi = 0$. When the bond angle is sufficiently large that the two patches are independent the reference system $e_{HS}(r_{23}) = 1$ for all configurations resulting in $\Psi = 1$.

When the condition $\alpha_{AB} - 2\theta_c < 0$ holds true it is possible for double bonding of colloids to occur; that is, according to the potential given by Eq. (2.1), when the vector connecting the centers of the two colloids $\vec{r}_{12}$ passes through both patches on both colloids and $r_{12} \le r_c$, the two colloids are considered to be double bonded. This situation is depicted in Fig. 5.3.

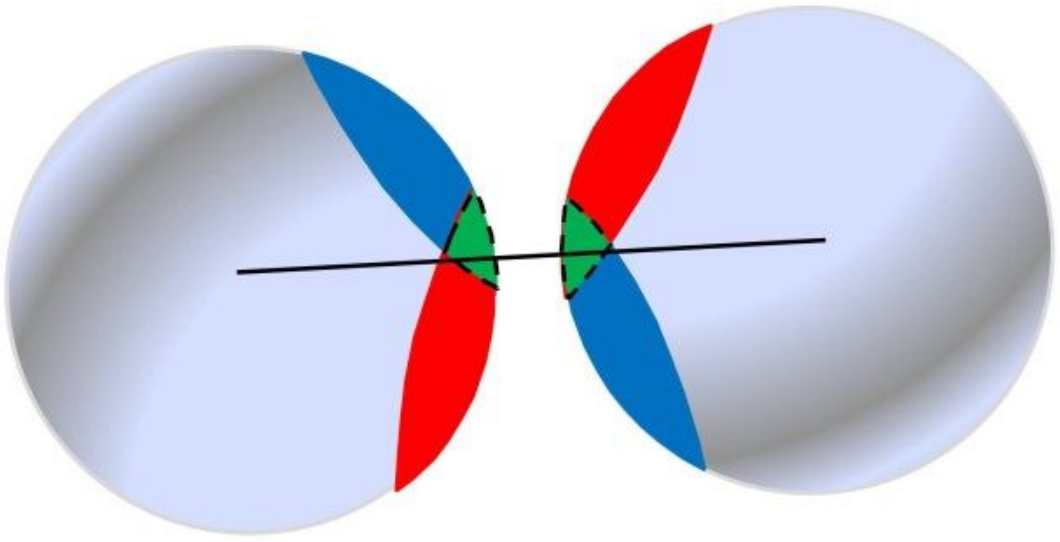

**Figure 5.3:** Diagram of double bonded colloids. The line connecting the centers of each colloid must pass through the solid angle of the overlap of both patches, outlined in dashed curve, for double bonding to occur

The contribution to the graph sum which accounts for double bonded dimers is given by SJ (we introduce different constants but the result is the same)[91]

$$\Delta c_d^{(o)} / V = \rho_o^2 f_{AB}^2 \xi I_d / 2 \tag{5.6}$$

Where $I_d$ is the probability that two colloids are oriented such that double bonding can occur. In the work of SJ[91] $I_d$ was not defined in this way and was not explicitly evaluated; instead it was written in terms of a parameter which allowed only qualitative discussion and could not explicitly be compared to simulation. With the identification of $I_d$ as the probability two colloids are oriented for double bonding, and the definition of the patchy potential in Eq. (2.1), we can easily evaluate $I_d$ as follows. Consider the two colloids in Fig. 5.3 whose bond angle satisfies the condition $\alpha_{AB} - 2\theta_c \le 0$. For one colloid to be in an orientation to double bond to the other, the vector connecting the centers of the two colloids must pass through the area of surface where the two patches overlap. This is the dashed outline area in Fig. 5.3. If all orientations of a colloid are equally likely, the probability of this occurring will simply be the ratio of the surface area of patch overlap to the total surface area of the sphere $S_{AB} / 4\pi$ where $S_{AB}$ is the solid angle of the overlap of the two patches. This solid angle $S_{AB}$ is simply the solid angle of the intersection of 2 cones of apex angle $\theta_c$ which share a common origin whose axes are at an angle of $\alpha_{AB}$ to each other. For association to occur both colloids must be oriented correctly, so we square the single colloid probability to obtain

$$I_d = \begin{cases} 0 & for \;\; \alpha_{AB} - 2\theta_c \ge 0 \\ S_{AB}^2 / 16\pi^2 & otherwise \end{cases} \tag{5.7}$$

The solid angle $S_{AB}$ has been solved elsewhere[92] and is given by

$$S_{AB} = 4\cos^{-1}\left(\frac{\sin\gamma_{AB}}{\sin\theta_c}\right) - 4\cos(\theta_c)\cos^{-1}\left(\frac{\tan\gamma_{AB}}{\tan\theta_c}\right) \tag{5.8}$$

where $\gamma_{AB}$ is obtained through the relation $\gamma_{AB} = \tan^{-1}\left((1-\cos\alpha_{AB})/\sin\alpha_{AB}\right)$.

Finally, we must account for the possibility of rings of associated colloids with the contribution

$$\Delta c_{ring}^{(o)} = \sum_{n=3}^{\infty} \Delta c_n^{ring} \tag{5.9}$$

where $\Delta c_n^{ring}$ is the contribution for rings of size $n$. The contributions $\Delta c_n^{ring}$ are given by

$$\Delta c_n^{ring}/V = \frac{(\rho_o f_{AB})^n}{n\left(8\pi^2\right)^{n-1}} \int \mathrm{O}_{AB}(12)\ldots\mathrm{O}_{AB}(n-1,n)\mathrm{O}_{AB}(1,n) g_{HS}(\vec{r}_1 \ldots \vec{r}_n) d\vec{r}_2 d\Omega_2 \ldots d\vec{r}_n d\Omega_n \tag{5.10}$$

where $\mathrm{O}_{AB}(1,2) = -\phi_{AB}(12)/\varepsilon_{AB}$ are the overlap functons. Equation (5.10) is more general than the ring graph of Sear and Jackson[86] with the introduction of the $n$ - body correlation function of the hard sphere reference system $g_{HS}(\vec{r}_1 \ldots \vec{r}_n)$ which we approximate as

$$g_{HS}(\vec{r}_1 \ldots \vec{r}_n) = \prod_{\substack{\text{bonded pairs} \\ \{i,j\}}} y_{HS}(r_{ij}) \prod_{\substack{\text{all pairs} \\ \{l,k\}}} e_{HS}(r_{lk}) \tag{5.11}$$

The superposition given by Eq. (5.11) gives a $g_{HS}(r_{ij}) = y_{HS}(r_{ij}) e_{HS}(r_{ij})$ to each pair of colloids which share an association bond and an $e_{HS}(r_{lk})$ to each unbounded pair which serves to prevent hard sphere overlap between non-adjacent spheres in the ring.

In the approach taken by SJ[87], Eq. (5.11) is replaced by a simple linear superposition of pair correlation functions and the integral in Eq. (5.10) is approximated as $\Delta c_n^{ring}/V =$ $\left(f_{AB}\rho_o g_{HS}(d)K\right)^n W_{n-1}/n$ where $K = 4\pi d^2(r_c - d)\kappa$ and $W_{n-1}$ is the probability that, in a freely jointed chain, the first and last sphere in the chain are in contact and is obtained using the expression of Treolar[87]. The SJ approximation of the ring integral is not useful in our current approach because the effect of bond angle has not been included. For instance, in the SJ approximation there will be a significant probability colloids which have a bond angle of $180^{\circ}$ will associate into triatomic rings while in reality this is geometrically impossible.

In this work we treat Eq. (5.10) in a more general way which allows for the inclusion of bond angle dependence. Using the approximation Eq. (2.16) with the fact that for $r \geq d$, $g_{HS}(r) = y_{HS}(r)$ allows us to rewrite $\Delta c_n^{ring}$ as

$$\Delta c_n^{ring} / V = \frac{\left(f_{AB}\rho_o g_{HS}(d) K\right)^n}{n d^3} I_r^{(n)} \tag{5.12}$$

Where the ring integral $I_r^{(n)}$ is given by

$$I_r^{(n)} = \frac{d^3}{\left(8\pi^2\right)^{n-1} K^n} \int \prod_{\substack{bonded\ pairs \\ \{i,j\}}} \left(\frac{d^p}{r_{ij}^p} O_{AB}(i,j)\right) \prod_{\substack{all\ pairs \\ \{l,k\}}} e_{HS}(r_{lk}) d\vec{r}_2 d\Omega_2 \ldots d\vec{r}_n d\Omega_n \tag{5.13}$$

The probability distribution function of a ring of size *n* is in a configuration (1…*n*) is given by

$$P_r^{(n)}(1 \ldots n) = \frac{\prod\limits_{\substack{bonded\ pairs \\ \{i,j\}}} O_{AB}(i,j) \prod\limits_{\substack{all\ pairs \\ \{l,k\}}} e_{HS}(r_{lk})}{Z_r^{(n)}} \tag{5.14}$$

Where $Z_r^{(n)}$ is the ring partition function given by

$$Z_r^{(n)} = \int \prod_{\substack{bonded\ pairs \\ \{i,j\}}} O_{AB}(i,j) \prod_{\substack{all\ pairs \\ \{l,k\}}} e_{HS}(r_{lk}) d\vec{r}_2 d\Omega_2 \ldots d\vec{r}_n d\Omega_n \tag{5.15}$$

Combining (5.13) – (5.15) we obtain

$$I_r^{(n)} = \Gamma^{(n)} \left\langle \prod_{\substack{bonded\ pairs \\ \{i,j\}}} \left(\frac{d^p}{r_{ij}^p}\right) \right\rangle \tag{5.16}$$

where

$$\Gamma^{(n)} = \frac{d^3 Z_r^{(n)}}{\left(8\pi^2\right)^{n-1} K^n} \tag{5.17}$$

and $\langle\ \rangle$ represents an average over the distribution function Eq. (5.14). Since $P_r^{(n)}$ is only nonzero when there is no hard sphere overlap we can accurately approximate this average as

$$\left\langle \prod_{\substack{bonded\ pairs \\ \{i,j\}}} \left( \frac{d^p}{r_{ij}^p} \right) \right\rangle \approx \frac{2^{np}}{\left(r_c / d + 1\right)^{np}} \tag{5.18}$$

which states that on average each colloid pair should be approximately located in the middle of the range $\{d \le r \le r_c\}$. Using this approach $\Delta c_n^{ring}$ is explicitly dependent on bond angle through the partition function which must be evaluated numerically. Combining these results we obtain

$$\Delta c_n^{ring} / V = \frac{\left(f_{AB} \rho_o \hat{g}_{HS} K\right)^n}{n d^3} \Gamma^{(n)} \tag{5.19}$$

where

$$\hat{g}_{HS} = \frac{2^p g_{HS}(d)}{\left(r_c / d + 1\right)^p} \tag{5.20}$$

Now that the graph sum has been fully specified the densities of colloids in each bonding state can be obtained. For the two patch case three densities describe all possible bonding states of the colloid: the monomer density $\rho_o$, density of colloids bonded at patch $A$ (or equivalently $B$) $\rho_A = \rho_B$ and the density of colloids bonded at patches $A$ and $B$ $\rho_{AB}$. We obtain $\rho_A$ and $\rho_{AB}$ through Eq. (1.37) as

$$\frac{\rho_A}{\rho_o} = \frac{\sigma_A}{\rho_o} - 1 = \frac{\sigma_A \kappa f_{AB} \xi}{1 + (1 - \Psi) \kappa f_{AB} \rho_o \xi} \tag{5.21}$$

and

$$\frac{\rho_{AB}}{\rho_o} = \frac{\partial}{\partial \rho_o} \frac{\sigma_B \sigma_A \kappa f_{AB} \xi}{1 + (1 - \Psi)\kappa f_{AB} \rho_o \xi} + \left( \frac{\sigma_A \kappa f_{AB} \xi}{1 + (1 - \Psi)\kappa f_{AB} \rho_o \xi} \right)^2 + \rho_o f_{AB}^2 \xi I_d + \sum_{n=3}^{\infty} \frac{\left(f_{AB} \rho_o \hat{g}_{HS} K\right)^n}{\rho_o d^3} \Gamma^{(n)}$$

(5.22)

The density of colloids bonded twice must satisfy the relation

$$\rho_{AB} = \rho_{2c} + \rho_d + \sum_{n=3}^{\infty} \rho_n^{ring} \tag{5.23}$$

Where $\rho_{2c}$ is the density of colloids which are bonded at both patches in a linear chain, $\rho_d$ is the density of colloids in double bonded dimers and $\rho_n^{ring}$ is the density of colloids in rings of size *n*. The first two terms on the right hand side (RHS) of Eq. (5.22) correspond to $\rho_{2c}$ and can be simplified as

$$\frac{\rho_{2c}}{\rho_o} = \Psi \left( \frac{\sigma_A \kappa f_{AB} \xi}{1 + (1 - \Psi)\kappa f_{AB} \rho_o \xi} \right)^2 \tag{5.24}$$

For complete blockage of one patch by the other $\Psi \to 0$ resulting in $\rho_{2c} \to 0$, while for independent patches $\Psi \to 1$ and the TPT1 result is obtained. Likewise, the third term on the RHS of Eq. (5.22) corresponds to $\rho_d$

$$\frac{\rho_d}{\rho_o} = \rho_o f_{AB}^2 \xi I_d \tag{5.25}$$

and the fourth term on the RHS of Eq. (5.22) gives the sum of the densities of colloids in rings of size *n*

$$\frac{\rho_n^{ring}}{\rho_o} = \frac{\left(f_{AB} \rho_o \hat{g}_{HS} K\right)^n}{\rho_o d^3} \Gamma^{(n)} \tag{5.26}$$

The total density is given as the sum over all of the bonding states of the colloids

$$\rho = \rho_o + 2\rho_A + \rho_{AB} \tag{5.27}$$

Since $\rho$ is known Eqns. (5.21) and (5.27) provide a closed set of equations to solve for $\rho_o$ and $\rho_A$ from which $\rho_{2c}$, $\rho_d$ and $\rho_n^{ring}$ immediately follow. Combining these results the free energy can be simplified to the following form

$$\frac{A - A^{HS}}{Nk_BT} = \ln X_o + 1 - X_A - \frac{X_d}{2} - \sum_{n=3}^{\infty} \frac{X_n^{ring}}{n} \tag{5.28}$$

Where we have introduced the fractions $X_A = \sigma_B / \rho$ is the fraction of colloids not bonded at patch A, $X_d = \rho_d / \rho$ is the fraction of colloids in double bonded dimers and $X_n^{ring} = \rho_n^{ring} / \rho$ is the fraction of colloids associated in rings of size *n*.

## 5.2: Geometric integrals

To apply the theory we must evaluate Eqns. (5.5) and (5.17) for the integrals $\Psi$ and $\Gamma^{(n)}$. Due to the highly discontinuous nature of these integrals they must be evaluated using Monte Carlo integration. Obviously we cannot evaluate the sum over ring fractions for all possible ring sizes, so we truncate the sum at *n* = 7. As will be seen, this is more than sufficient to describe the conditions studied in this chapter. The integrals $\Psi$ are very easily evaluated using Monte Carlo integration as

$$\Psi = \left\{ \begin{array}{l} \text{The probability that if the positions of two colloids are generated such} \\ \text{that they are correctly positioned to associate with a third colloid, that} \\ \text{there is no core overlap between the two generated colloids} \end{array} \right\} \tag{5.29}$$

The partition function $Z_r^{(n)}$ in Eq. (5.17) is obtained by generating chains of associated colloids and calculating the probability that a chain is in a valid ring configuration. We find that the integrals $\Gamma^{(n)}$ are well correlated as a function of bond angle using the skewed Gaussian function

$$\Gamma^{(n)} = A_n \exp\left(-B_n\left(\alpha_{AB} - C_n\right)^2\right)\left(1 + erf\left(-D_n\left(\alpha_{AB} - C_n\right)\right)\right). \tag{5.30}$$

The constants $A_n, B_n, C_n$ and $D_n$ depend on ring size *n* and are given for ring sizes *n* = 3 – 10 in table 5.1. In Eq. (5.30) $\alpha_{AB}$ must be given in degrees. Both Eqns. (5.29) and (5.30) are independent of density and temperature, so they only need to be performed once for each $\alpha_{AB}$.

**Table 5.1:** Constants for ring integral correlation Eq. (5.30) for ring sizes *n* = 3 – 10

| *n* | $A_n$ | $B_n$ | $C_n$ | $D_n$ |
|---|---|---|---|---|
| 3 | 0.681 | 0.00514 | 60.0 | 0.00116 |
| 4 | 0.0651 | 0.00111 | 94.4 | 0.0907 |
| 5 | 0.0231 | 0.00159 | 105.8 | 0.0668 |
| 6 | 0.0111 | 0.00180 | 112.0 | 0.0461 |
| 7 | 0.00363 | 0.00278 | 107.5 | 0.0452 |
| 8 | 0.00248 | 0.00271 | 109.8 | 0.0452 |
| 9 | 0.00208 | 0.00291 | 112.2 | 0.0452 |
| 10 | 0.00165 | 0.00276 | 113.6 | 0.0452 |

Figure 5.4 shows numerical calculations for $\Psi$ and $\Gamma^{(n)}$ for *n* = 3 – 7. We have also included the analytical solution of $I_d$ Eq. (5.7) for comparison. All calculations were performed for the case $\theta_c = 27^\circ$ and $r_c = 1.1d$ . As expected $\Psi$ vanishes for small $\alpha_{AB}$ due to steric hindrance and becomes unity for large $\alpha_{AB}$ when association at one patch no longer interferes with the ability of the other patch to bond. The ring integrals $\Gamma^{(n)}$ are peaked around an optimum bond angle for ring formation and the maximums of $\Gamma^{(n)}$ decrease and shift to larger bond angles as $n$ increases. The double bonding integral $I_d$, which represents the probability two colloids are oriented such that double bonding can occur, vanishes for $\alpha_{AB} > 54^{\circ}$. In the limiting case $\alpha_{AB} = 0^{\circ}$ the integral $I_d \to \kappa$ due to the fact that since both patches are superimposed the probability two colloids are oriented for double bonding is just the probability that two monomers are oriented such that a specific patch on one colloid can bond to a specific patch on the other. By inspection of the integrals in Fig. 5.4 we should expect, in strongly associating fluids, double bonded dimers to dominate for small $\alpha_{AB}$, rings to dominate for moderate $\alpha_{AB}$ and chains to dominate for large $\alpha_{AB}$. It will be shown that this is indeed the case.

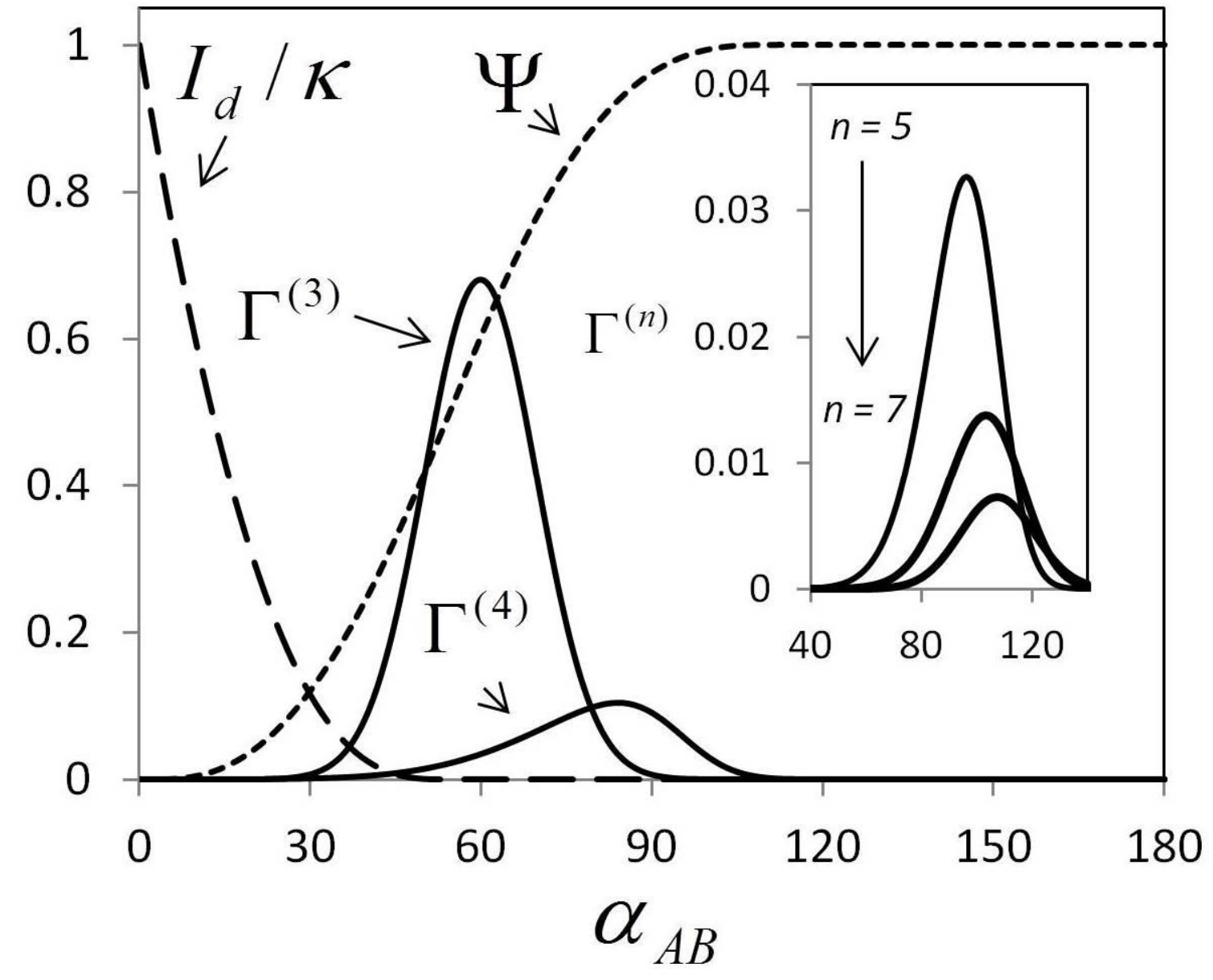

**Figure 5.4:** Geometric integrals used in the application of the theory. $I_d$ is the probability two colloids are oriented such that double bonding can occur, $\Psi$ is the blockage parameter which accounts for the fact bonding at one patch can interfere with bonding at the other, $\Gamma^{(n)}$ are the ring integrals which are proportional to the total number of ways $n$ colloids can associate into rings of size $n$. Inset of figure gives $\Gamma^{(n)}$ for $n$ = 5 – 7 with 5 being the largest peak and 7 being the smallest

## 5.3: Numerical calculations

To test the theory we perform *NVT* (constant *N*, *V* and *T*) and *NPT* (constant pressure *P*, *V* and *T*) Monte Carlo simulations. The colloids interact with the potential given by Eq. (5.1) with $r_c = 1.1d$ and $\theta_c = 27^\circ$. The simulations were allowed to equilibrate for $10^6$ cycles and averages were taken over another $10^6$ cycles. A cycle is defined as *N* attempted trial moves where a trial move is defined as an attempted relocation and reorientation of a colloid. For the *NPT* simulations a volume change was attempted once each cycle. For the majority of simulations performed in this work, small clusters of associated colloids (double bonded dimers, trimer rings etc…) dominate the fluid even at low temperatures. For this reason a choice of *N* = 256 colloids is sufficient to obtain good statistics. For larger bond angles where colloids can polymerize into longer chains at low temperatures,[93] we performed additional simulations using *N* = 864 colloids. Increasing the number of colloids had no significant effect on the simulated quantities.

We begin with a discussion of the effect of bond angle on the fraction of colloids which are monomers $X_o$, bonded once $X_1$ and bonded twice $X_2 = 1 - X_o - X_1$ at constant packing fraction $\eta$ and association energy $\varepsilon^* = \varepsilon_{AB} / k_B T$. Figure 5.5 gives these fractions at a relatively low $\varepsilon^*$ = 5 and for the higher association energy case $\varepsilon^*$ = 8. For each $\varepsilon^*$ we consider low density $\eta = 0.1$ and high density $\eta$ = 0.4 fluids. In all cases $X_2$ dominates for small $\alpha_{AB}$ and decreases to some limiting value as the bond angle dependence saturates, around $\alpha_{AB} = 60^\circ$ for $\varepsilon^* = 5$ and $\alpha_{AB} = 90^\circ$ for $\varepsilon^* = 8$. It is at these bond angles that TPT1 becomes accurate. The fractions $X_o$ and $X_1$ are a maximum at large $\alpha_{AB}$ and then decrease rapidly as $X_2$ increases at smaller $\alpha_{AB}$. As expected, the general trend observed for all $\alpha_{AB}$ is that association between the colloids increases with increasing $\varepsilon^*$ and $\eta$. Overall the theory and simulation are in excellent agreement.

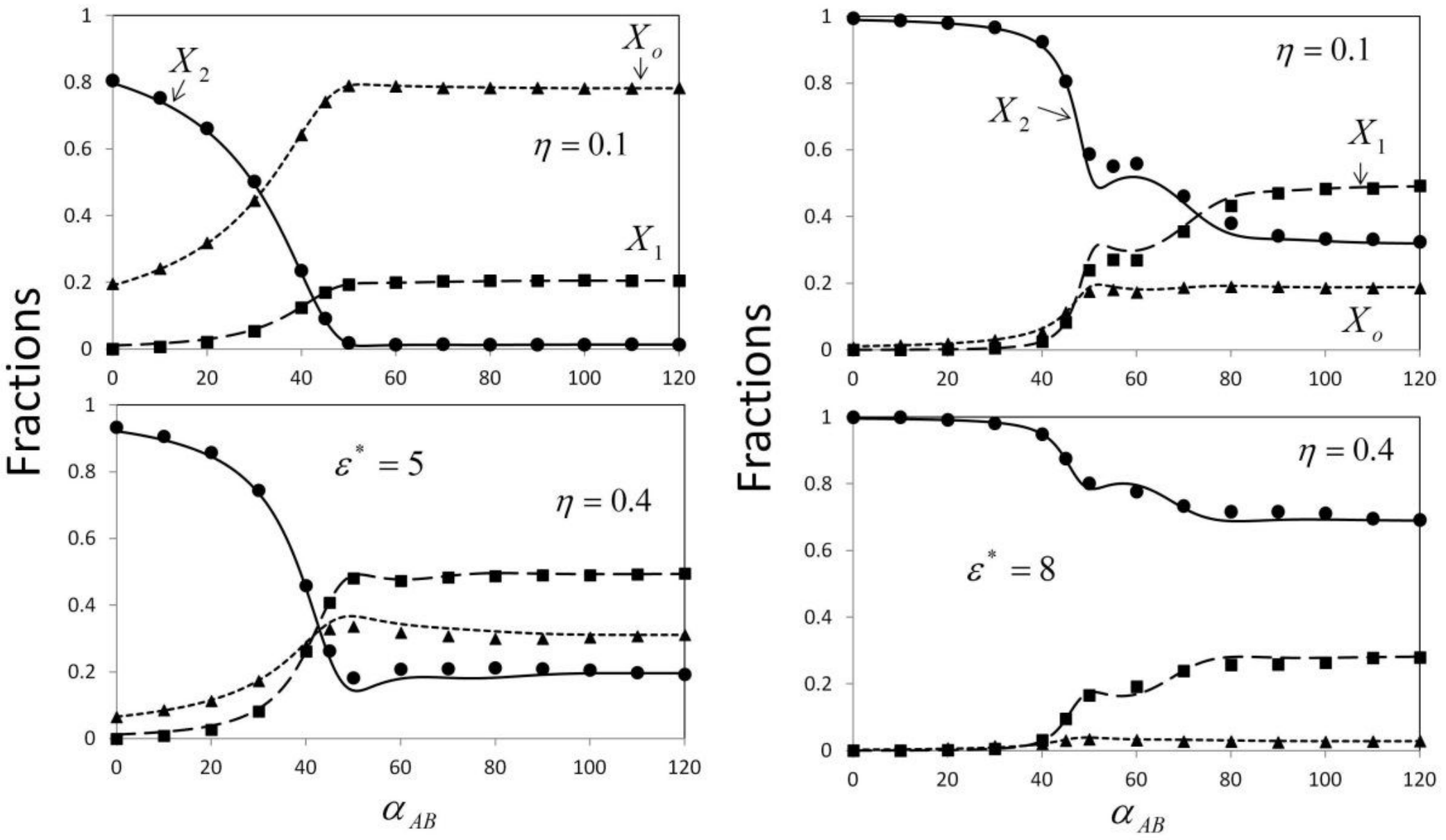

**Figure 5.5:** Monomer fractions $X_o$ (short dashed line – theory, triangles – simulation), fractions of colloids bonded once $X_1$ (long dashed line – theory, squares – simulation) and fractions of colloids bonded twice $X_2$ (solid line – theory, circles – simulation) as a function of bond angle $\alpha_{AB}$. Association energies are $\varepsilon^* = 5$ (left panel) and $\varepsilon^* = 8$ (right panel) for packing fractions $\eta = 0.1$ (top) and $\eta = 0.4$ (bottom)

To better explain the behavior observed in Fig. 5.5 we show the fraction of colloids in doubly bonded dimers $X_d$, fraction in rings of size $n$ $X_n^{ring}$ and fraction of colloids bonded at both patches in a chain (not a ring or double bonded dimer) $X_{2c} = \rho_{2c} / \rho$ for the case $\varepsilon^* = 8$ in Fig. 5.6. For small $\alpha_{AB}$, double bonded dimers dominate. For these small bond angles $I_d$ is maximum, ring formation is impossible due to vanishing $\Gamma^{(n)}$, $\Psi \rightarrow 0$ meaning steric hindrance between patches is nearly complete, and finally there are more mutual orientations where colloids can form a double bond than there are mutual orientations where a single bond is formed (at $\alpha_{AB} = 0^{\circ}$ single bonding of a patch becomes impossible). The tendency of the colloids to double bond is the genesis of the $X_2$ dominance for small $\alpha_{AB}$. As bond angle is increased

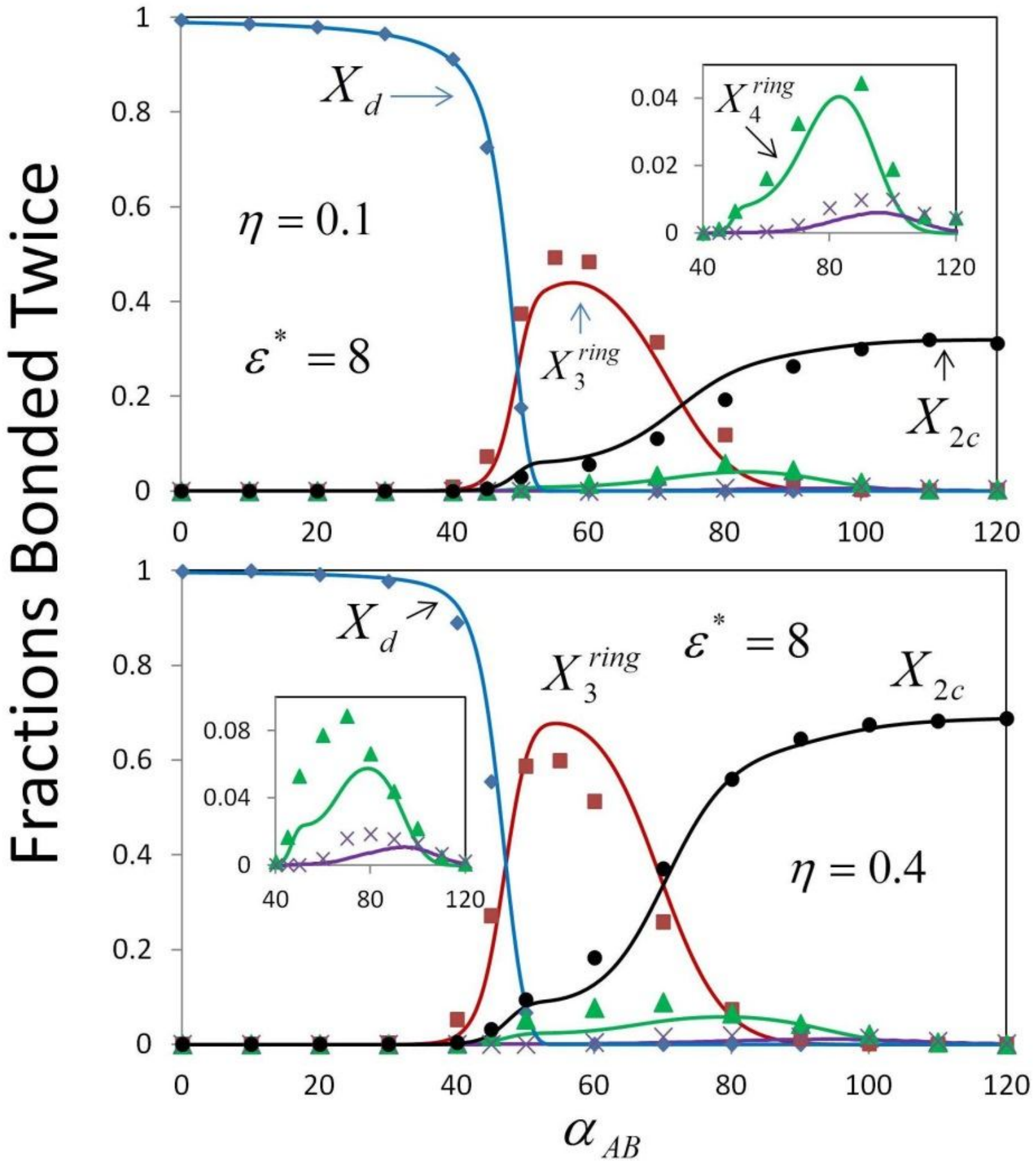

**Figure 5.6:** Fractions of colloids bonded twice in dimers $X_d$ , twice in chains $X_{2c}$ and twice in rings of size $n$ for $n$ = 3 – 5 for an association energy of $\varepsilon^*$ = 8 at packing fractions $\eta$ = 0.1 (top) and $\eta$ = 0.4 (bottom). Curves give theoretical predictions and symbols give simulation results. Insets show ring fractions for $n$ = 4 (triangles) and $n$ = 5(crosses)

the solid angle available for double bonding decreases and vanishes completely at $\alpha_{AB} = 54^{\circ}$.

In the region $50^{\circ} \le \alpha_{AB} \le 70^{\circ}$ rings become the dominant type of associated cluster in the fluid.

The reason for this can be seen in the geometric integrals given in Fig. 5.4. In this region $\Psi$ is depleted and the ring integrals $\Gamma^{(3)}$ and $\Gamma^{(4)}$ are significant. The maximum of $X_3^{ring}$ is significantly larger than the maximum of $X_4^{ring}$, which in turn is much larger than that seen in $X_5^{ring}$. For $n > 5$ $X_n^{ring}$ ring is small for all $\alpha_{AB}$ at these conditions. For $\alpha_{AB} > 70^{\circ}$, $X_{2c}$ becomes the dominant contribution to $X_2$ due to decreasing $\Gamma^{(n)}$ and the fact that there is little steric hindrance between patches. The theory does an excellent job in describing the various bonding fractions of the system.

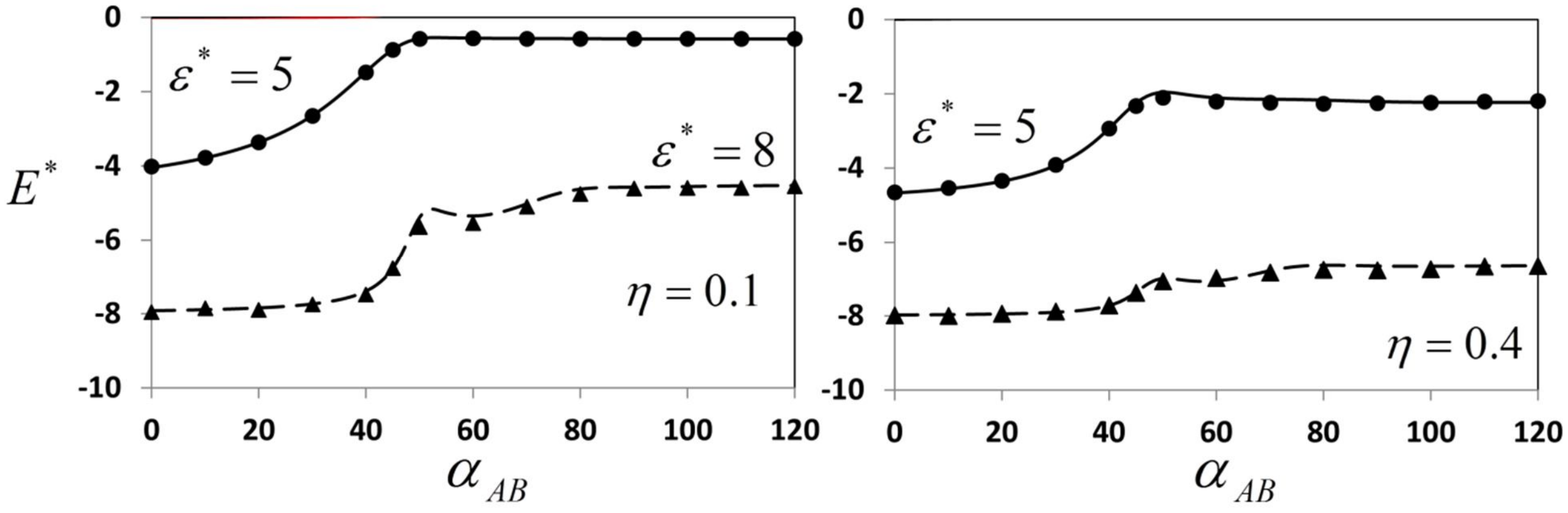

**Figure 5.7:** Bond angle dependence of the excess internal energy for $\varepsilon^*$ = 5(solid curve – theory, circles – simulation) and $\varepsilon^*$ = 8 (dashed curve – theory, triangles – simulation) for $\eta$ = 0.1(left) and $\eta$ = 0.4 (right)

Figure 5.7 shows the $\alpha_{AB}$ dependence of the excess internal energy $E^* = E_{AS} / Nk_BT$. For each case $|E^*|$ is largest for small $\alpha_{AB}$. This is due the fact that at these bond angles double bonded dimers dominate which give the energetic benefit of forming a double bond for the entropic penalty of forming a single bond in the large $\alpha_{AB}$ case. Increasing $\alpha_{AB}$ decreases $X_d$ resulting in a corresponding decrease in $|E^*|$. For $\varepsilon^* = 8$, $E^*$ shows oscillatory behavior in the region $40^{\circ} \le \alpha_{AB} \le 80^{\circ}$ as the system switches between various modes of association. The energy reaches a limiting value near $\alpha_{AB} = 80^{\circ}$ at which point TPT1 would give accurate predictions.

As can be seen, the current theory is in excellent agreement with simulation over the full range of bond angles.

Now we wish to explore the effect of $\alpha_{AB}$ on pressure. To determine the performance of the new theory in the prediction of pressure we performed *NPT* simulations for 3 isotherms. Figure 5.8 compares theory and simulation predictions for $\alpha_{AB} = 45^\circ$ at $\varepsilon^* = 4$ and 8 and $\alpha_{AB} = 180^\circ$ at $\varepsilon^* = 8$. For $\alpha_{AB} = 45^\circ$ increasing $\varepsilon^*$ decreases pressure due to the fact that more colloids are associating into clusters. The system has a significantly lower pressure for $\alpha_{AB} = 180^\circ$ than $\alpha_{AB} = 45^\circ$ at $\varepsilon^* = 8$. The reason for this can be seen in the types of associated clusters observed in Fig. 5.6. For $\alpha_{AB} = 45^\circ$ the system is dominated by small clusters such as double bonded dimers and triatomic rings, while for $\alpha_{AB} = 180^\circ$ the system is dominated by larger clusters of associated linear chains. Overall the theory does a good job in predicting the pressure isotherms.

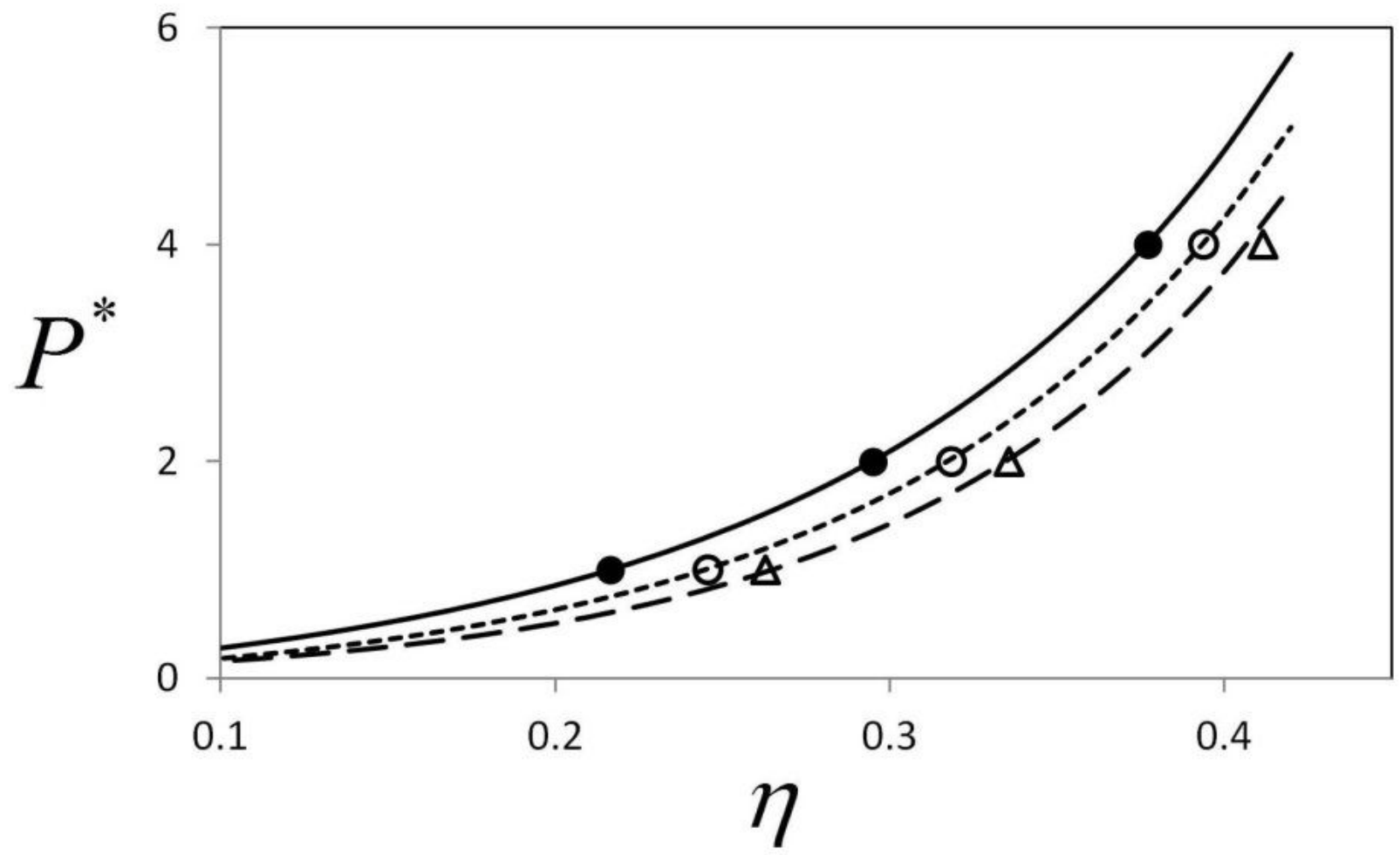

**Figure 5.8:** Pressure $P^* = Pd^3/k_BT$ isotherms for $\alpha_{AB} = 45^\circ$ at $\varepsilon^* = 4$(solid line – theory, filled circles – simulation) and $\varepsilon^* = 8$ (short dashed line – theory, open circles – simulation). Long dashed line and open triangles give theory and simulation predictions for the case $\alpha_{AB} = 180^\circ$ at $\varepsilon^* = 8$

In Fig. 5.9 we hold density and $\varepsilon^*$ constant and plot reduced pressure over the full bond angle range. For $\eta = 0.1$ and $\varepsilon^* = 5$ the pressure is at a minimum for $\alpha_{AB} = 0^\circ$ where the majority of colloids are associated into double bonded dimers and then increases to a limiting

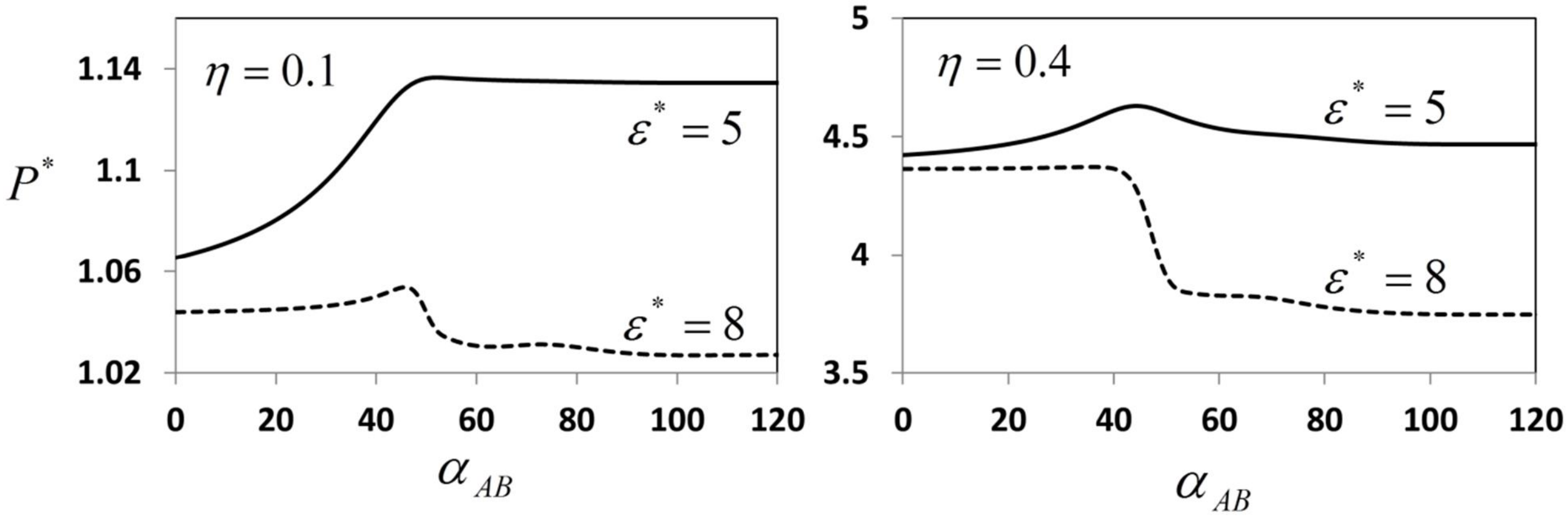

**Figure 5.9:** Bond angle dependence of pressure at association energies $\varepsilon^* = 5$ (solid curve) and $\varepsilon^* = 8$ (dashed curve) for packing fractions $\eta = 0.1$ (left) and $\eta = 0.4$ (right)

value near $\alpha_{AB} = 50^\circ$ where most colloids are monomers (Fig. 5.5) at these conditions. Increasing to $\varepsilon^* = 8$ at $\eta = 0.1$, we see the opposite behavior; now pressure is a maximum for $\alpha_{AB} = 0^\circ$ where nearly all colloids are bonded in double bonded dimers, remains relatively constant until $\alpha_{AB} \sim 40^\circ$, goes through a maximum near $\alpha_{AB} \sim 45^\circ$ where $X_2$ goes through a minimum (Fig. 5.5) and then decreases to a limiting value as the bond angle dependence saturates and the system becomes dominated by linear chains. Increasing the packing fraction to $\eta = 0.4$ we see similar behavior with the exception that for $\varepsilon^* = 5$ the pressure for $\alpha_{AB} = 180^\circ$ is only slightly higher than the $\alpha_{AB} = 0^\circ$ case. This is due to the increase in association at this density resulting in a maximum in pressure near $\alpha_{AB} \sim 45^\circ$ while the maximum for the $\varepsilon^* = 8$ case disappears.

## 5.4: Summary and conclusions

We have extended Wertheim's theory to model two patch colloids where the patches can be separated by any bond angle $\alpha_{AB}$. We used Wertheim's resummed perturbation theory[37] to account for blockage effects in chain formation, Sear and Jacksons graph for double bonded dimers[91] and modified the ring graph of Sear and Jackson[86] to account for the association of colloids into non overlapping rings. We obtained an analytical solution for the double bonding integral $I_d$ which represents the probability that two colloids are oriented such that double bonding can occur, this quantity was treated as a parameter in previous studies[91]. The integrals $\Psi$ (which account for the fact that bonding at one patch may block bonding at the other) and $\Gamma^{(n)}$ (which are proportional to the number of ways *n* colloids can position themselves to form rings of size *n*) were evaluated using Monte Carlo integration as a function of $\alpha_{AB}$. This is the first application of Wertheim's theory to associating fluids which explicitly accounts for the effect of bond angle.

The new theory was extensively tested against new Monte Carlo simulation data and found to be very accurate. It was shown that $\alpha_{AB}$ plays a crucial role in the thermodynamics of these fluids. In systems which exhibit significant association, double bonded dimers dominate for small $\alpha_{AB}$. Increasing $\alpha_{AB}$ further, there is a transition to a ring dominated fluid; increasing $\alpha_{AB}$ even further, ring formation becomes unlikely and the system becomes dominated by associated chains of colloids. In the region $40^\circ \le \alpha_{AB} \le 90^\circ$ there is a vicious competition between the various modes of association. The new theory was shown to successfully account for this full range of interactions and accurately predict the fraction of colloids in each type of associated cluster, internal energy and pressure.

The analysis presented in this chapter is restricted to 2 patch colloids. However, it is known[94] that to have a liquid – vapor phase transition when only 1 bond per patch is allowed a colloid must have a minimum of three patches. In addition, lattice simulations have shown[88] that bond angle can have significant effect on the phase diagram of patchy colloids. To allow for the study of the effect of bond angle on liquid – vapor equilibria the approach developed in this chapter must be extended to allow for more than two patches. This will be the subject of chapter 6.

# CHAPTER 6

# Bond angle dependence in associating fluids with more than two sites

In chapter 5 we extended Wertheim's thermodynamic perturbation theory[35, 37] (TPT) to include the effect of bond angle $\alpha_{AB}$ in the equation of state for associating fluids with two association sites (an *A* association site and a *B* association site). It was found that various modes of association became dominant in various bond angle ranges. In strongly associating systems with large $\alpha_{AB}$ chains were the dominant type of associated cluster, for moderate $\alpha_{AB}$ rings became dominant, and for small $\alpha_{AB}$ double bonded molecules were overwhelmingly favored. The theory accounted for each association possibility and the effect of bond angle was included in each contribution. The theory was tested against Monte Carlo simulation data and found to be highly accurate.

While shown to be very accurate, this approach is restricted to the two site case.

Of course, it would be desirable to extend the theory to the case of > 2 association sites. There are numerous example of hydrogen bonding fluids with > 2 sites, water being the most famous example. Also, to achieve phase equilibria and percolation in patchy colloid fluids, a minimum of three patches is required.[94] In this chapter we will extend this theory to allow for small bond angle effects in molecules / colloids with more than two association sites. Instead of tackling the more general case in which we allow for small bond angle effects between each pair of association sites, we will restrict our analysis to the case where small bond angle effects are only accounted for a single pair of association sites. This will allow for a tractable and logical extension of the results of chapter 5. To validate the theory we compare to new Monte Carlo simulation data for the three patch case. Once validated, we show that bond angle has a huge effect on the liquid – vapor equilibria of three patch colloids. Throughout the chapter we refer to associating molecules as colloids and association sites as patches; however, the results in this paper are equally applicable as a primitive model for hydrogen bonding fluids.

## 6.1: Theory

In this section the theory for colloids of diameter *d* with a set of patches $\Gamma = \{A, B, C...\}$ will be developed. The center of each pair of patches is separated by a bond angle $\alpha_{SP}$. A diagram of this type of colloid can be found in Fig. 6.1 for the three patch case $\Gamma = \{A, B, C\}$. The potential of interaction between two colloids is given by the sum of a hard sphere potential $\phi_{HS}(r_{12})$ and orientation dependent association potential

$$\phi(12) = \phi_{HS}(r_{12}) + \sum_{S\in\Gamma}\sum_{P\in\Gamma}\phi_{SP}(12) \tag{6.1}$$

Again, we consider conical association sites as described in Chapter 2.1. For a uniform system Eq. (1.39) simplifies to

$$\frac{A-A_{HS}}{Vk_BT}=\rho\ln\left(\frac{\rho_o}{\rho}\right)+Q+\rho-\Delta c^{(o)}/V \tag{6.2}$$

In the two patch case it was shown that the bond angle dependence saturates near $\alpha_{AB}=90^{\circ}$; meaning for $\alpha_{AB}>90^{\circ}$ TPT1 is adequate. In order to make the derivation tractable and easily followed we will assume that the only small bond angle is $\alpha_{AB}$. This means that all interactions which do not involve patches $A$ or $B$ will be treated in TPT1. As in chapter 5, we will assume that patches $A$ and $B$ do not self-attract, that is $\varepsilon_{AA}=\varepsilon_{BB}=0$, and that each association site is singly bondable. Also, we will assume that the potential parameters $\theta_c$ and $r_c$ are the same for all patches.

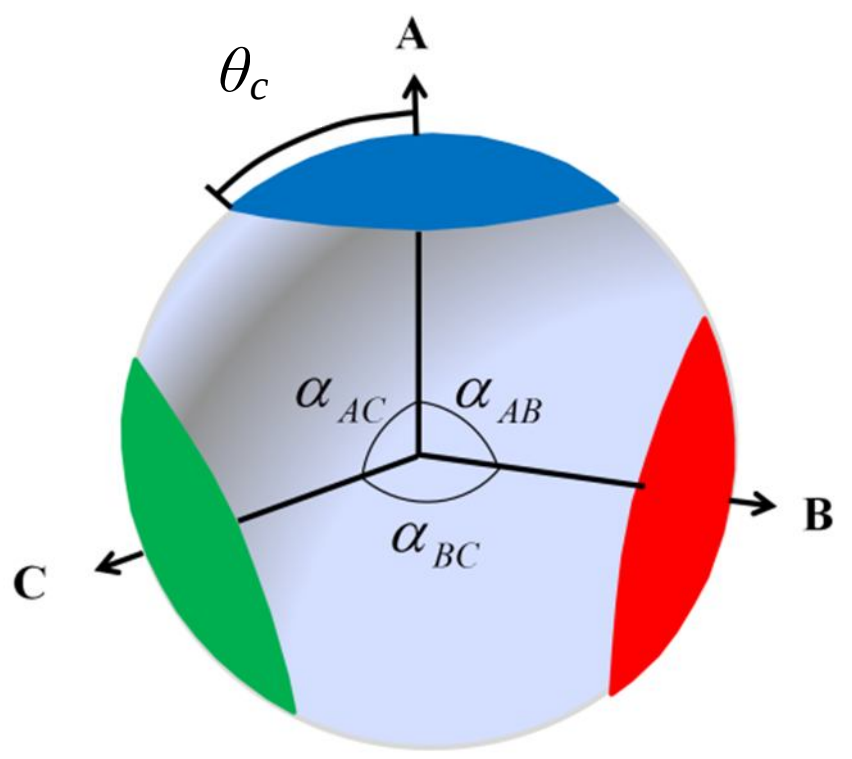

**Figure 6.1:** Diagram of patchy colloid model

We will split the fundamental graph sum into the TPT1 contribution $\Delta c_{TPT1}^{(o)}$ and the higher order contribution $\Delta c_{HOAB}^{(o)}$ which corrects for the small $\alpha_{AB}$.

$$\Delta c^{(o)}\approx\Delta\hat{c}_1^{(o)}=\Delta c_{\mathrm{TPT1}}^{(o)}+\Delta c_{HOAB}^{(o)} \tag{6.3}$$

The TPT1 contribution is given by

$$\Delta c_{\text{TPT1}}^{(o)} / V = \frac{1}{2} \sum_{S \in \Gamma} \sum_{P \in \Gamma} \sigma_{\Gamma - S} \xi \kappa f_{SP} \sigma_{\Gamma - P} \tag{6.4}$$

where $f_{SP} = \exp\left(\varepsilon_{SP} / k_B T\right) - 1$ is the magnitude of the association Mayer function, $\kappa$ is given by Eq. (2.21) and $\xi$ is given by Eq. (2.17). For the higher order $\Delta c_{HOAB}^{(o)}$ we have contributions which account for steric hindrance between patches $A$ and $B$, $\Delta c_{ch:AB}^{(o)}$, cycles of $n$ association bonds , $\Delta c_n^{cycle}$, and colloids double bonded at patches $A$ and $B$ , $\Delta c_d^{(o)}$

$$\Delta c_{HOAB}^{(o)} = \Delta c_{ch:AB}^{(o)} + \sum_{n=3}^{\infty} \Delta c_n^{cycle} + \Delta c_d^{(o)} \tag{6.5}$$

Figure 6.2 illustrates the various types of associated clusters which may exist.

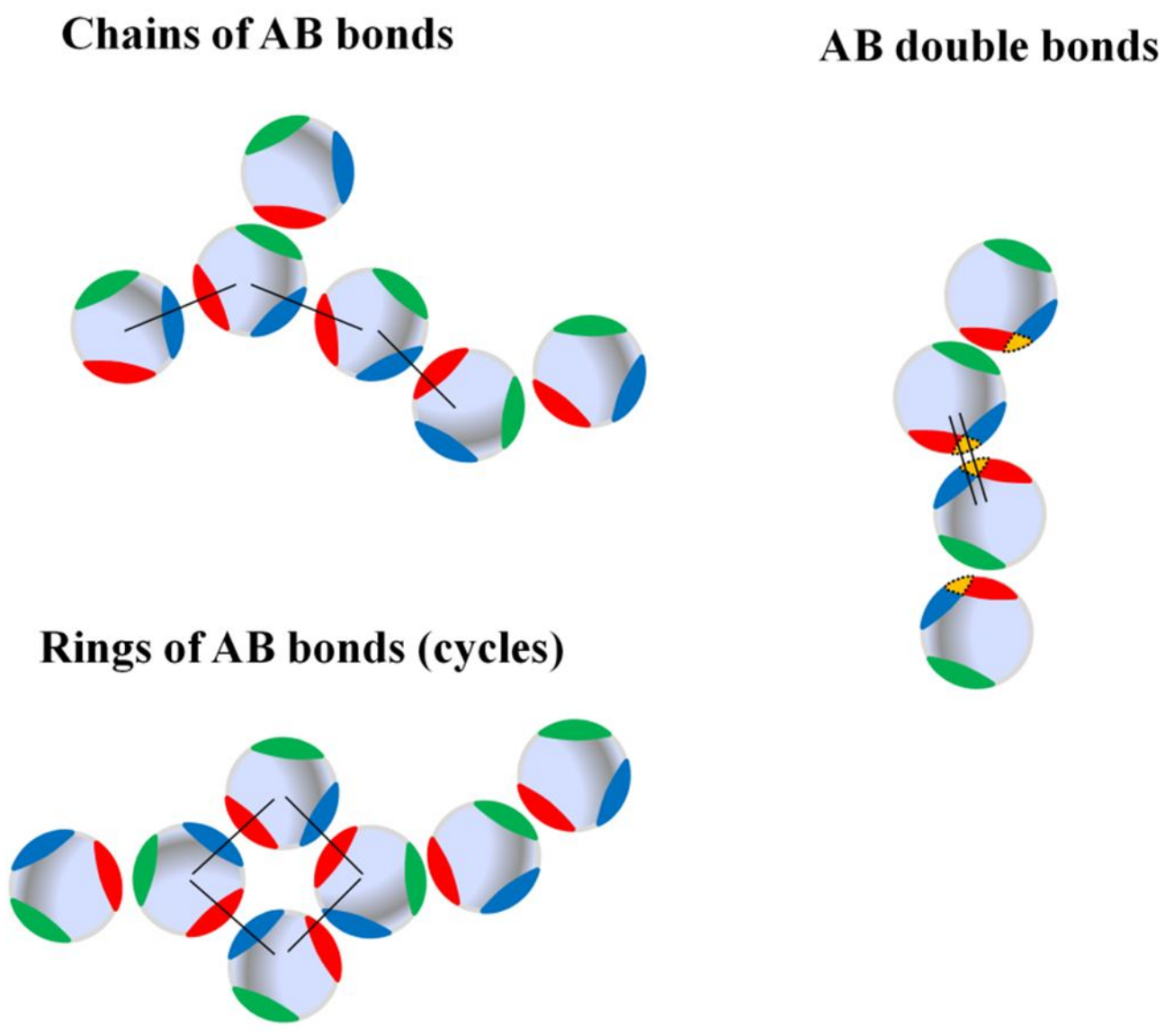

**Figure 6.2:** Examples of associated clusters with paths of *AB* bonds. Black lines represent an *AB* bond. Here we show 4-mer cycles, but cycles of all sizes are accounted for. Orange area gives surface of overlap of the *A* and *B* patches

The contribution $\Delta c_{ch:AB}^{(o)}$ is obtained by extension of Wertheim's resummed perturbation theory[37] to the multi – patch case where we only consider blocking effects between the *A* and *B* patches. For this case $\Delta c_{ch:AB}^{(o)}$ is approximated as the infinite series of chain diagrams

$$\frac{\Delta c_{ch:AB}^{(o)}}{V} = \sum_{S \in \Gamma} \sum_{P \in \Gamma} \sigma_{\Gamma - S} \sigma_{\Gamma - P} \sum_{n=1}^{\infty} \sigma_{\Gamma - AB}^{n} I_{n+1} \tag{6.6}$$

Where the integrals $I_n$ are given by

$$I_n = \frac{1}{\Omega^n} \int f_{PB}(12) f_{AB}(23) .... f_{AB}(n-1, n) f_{AS}(n, n+1) G_{HS}(1...n+1) d(2)...d(n+1) \tag{6.7}$$

In first order resummed perturbation theory (RTPT1) the function $G_{HS}(1...n)$ is approximated as[73]

$$G_{HS}(1...n) = g_{HS}(r_{n-1,n}) \prod_{k=1}^{n-2} g_{HS}(r_{k,k+1}) f_{HS}(r_{k,k+2}) \tag{6.8}$$

Since we have assumed the potential parameters $r_c$ and $\theta_c$ are the same for each patch, the Mayer functions $f_{AS}(12)$ can be written in terms of the overlap function $O_{AB}(12) = -\phi_{AB}(12) / \varepsilon_{AB}$

$$f_{AS}(12) = \frac{f_{AS}}{f_{AB}} f_{AB} O_{AB}(12) \tag{6.9}$$

Combing these results

$$\frac{\Delta c_{ch:AB}^{(o)}}{V} = \sum_{S \in \Gamma} \sum_{P \in \Gamma} \sigma_{\Gamma - S} \sigma_{\Gamma - P} \frac{f_{AS} f_{PB}}{f_{AB}^2} \sum_{n=0}^{\infty} \sigma_{\Gamma - AB}^{n} \hat{I}_{n+1} - \sum_{S \in \Gamma} \sum_{P \in \Gamma} \sigma_{\Gamma - S} \sigma_{\Gamma - P} \frac{f_{AS} f_{PB}}{f_{AB}^2} \hat{I}_1 \tag{6.10}$$

where $\hat{I}_n$ is given by

$$\hat{I}_n = \frac{f_{AB}^n}{\Omega^n} \int O_{AB}(12) O_{AB}(23) .... O_{AB}(n-1, n) O_{AB}(n, n+1) G_{HS}(1...n+1) d(2)...d(n+1) \tag{6.11}$$

and we have added and subtracted the contribution due to $\hat{I}_1$. Now, the infinite sum in Eq. (6.10) can be approximated similar to as described in chapter 2, Eq. (2.11), to yield

$$\frac{\Delta c_{ch:AB}^{(o)}}{V} = \sum_{S\in\Gamma}\sum_{P\in\Gamma}\sigma_{\Gamma-S}\sigma_{\Gamma-P}\frac{f_{AS}f_{PB}}{f_{AB}}\kappa\xi\left(\lambda-1\right) \tag{6.12}$$

Where $\lambda$ is given by

$$\lambda = \frac{1}{1+\left(1-\Psi\right)\kappa\, f_{AB}\sigma_{\Gamma-AB}\xi} \tag{6.13}$$

and $\Psi = 1+\hat{I}_2/\hat{I}_1^2$ is the blockage integral Eq. (5.5), which was evaluated in chapter 5. When the angle $\alpha_{AB}$ is large enough that patches *A* and *B* are independent, $\Psi \to 1$, which results in $\lambda \to 1$ and $\Delta c_{ch:AB}^{(o)} \to 0$.

For the two patch case it was shown that rings of association bonds (cycles) played a crucial role for $40^\circ \le \alpha_{AB} \le 90^\circ$. The case here is much more complex since there are many more cycle forming possibilities. For instance, a cycle containing three colloids, of the type given in Fig. 6.1, could contain two colloids bonded at both patches *A* and *B* and one colloid bonded at patches *A* and *C*; another possibility would be a cycle with 3 colloids bonded at patches *A* and *B etc*.. To account for all cycle forming possibilities we would have to enumerate each possible cycle "composition" for each cycle size. This would be a doable, but tedious task. Instead, since we are assuming $\alpha_{AB}$ is the only small bond angle, we will only account for cycles formed from *AB* bonds only. For this case the contributions $\Delta c_n^{ring}$ are obtained by a simple extension of the two patch case Eq. (5.19)

$$\Delta c_n^{cycle}/V = \frac{\left(f_{AB}\sigma_{\Gamma-AB}\hat{g}_{HS}K\right)^n}{nd^3}\Gamma^{(n)} \tag{6.14}$$

Equation (6.14) was obtained from Eq. (5.19) by the substitution $\rho_o \to \sigma_{\Gamma-AB}$. It should be noted that $\Delta c_n^{cycle}$ accounts for cycles of *AB* association bonds in clusters and is in no way limited to stand alone associated "rings" of colloids.

The last contribution to consider in Eq. (6.5) is for colloids which are double bonded at patches *A* and *B*. The contribution $\Delta c_d^{(o)}$ is similar to the two patch case Eq. (5.6) and is given by

$$\Delta c_d^{(o)} / V = \left(f_{AB}\sigma_{\Gamma-AB}\right)^2 \xi I_d / 2 \tag{6.15}$$

The only difference between Eq. (6.15) and the two patch case Eq. (5.6) is the substitution $\rho_o \to \sigma_{\Gamma-AB}$. Equation (6.15) accounts for the double bonding of patches *A* and *B* in larger associated clusters as well as double bonded dimers.

From Eqns. (1.36) and (6.3) we see that $c_\beta = 0$ for $n(\beta) > 2$ and $c_{SP} = 0$ for $SP \neq AB$; this results in the following rule for the densities obtained from Eq. (1.37)

$$\frac{\rho_\gamma}{\rho_o} = \begin{cases} \prod_{S\in\gamma} c_S \left(1 + \dfrac{c_{AB}}{c_A c_B}\right) & \text{for} \quad AB \in \gamma \\ \prod_{S\in\gamma} c_S & \text{otherwise} \end{cases} \tag{6.16}$$

Equation (6.16) can be further simplified as

$$\frac{\rho_\gamma}{\rho_o} = \begin{cases} \dfrac{\rho_{AB}}{\rho_o} \prod_{S\in\gamma-AB} \dfrac{\rho_S}{\rho_o} & \text{for} \quad AB \in \gamma \\ \prod_{S\in\gamma} \dfrac{\rho_S}{\rho_o} & \text{otherwise} \end{cases} \tag{6.17}$$

Using the site operators of Wertheim[35], Eq. (6.17) can be rewritten as

$$\hat{\sigma}_\gamma = \begin{cases} \hat{\sigma}_{AB} \prod_{S\in\gamma-AB} \hat{\sigma}_S & \text{for} \quad AB \in \gamma \\ \prod_{S\in\gamma} \hat{\sigma}_S & \text{otherwise} \end{cases} \tag{6.18}$$

where $\hat{\sigma}_\gamma = \sigma_\gamma / \rho_o$. For $\sigma_{\Gamma-AB}$ we obtain the simple relation from Eq. (6.18)

$$\hat{\sigma}_\Gamma = \hat{\sigma}_{AB}\hat{\sigma}_{\Gamma-AB} \tag{6.19}$$

Defining the fraction of colloids *not* bonded at both patches *A* and *B*, $X_{AB} = \sigma_{\Gamma-AB} / \rho$ we obtain from Eq. (1.37) and (6.19)

$$X_{AB} = \frac{1}{\hat{\sigma}_{AB}} = \frac{1}{c_{AB} + (1 + c_A)(1 + c_B)} \tag{6.20}$$

Equation (6.19) allows us to obtain $\sigma_{\Gamma-S}$ from Eq. (6.18) as

$$\hat{\sigma}_{\Gamma-S} = \begin{cases} \hat{\sigma}_\Gamma / \hat{\sigma}_S & for \quad S \neq A \, or \, B \\ \prod_{P\in\Gamma-S} \hat{\sigma}_P & otherwise \end{cases} \tag{6.21}$$

Combining (6.18) and (6.21) we get the fraction of colloids not bonded at patch *S*, $X_S =$

$\sigma_{\Gamma-S} / \rho$

$$X_S = \begin{cases} \dfrac{1}{1 + c_S} & for \quad S \neq A \, or \, B \\ (1 + c_L) X_{AB} & otherwise \end{cases} \tag{6.22}$$

In Eq. (6.22) when *S = A*, *L = B* and when *S = B*, *L = A*. To obtain Eq. (6.22) we used the following relationship which was developed using Eq. (6.19)

$$\prod_{P\in\Gamma-A} \hat{\sigma}_P = \frac{\hat{\sigma}_\Gamma \prod_{P\in\Gamma-A} \hat{\sigma}_P}{\hat{\sigma}_{AB}\hat{\sigma}_{\Gamma-AB}} = \frac{\hat{\sigma}_\Gamma \prod_{P\in\Gamma-A} \hat{\sigma}_P}{\hat{\sigma}_{AB} \prod_{P\in\Gamma-AB} \hat{\sigma}_P} = \frac{\hat{\sigma}_\Gamma \hat{\sigma}_B}{\hat{\sigma}_{AB}} \tag{6.23}$$

The last density relation we need is for $\hat{\sigma}_\Gamma = \rho / \rho_o$, which we obtain using Eqns. (6.19) – (6.20)

$$\frac{1}{\hat{\sigma}_\Gamma} = X_o = \frac{X_{AB}}{X_A X_B} \prod_{S \in \Gamma} X_S \tag{6.24}$$

Where $X_o$ is the fraction of colloids which are monomers. The condition that $c_{SP} = 0$ for $SP \neq AB$ allows for simplification of the $Q$ function obtained from Eq. (1.35) as

$$Q = -\rho + \sum_{S \in \Gamma} c_S \sigma_{\Gamma - S} + c_{AB} \sigma_{\Gamma - AB} \tag{6.25}$$

Which using the results above can be further simplified as

$$Q / \rho = \sum_{S \in \Gamma} (1 - X_S) + \frac{X_A X_B}{X_{AB}} - 2 \tag{6.26}$$

The $c_S$ are obtained from Eq. (1.36) as

$$c_S = \sum_{P \in \Gamma} \Delta_{SP} X_P \left( 1 + \left\{ \frac{f_{SA}}{f_{SP}} \frac{f_{PB}}{f_{AB}} + \frac{f_{PA}}{f_{SP}} \frac{f_{SB}}{f_{AB}} \right\} (\lambda - 1) \right) \tag{6.27}$$

where $\Delta_{SP} = \xi \kappa f_{SP} \rho$ . Likewise $c_{AB}$ is obtained as

$$\begin{aligned} c_{AB} = &-(1 - \Psi) \sum_{S \in \Gamma} \sum_{P \in \Gamma} X_S X_P \Delta_{SA} \Delta_{PB} \lambda^2 + \rho X_{AB} f_{AB}^2 \xi I_d \\ &+ \sum_{n=3}^{\infty} \left( f_{AB} \hat{g}_{HS} K \right)^n (\rho X_{AB})^{n-1} d^{-3} \Gamma^{(n)} \end{aligned} \tag{6.28}$$

Combining these results we find

$$\frac{\Delta c_{\text{TPT1}}^{(o)} + \Delta c_{ch:AB}^{(o)}}{N} = \frac{1}{2} \sum_{S \in \Gamma} (1 - X_S) + \frac{X_A X_B}{X_{AB}} - 1 \tag{6.29}$$

Now the free energy can be simplified as

$$\frac{A - A_{HS}}{N k_B T} = \sum_{S \in \Gamma} \left( \ln X_S - \frac{X_S}{2} + \frac{1}{2} \right) + \frac{\Delta A_{AB}}{N k_B T} \tag{6.30}$$

Where the summation gives the standard TPT1 free energy and $\Delta A_{AB}$ represents the correction for the fact that sites *A* and *B* interact beyond first order and is given by

$$\frac{\Delta A_{AB}}{Nk_B T} = \ln\left(\frac{X_{AB}}{X_A X_B}\right) - \sum_{n=3}^{\infty}\frac{\Lambda_n}{n} - \frac{\Xi_d}{2} \tag{6.31}$$

Where we have defined the quantities

$$\Lambda_n = \left(f_{AB} X_{AB} \hat{g}_{HS}\right)^n K^n \rho^{n-1} \Gamma^{(n)} d^{-3} \tag{6.32}$$

and

$$\Xi_d = \left(f_{AB} X_{AB}\right)^2 \rho \xi I_d \tag{6.33}$$

To evaluate Eq. (6.30) the fractions $X_{AB}$ and $X_S$ must be known. The fractions $X_S$ are obtained from Eqn. (6.22). Solving Eqns. (6.20) and (6.22) we obtain for $X_{AB}$

$$X_{AB} - X_{AB}^2 c_{AB} - X_A X_B = 0 \tag{6.34}$$

Equation (6.34) concludes our analysis for colloids with a set of patches Γ. In the development of the theory we have assumed small bond angle effects only play a significant role for the bond angle $\alpha_{AB}$. In the following section we specialize the theory to the 3 patch case $\Gamma = \{A, B, C\}$.

## 6.2: Specialization to the 3 patch case

In this section we apply the results of section 6.1 to the 3 patch case $\Gamma = \{A, B, C\}$. This type of colloid is depicted in Fig. 6.1. From Eq. (6.27) we obtain the $c_K$'s as

$$\begin{aligned} c_A &= \left(\Delta_{AB} X_B + \Delta_{AC} X_C\right)\lambda \qquad c_B = \left(\Delta_{AB} X_A + \Delta_{BC} X_C\right)\lambda \\ c_C &= \left(\Delta_{CA} X_A + \Delta_{CB} X_B\right)\lambda + X_C \tilde{\Delta} \end{aligned} \tag{6.35}$$

Where the quantity $\tilde{\Delta}$ is given by

$$\tilde{\Delta} \equiv \Delta_{CC} + \Delta_{CA}(\lambda - 1)\frac{f_{CB}}{f_{AB}} + \Delta_{CB}(\lambda - 1)\frac{f_{CA}}{f_{AB}} \tag{6.36}$$

Now the required fractions can be determined as

$$X_A = \frac{X_{AB}\left(1+\Delta_{BC}\lambda X_C\right)}{1-\Delta_{AB}\lambda X_{AB}} \qquad X_B = \frac{X_{AB}\left(1+\Delta_{AC}\lambda X_C\right)}{1-\Delta_{AB}\lambda X_{AB}}$$

$$\frac{1}{X_C} = 1+\Delta_{CA}X_A\lambda + \Delta_{CB}X_B\lambda + X_C\tilde{\Delta} \tag{6.37}$$

Combining Eqns. (6.34) and (6.37) gives a closed equation for $X_{AB}$. Once the fraction $X_{AB}$ is determined the remaining fractions can be calculated using Eqn. (6.37). To compare to simulations we will use the fractions of colloids bonded *i* times $X_i$ $(i = 0-3)$ which are obtained as

$$X_o = X_{AB}X_C \qquad X_1 = X_o\left(c_A + c_B + c_C\right) \tag{6.38}$$

$$X_2 = X_o\left(c_{AB} + c_A c_B + c_A c_C + c_B c_C\right) \qquad X_3 = X_o\left(c_{AB}c_C + c_A c_B c_C\right)$$

Lastly, we will calculate the fraction of colloids which are bonded at both patches *A* and *B* in linear chains of *AB* bonds, rings of *AB* bonds and double bonds. The total density of colloids bonded at both patches *A* and *B*, $\tilde{\rho}_{AB}$ contains contributions for colloids which are bonded at *A* and *B* only, as well as a contribution from colloids which are fully bonded. That is,

$$\tilde{\rho}_{AB} = \rho_{AB} + \rho_{ABC} \tag{6.39}$$

Using Eqns. (6.16) and (6.39) and defining the fraction $\tilde{\chi}_{AB} = \tilde{\rho}_{AB}/\rho$ we obtain

$$\tilde{\chi}_{AB} = X_o\left(c_{AB}c_C + c_{AB} + c_A c_B + c_A c_B c_C\right) \tag{6.40}$$

We also know that $\tilde{\chi}_{AB}$ must satisfy the relation

$$\tilde{\chi}_{AB} = \tilde{\chi}_{AB}^{ch} + \tilde{\chi}_{AB}^{d} + \sum_{n=3}^{\infty} \tilde{\chi}_{AB}^{(n)} \tag{6.41}$$

where $\tilde{\chi}_{AB}^{ch}$ is the fraction of colloids bonded at both *A* and *B* which are in a chain of *AB* bonds, $\tilde{\chi}_{AB}^{d}$ is the fraction of colloids double bonded at *A* and *B*, and finally $\tilde{\chi}_{AB}^{(n)}$ is the fraction of colloids bonded at both *A* and *B* in a cycle of *n AB* bonds. These contributions are depicted pictorially in Fig. 6.2. Comparing Eqns. (6.40) and (6.41) the following relations can be deduced

$$\frac{\tilde{\chi}_{AB}^{d}}{X_o} = \rho X_{AB} f_{AB}^{2} \xi I_d \left(1 + c_C\right) \qquad \frac{\tilde{\chi}_{AB}^{(n)}}{X_o} = \left(f_{AB} \hat{g}_{HS} K\right)^n \left(\rho X_{AB}\right)^{n-1} d^{-3} \Gamma^{(n)} \left(1 + c_C\right)$$

$$\frac{\tilde{\chi}_{AB}^{ch}}{X_o} = \Psi \sum_{S \in \Gamma} \sum_{P \in \Gamma} X_S X_P \Delta_{SA} \Delta_{PB} \lambda^2 \left(1 + c_C\right) \tag{6.42}$$

For the case of total steric hindrance between patches *A* and *B*, bonding at both patches in a chain of *AB* bonds to two other colloids becomes impossible, resulting in $\Psi \to 0$. From Eq. (6.42) we see that for this case $\tilde{\chi}_{AB}^{ch} \to 0$, showing that the resummed perturbation theory was indeed successful.

## 6.3: Model and simulation

To validate the theory we perform new Monte Carlo simulations for three patch colloids of the type depicted in Fig. 6.1. In a spherical coordinate system ($\theta$ is the polar angle, $\phi$ is the azimuthal angle) the center of patch *C* is located on the *z* axis at $\theta = 0$, the center of patch *A* is at $\phi = 0$ and $\theta = \pi - \alpha_{AB}/2$, and finally the center of patch *B* is located at $\phi = \pi$ and $\theta = \pi - \alpha_{AB}/2$.

We consider 2 specific cases. In case I patch *C* is of the same type as patch *A*; that is $\varepsilon_{CB} = \varepsilon_{AB}$ and $\varepsilon_{CA} = \varepsilon_{CC} = 0$. Case I could be a primitive model for a hydrogen bonding fluid with 2 hydrogen and 1 oxygen sites (or vice versa). Also this can be used as a model for patchy colloids. If the only attractions are due to association it seems unlikely this type of colloid could undergo a liquid vapor transition due to the fact that the *B* type patches are limiting. To study the effect of bond angle on phase equilibria we consider another type, case II. In case II, the *C* patch is attracted to all three patch types $\varepsilon_{CB} = \varepsilon_{CA} = \varepsilon_{CC} = \varepsilon_{AB}$.

To test the theory we perform *NVT* (constant *N*, *V* and *T*) and *NPT* (constant pressure *P*, *V* and *T*) Monte Carlo simulations. The colloids interact with the potential given by Eq. (6.1) with $r_c = 1.1d$ and $\theta_c = 27^\circ$. The simulations were allowed to equilibrate for $10^6$ cycles and averages were taken over another $10^6$ cycles. A cycle is defined as *N* attempted trial moves where a trial move is defined as an attempted relocation and reorientation of a colloid. For the *NPT* simulations a volume change was attempted once each cycle. In this chapter we used *N* = 864 colloids.

## 6.4: Numerical results

In this section we will compare theoretical and simulation predictions for three patch colloids. *NVT* simulations were performed for the bond angle range $0^\circ \le \alpha_{AB} \le 90^\circ$ at two states, $\eta = \pi\rho d^3/6 = 0.1$, $\varepsilon^* = \varepsilon_{AB}/k_BT = 8$ and $\eta = 0.35$, $\varepsilon^* = 7.5$. Figure 6.3 shows the fractions of colloids bonded *k* times for both case I and case II discussed in section 6.3. First considering case I, we see for both states the fractions remain relatively constant in the bond angle range $60^\circ < \alpha_{AB} < 90^\circ$ and oscillate in the range $45^\circ < \alpha_{AB} < 60^\circ$. For $\alpha_{AB} < 45^\circ$ all fractions decrease as $\alpha_{AB}$ is decreased, with the exception of $X_2$ which increases rapidly as $\alpha_{AB}$ is decreased. Theory and simulation are in excellent agreement. For case II the bond angle dependence is stronger than for case I, with the fractions varying over the full range of $\alpha_{AB}$. In comparing case I and case II the most notable difference is the $\alpha_{AB}$ dependence of $X_3$ for $\alpha_{AB} < 45^{\circ}$. In case I, $X_3$ decreases with decreasing $\alpha_{AB}$ in this region, while for case II the opposite is true.

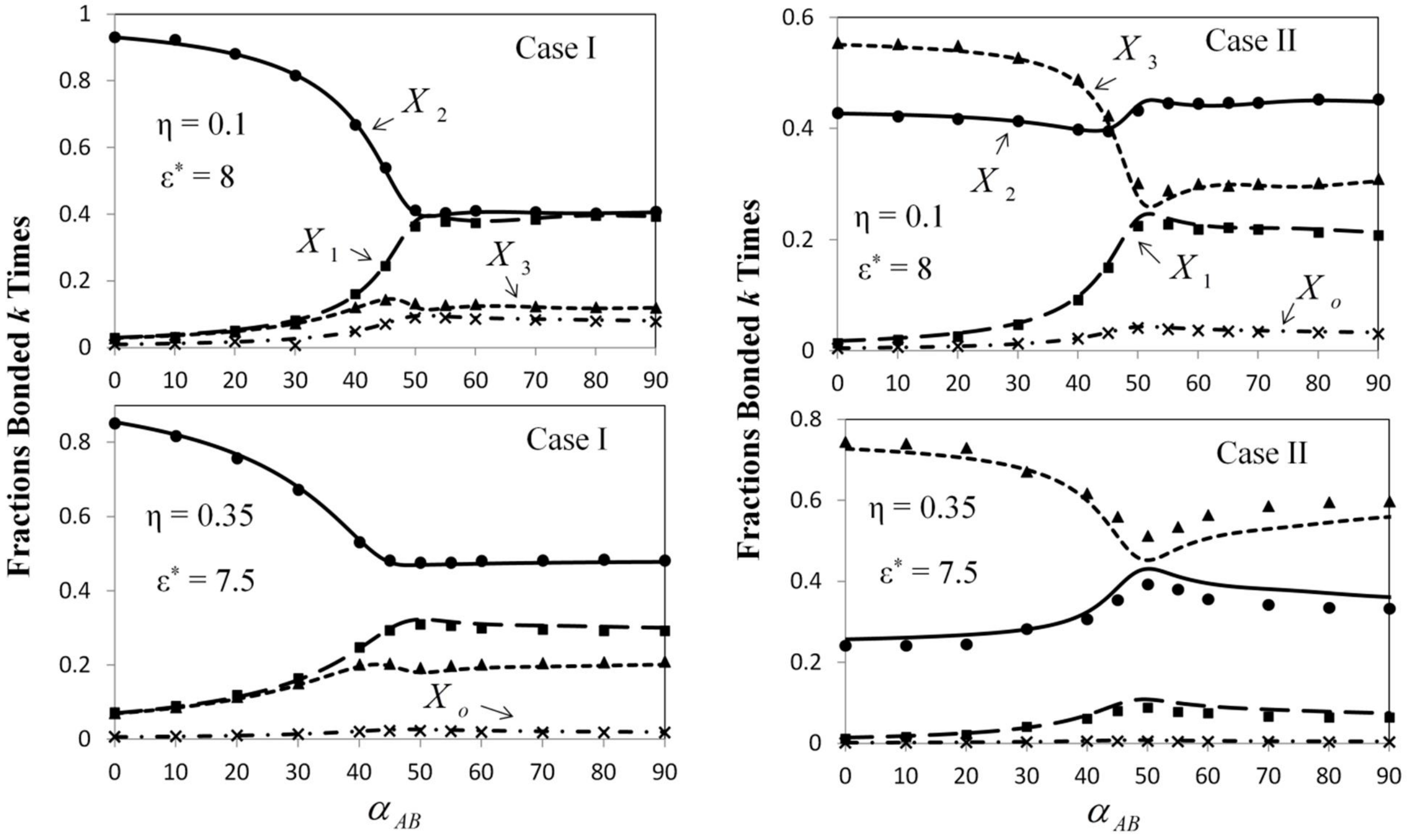

**Figure 6.3:** The fractions of colloids bonded $k$ times at $\eta$ = 0.1 and $\varepsilon^*$ = 8 (top) and $\eta$ = 0.35 and $\varepsilon^*$ = 7.5 (bottom). Left panel gives results for case I and right panel for case II. Curves give theory predictions and symbols give simulation results

To shed light on this behavior, Fig. 6.4 gives the fractions of colloids bonded at both patches $A$ and $B$ in the various cluster types, see Fig. 6.2, for the state $\eta = 0.1$ and $\varepsilon^* = 8$. The fractions $\tilde{\chi}_{AB}^{d}$ were easily measured using *NVT* simulations, so we report these simulated fractions in addition to the theoretical predictions. In the region $65^{\circ} < \alpha_{AB} < 120^{\circ}$ chains of $AB$ bonds dominate. In the region $40^\circ < \alpha_{AB} < 80^\circ$ triatomic cycles contribute significantly, becoming the dominant contribution to $\tilde{\chi}_{AB}$ in the range $50^\circ < \alpha_{AB} < 65^\circ$. For bond angles $\alpha_{AB} < 54^\circ$ double bonding becomes a possibility and increases rapidly with decreasing $\alpha_{AB}$. For bond angles $\alpha_{AB} < 50^\circ$, $\tilde{\chi}_{AB}^{d}$ becomes the dominant contribution to $\tilde{\chi}_{AB}$. As $\alpha_{AB}$ becomes small, $\tilde{\chi}_{AB}^{d}$ approaches unity. For small $\alpha_{AB}$, double bonding is strongly favored due to the fact that you get the energetic benefit of forming a double bond for the same entropic penalty as a single bond. Theory and simulation are in excellent agreement. The insets of Fig. 6.4 give the fractions in 4-mer and 5-mer cycles of *AB* bonds. As can be seen, these contributions (as well as for all larger *AB* cycles) are small at these conditions. Both cases give similar results in Fig. 6.4; although, ring formation is slightly stronger for case I.

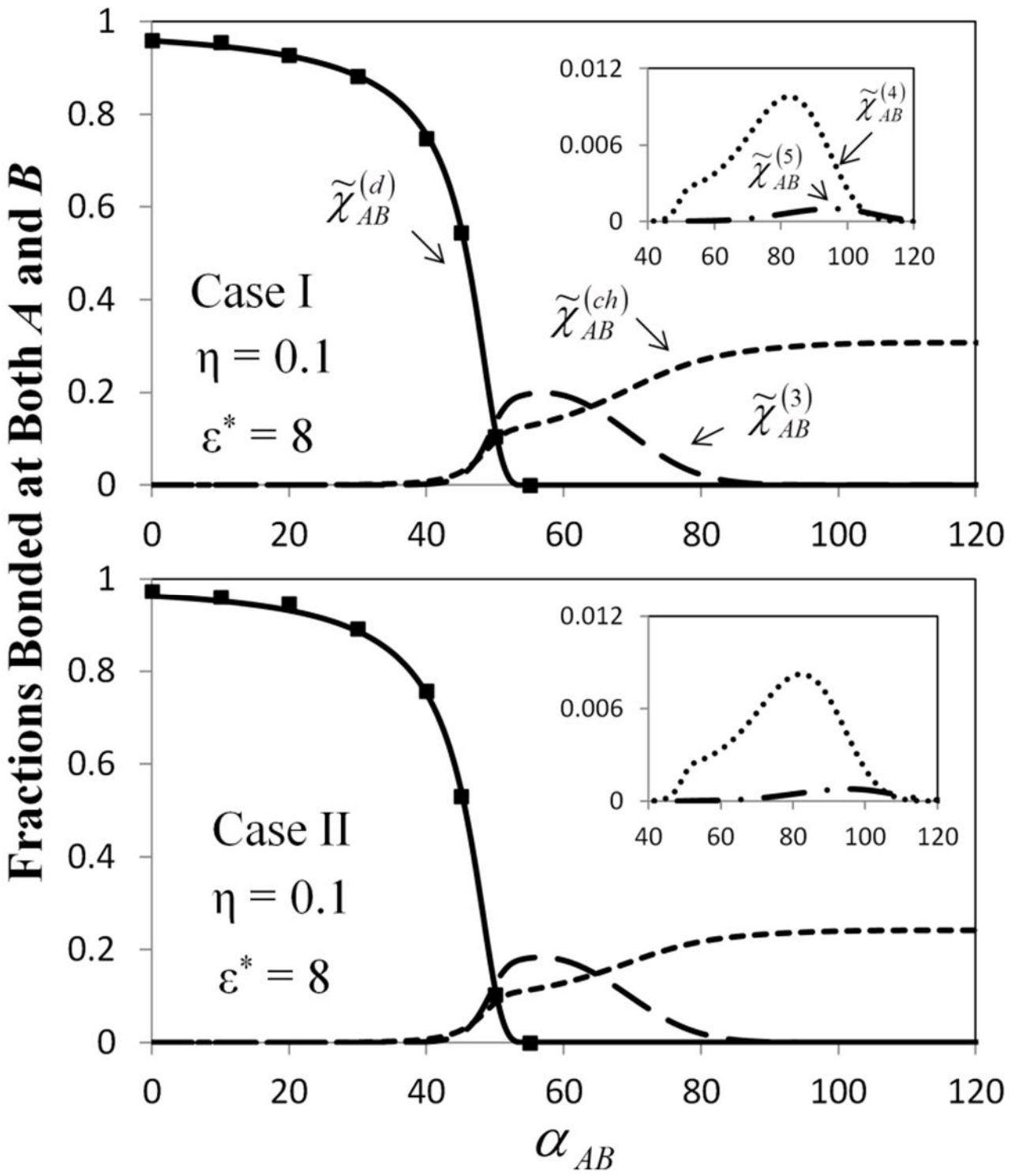

**Figure 6.4:** Fractions bonded at both patches A and B in the various cluster types, Eq. (6.42), at $\eta$ = 0.1 and $\varepsilon^*$ = 8 for case I (top) and case II (bottom). Curves give theory predictions and symbols give simulation values for fraction of colloids double bonded. Insets give cycle fractions for $n$ = 4 and 5. Larger cycle sizes are non-zero but negligible and are not shown

Now we can explain the difference in the $\alpha_{AB}$ dependence of $X_3$ for $\alpha_{AB} < 45^\circ$ between cases I and II observed in Fig. 6.3. In this range, double bonding of patches *A* and *B* dominates. For case I, patch *C* is of the same type as patch *A* with $\varepsilon_{CA} = \varepsilon_{CC} = 0$. Since double bonding is favored at these small bond angles, the majority of *B* patches are occupied in *AB* double bonds, which means there are very few *B* patches available to bond with *C* patches. This results in a decrease in $X_3$ as $\alpha_{AB}$ is decreased and double bonding increases. The situation for case II is different. For this case, patch *C* bonds to all three patch types. When $\tilde{\chi}_{AB}^{d}$ approaches unity at small $\alpha_{AB}$, the colloids can still bond three times by filling in with *CC* bonds. For this reason we note the behavior for case II, that decreasing $\alpha_{AB}$ results in an increase in $X_3$ for $\alpha_{AB} < 45^\circ$. This is the opposite behavior observed in case I.

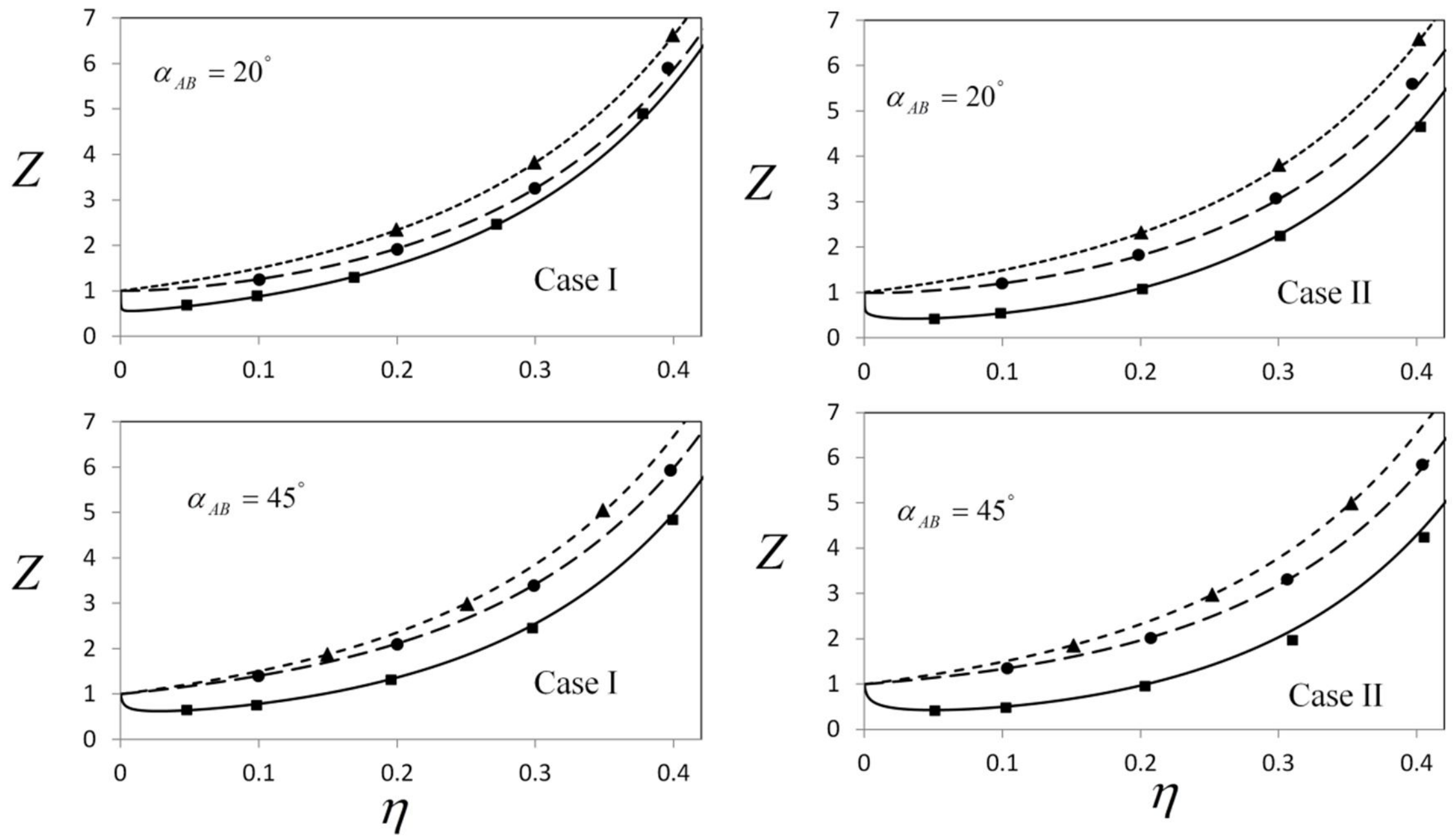

**Figure: 6.5:** Compressibility factor for colloids at bond angles $\alpha_{AB} = 20^\circ$ (top) and $\alpha_{AB} = 45^\circ$ (bottom) at association energies $\varepsilon^* = 2$(short dashed line - theory, triangles -simulation), $\varepsilon^* = 4$ (long dashed line - theory, circles - simulation) and $\varepsilon^* = 8$ (solid line - theory, squares - simulation). Left panel gives case I results and right panel case II

To further test the accuracy of the theory, Fig. 6.5 compares theoretical predictions and *NPT* simulation results of the compressibility factor $Z = P / \rho k_B T$ for both cases I and II. For each case we consider bond angles $\alpha_{AB} = 20^\circ$ and $45^\circ$ at association energies $\varepsilon^* = 2,4,8$. Overall, the theory does a good job in predicting the temperature and bond angle dependence of the compressibility factor. To better access the effect of bond angle on *Z,* we plot *Z* versus bond angle for the states $\eta = 0.1$, $\varepsilon^* = 8$ and $\eta = 0.35$ , $\varepsilon^* = 7.5$ in Fig. 6.6. For bond angles $\alpha_{AB} > 100^\circ$ there are essentially no cycles (or double bonds). It is in this region that Z is a minimum. In the range $54^\circ < \alpha_{AB} < 100^\circ$ there is a steady increase in cycles as $\alpha_{AB}$ is decreased. When *AB* cycles are formed, longer chains of *AB* bonds must be broken, this results in an increase in the compressibility factor. For $\alpha_{AB} < 54^\circ$ double bonding of patches *A* and *B* is possible, and rapidly becomes the dominant contribution to $\tilde{\chi}_{AB}$. The formation of double bonds will break larger clusters of associated colloids resulting in an increase in the compressibility factor of the system. For this reason *Z* increases rapidly in this region becoming a maximum at $\alpha_{AB} = 0^\circ$. The compressibility factor of case II is always lower than that of case I due to the increased amount of association. In the limit of strong association and $\alpha_{AB} = 0^{\circ}$, a fluid of case I colloids will be composed of dimers while a fluid of case II colloids will be composed of longer linear chains.

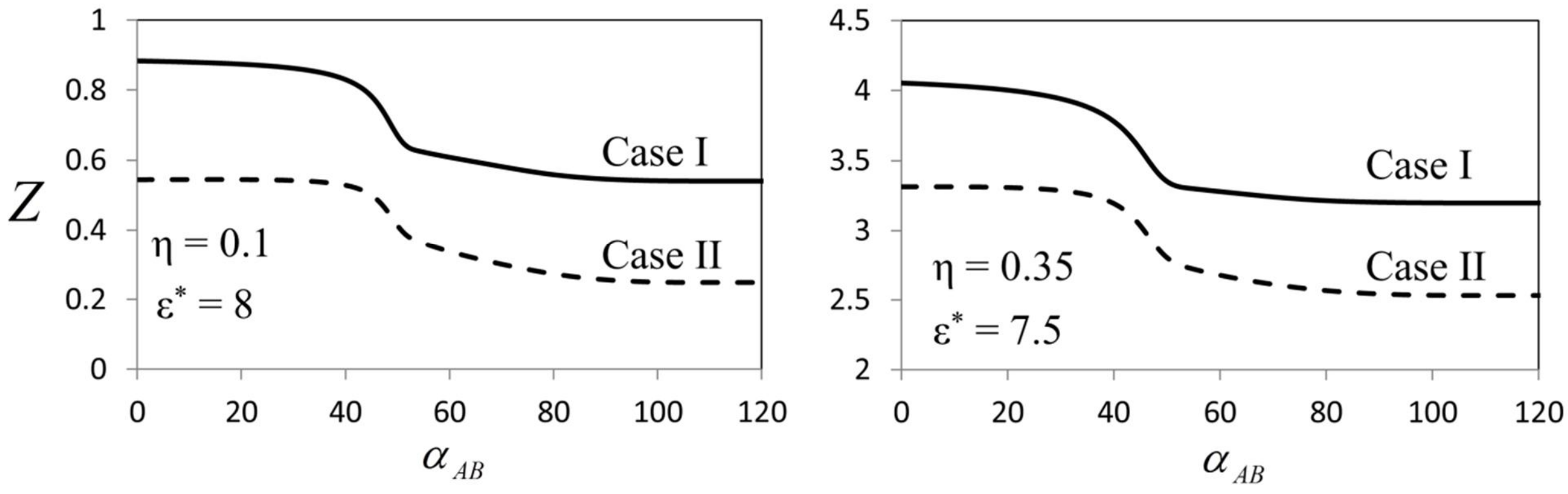

**Figure 6.6:** Compressibility factor versus $\alpha_{AB}$ at $\eta = 0.1$ and $\varepsilon^* = 8$ (left) and $\eta = 0.35$ and $\varepsilon^* = 7.5$ (right)

Lastly we consider the effect of the bond angle $\alpha_{AB}$ on the phase diagram of case II colloids in Fig. 6.7. As can be seen, the phase diagram is strongly dependent on $\alpha_{AB}$. Comparing the two cases $\alpha_{AB} = 60^{\circ}$ and $80^{\circ}$, it is clear that cycle formation decreases both the critical temperature $T_c$ and critical density $\rho_c$. Lattice simulations[88] have also shown that cycle formation has a substantial influence on phase equilibria. Decreasing $\alpha_{AB}$ further to $\alpha_{AB} = 50^{\circ}$, double bonding becomes significant which results in a further decrease in $T_c$ and $\rho_c$ as compared to the $\alpha_{AB} = 60^{\circ}$ case. Decreasing $\alpha_{AB}$ below $50^{\circ}$, results in a rapid increase in *AB* double bonds (Fig. 6.4). This rapid increase in $\tilde{\chi}_{AB}^{d}$ with decreasing $\alpha_{AB}$, results in a rapidly decreasing $T_c$ and $\rho_c$, as larger extended clusters must be sacrificed to accommodate double bonds. When double bonding becomes dominant, it is impossible to form a liquid phase due to the fact that a colloid which is double bonded can bond to a maximum of two colloids. Liquid – vapor phase equilibria in this case is impossible.[94]

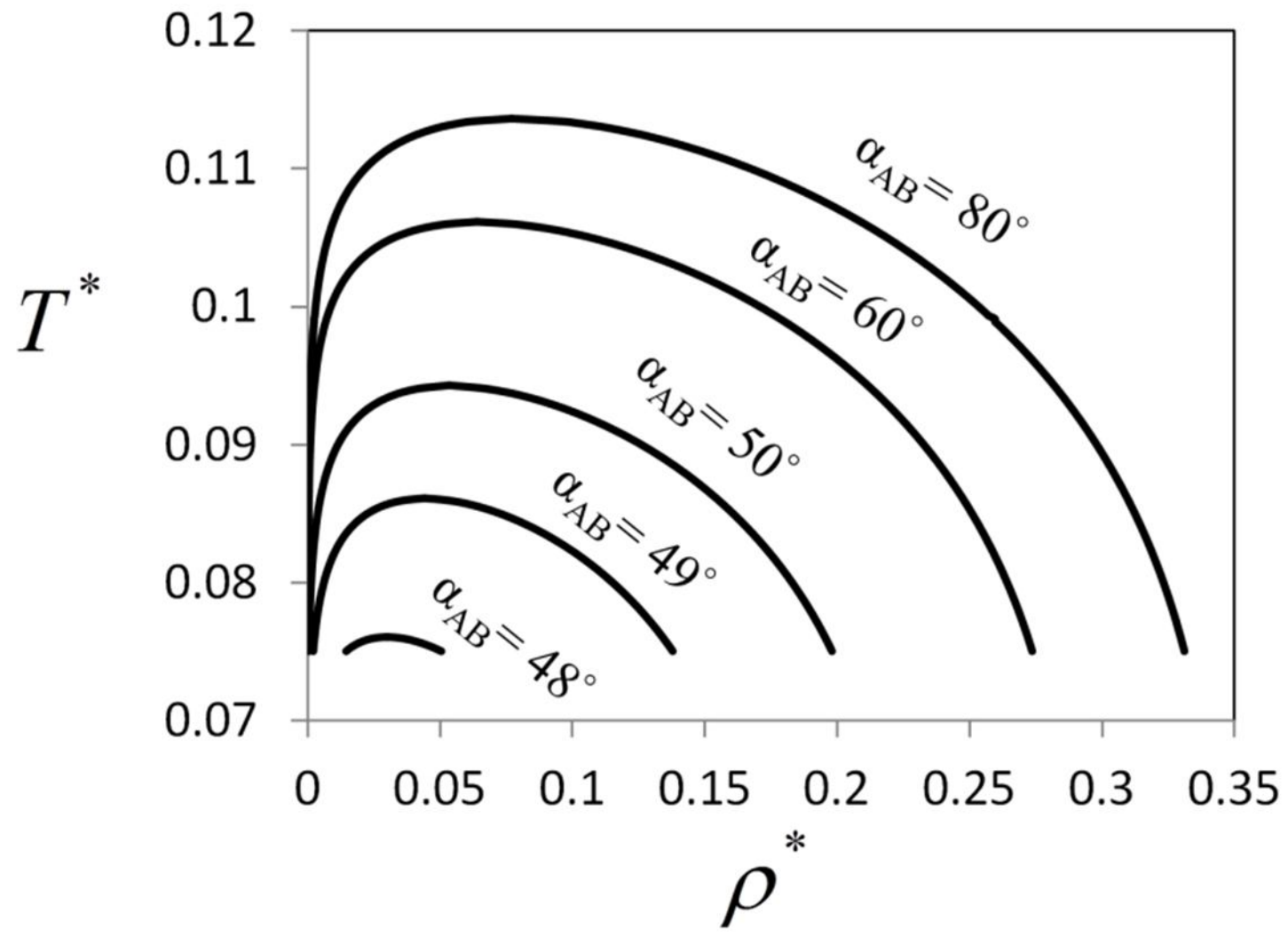

**Figure 6.7:** Phase diagrams $T^{*} = 1 / \varepsilon^{*}$ versus $\rho^{*} = \rho d^{3}$ for case II colloids at various bond angles $\alpha_{AB}$

## 6.5: Summary and conclusions

We have extended the theory developed in chapter 5 for the case of two patch associating fluids with small bond angles to the many patch case. Our model is restricted by the fact that we only account for small bond angle effects for the bond angle $\alpha_{AB}$. We have also assumed in the derivation that there is no self-attraction between $A$ and $B$ patches (ie: $\varepsilon_{AA} = \varepsilon_{BB} = 0$). These assumptions allowed for a manageable derivation of the equation of state. The new equation of state was tested against Monte Carlo simulations for two cases of 3 patch colloids and was found to be accurate. Once validated the new theory was used to predict the effect of the bond angle $\alpha_{AB}$ on the phase diagram of three patch colloids with $\varepsilon_{CB} = \varepsilon_{CA} = \varepsilon_{CC} = \varepsilon_{AB}$. It was found that decreasing bond angle decreases both the critical temperature and density. Once $\alpha_{AB}$ is sufficiently small, and double bonding dominates, the phase transition is quenched and phase equilibria is no longer possible. Extension of these results to the case that there are small bond angle effects between more than two association sites is discussed in chapter 9.

# CHAPTER 7

# Bond cooperativity

In the previous chapters we have shown how thermodynamic perturbation theory can be extended to describe a wide variety of associating fluids. In each case the distribution of associated clusters and the resulting equation of state were strongly dependent on a delicate balance between the energetic benefits of association and the resulting entropic penalty. In each case it was assumed that the total system energy is pairwise additive, there is no bond cooperativity. In nature, hydrogen bond cooperativity arises from the fact that when a multi – functional hydrogen bonding molecule forms hydrogen bonds, the polarization of the molecule is necessarily increased.[66] As has become increasingly apparent in recent years, hydrogen bond cooperativity plays a significant role in many physical processes. Both hydrogen fluoride (HF)[65] and alcohols[95] have been shown to exhibit strong hydrogen bond cooperativity. In addition, hydrogen bond cooperativity has been shown to stabilize peptide hydrogen bonds.[96] Indeed, it is believed that all polyfunctional hydrogen bonding molecules exhibit some degree of bond cooperativity.[66]

To account for bond cooperativity in an equation of state a number of lattice theories[97, 98] have been developed. In an alternative approach, Sear and Jackson (SJ)[99] considered a potential model of a two site associating fluid where the first bond in an associating cluster was given an association energy $-\varepsilon^{(1)}$ and each remaining bond was given an association energy $-\varepsilon^{(2)}$. This model approximates chains of hydrogen bonds in HF, where it has been shown that the binding energy per hydrogen bond decreases as cluster size increases. This continues until a chain length of approximately six, at which point the bond energy stabilizes as chain length increases.[65] Using this model SJ[99] developed a theory in the associating ideal gas limit and then used this ideal form of the theory to construct, in an adhoc fashion, the equation of state at higher densities. This theory has never been validated against molecular simulations and the effect of varying the magnitudes and ratios of the energies $\varepsilon^{(1)}$ and $\varepsilon^{(2)}$ was never extensively explored. For instance, do spheres associate when $\varepsilon^{(1)}$ is small and $\varepsilon^{(2)}$ is large?

With the wide ranging success of Wertheim's multi – density approach in modeling bulk and interfacial hydrogen bonding fluids which exhibit pairwise additivity, it would be prudent to incorporate hydrogen bond cooperativity into the formalism. At first sight, this seems like a daunting task due to the fact that Wertheim's approach is founded on the assumption of pairwise additivity [Eq. (1.2)]; however, as will be shown, incorporation of bond cooperativity in the form of a perturbation theory seems to be naturally accommodated by the theory. The purpose of this chapter is threefold. First we will extend Wertheim's multi – density formalism to account for bond cooperativity. As a first step we will consider the same two site case considered by SJ.[99]

We derive the theory in an intuitive manner using a new perturbation theory where we consider bond cooperativity as a perturbation. Second, we will test this new theory and the theory of SJ against Monte Carlo simulation results, and third we will discuss extensions of the theory to other, more complex, systems.

## 7.1: Theory

In this section we develop the thermodynamic perturbation theory for bond cooperativity in two site associating fluids with a single type *A* and type *B* association site. We restrict association such that there are *AB* attractions but no *AA* or *BB* attractions. We follow SJ and consider a fluid composed of $N$ hard spheres of diameter *d* with two association sites *A* and *B* with a total energy composed of pairwise and triplet contributions[99]

$$U(1 \ldots N) = \frac{1}{2}\sum_{i,j}\left(\phi_{HS}(r_{ij}) + \phi_{as}^{(2)}(ij)\right) + \frac{1}{6}\sum_{i,j,k}\phi_{as}^{(3)}(ijk) \tag{7.1}$$

where $(1) = \{\vec{r}_1, \Omega_1\}$ represents the position $\vec{r}_1$ and orientation $\Omega_1$ of sphere 1 and $\phi_{HS}$ is the hard sphere reference potential. The terms $\phi_{as}^{(2)}(ij)$ and $\phi_{as}^{(3)}(ijk)$ are the pairwise and triplet association contributions and are given by[99]

$$\phi_{as}^{(2)}(ij) = -\varepsilon^{(1)}\left(\text{O}_{AB}(ij) + \text{O}_{BA}(ij)\right) \tag{7.2}$$

$$\begin{aligned}\phi_{as}^{(3)}(ijk) = -\left(\varepsilon^{(2)} - \varepsilon^{(1)}\right)\big(&\text{O}_{AB}(ij)\text{O}_{BA}(ik) + \text{O}_{BA}(ij)\text{O}_{AB}(ik) \\ &+ \text{O}_{AB}(ji)\text{O}_{BA}(jk) + \text{O}_{BA}(ji)\text{O}_{AB}(jk) \\ &+ \text{O}_{AB}(ki)\text{O}_{BA}(kj) + \text{O}_{BA}(ki)\text{O}_{AB}(kj)\big)\end{aligned} \tag{7.3}$$

Where $\text{O}_{AB}(ij)$ is the overlap function which we obtain using the conical square well association sites of section 2.1

$$\mathrm{O}_{AB}(ij) = \begin{cases} 1 & r_{12} \le r_c \text{ and } \theta_{\mathrm{A}} \le \theta_c \text{ and } \theta_{\mathrm{B}} \le \theta_c \\ 0 & \text{otherwise} \end{cases} \tag{7.4}$$

which states that if spheres *i* and *j* are within a distance $r_c$ of each other and each sphere is oriented such that the angles between the site orientation vectors and the vector connecting the two spheres, $\theta_A$ for sphere *i* and $\theta_B$ for sphere *j*, are both less than the critical angle $\theta_c$ the two sites are considered bonded. See Fig. 5.1 for an illustration. The triplet contribution $\phi_{as}^{(3)}$ serves to add a correction $-\left(\varepsilon^{(2)} - \varepsilon^{(1)}\right)$ for each sphere bonded twice.

Before considering the case with bond cooperativity, we will review Wertheim's $M^{th}$ order perturbation theory (TPT*M*)[37] for pairwise additive association. For this two site case, in a uniform fluid, the Helmholtz free energy in Wertheim's theory with the pairwise additivity assumption is obtained from Eq. (1.37) as

$$\frac{A - A_{HS}}{k_B T V} = \rho \ln \frac{\rho_o}{\rho} - \sigma_A - \sigma_B + \frac{\sigma_A \sigma_B}{\rho_o} + \rho - \frac{\Delta c^{(o)}}{V} \tag{7.5}$$

In this chapter we will consider associating spheres as shown in **a** of Fig. 5.1, where the association sites are located on opposite poles of the sphere. Since the separation between sites is large, we can neglect the possibility of ring formation and double bonding. Also, we will consider potential parameters $\theta_c$ and $r_c$ such that multiple bonding of an association site cannot occur due to steric hindrance. In TPT*M*, $\Delta c^{(o)}$ is approximated by considering all chain diagrams which contain a single chain of *M* or less association bonds and is given as[37]

$$\Delta c^{(o)} \approx \Delta \hat{c}_1^{(o)} = \sum\nolimits_{n=1}^{M} \Delta c_n \tag{7.6}$$

where $\Delta c_n$ is the $n^{\text{th}}$ order contribution (involves chains of *n* association bonds) and is given by

$$\frac{\Delta c_n}{V} = \sigma_A \sigma_B \rho_o^{n-1} I_n \tag{7.7}$$

The integrals $I_n$ are given by

$$I_n = \frac{1}{\Omega^n} \int f_{AB}(12)\ldots f_{AB}(n, n+1) G_{HS}(1\ldots n+1) d(2)\ldots d(n+1) \tag{7.8}$$

Where $\Omega = 4\pi$ (for our axisymmetric case) and the $f_{AB}(ij)$ are the association Mayer functions. Wertheim defines the functions $G_{HS}(1\ldots n+1)$ as, "the subset of graphs in $g_{HS}(1\ldots n+1)$ such that combining them with the chain produces an irreducible graph; $g_{HS}(1\ldots s)$ denotes the *s* particle correlation function of the *reference* system".[37] This means, for instance, that in a second order perturbation theory the contribution $\Delta c_2$ will include the triplet correlation function $g_{HS}(123)$, but one must subtract off the contribution from the first order term $\Delta c_1$ to keep from double counting. We then obtain the $G_{HS}(1\ldots s)$ by summing $g_{HS}(1\ldots s)$ and all products of $g_{HS}$'s obtained by partitioning 1…s into subsequences which share the switching point and associating a negative 1 with each switching point.[37] A few examples include

$$G_{HS}(12) = g_{HS}(12)$$

$$G_{HS}(123) = g_{HS}(123) - g_{HS}(12) g_{HS}(23) \tag{7.9}$$

$$G_{HS}(1234) = g_{HS}(1234) - g_{HS}(123) g_{HS}(34) - g_{HS}(12) g_{HS}(234) + g_{HS}(12) g_{HS}(23) g_{HS}(34)$$

The general idea of TPT*M* is then to build up chains by adding in higher order contributions and subtracting off lower order contributions. What does this have to do with bond cooperativity?

Now we wish to approximate the associative graph sum for fluids which exhibit bond cooperativity as given in Eqns. (7.1) and (7.2). For this case an associated chain of *s* spheres will have a cluster energy $\varepsilon_s = -\varepsilon^{(1)} - (s-2)\varepsilon^{(2)}$. To develop the new theory we must partition this cluster energy among the various bonds in the cluster. A convenient approach to partition the cluster energy, is to give the first bond in an associated cluster an effective energy $-\varepsilon^{(1)}$, while each remaining bond receives an energy $-\varepsilon^{(2)}$. This effective bond energy distribution is illustrated in Fig. 7.1.

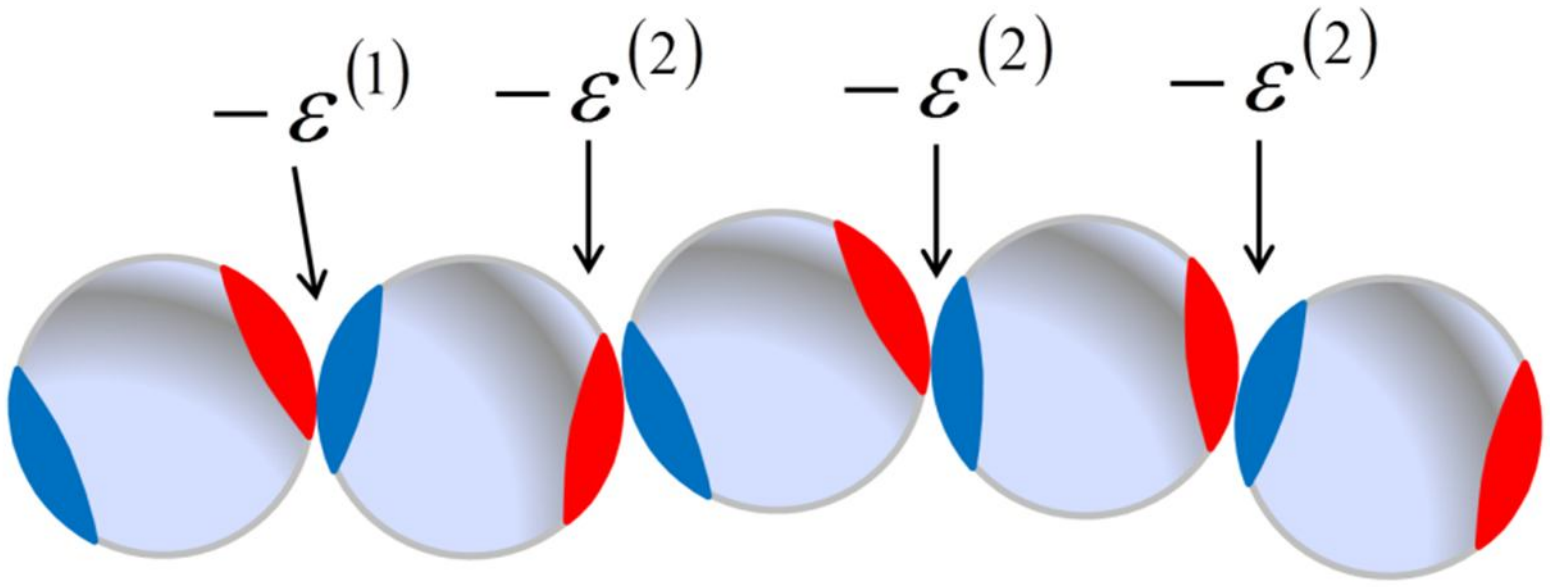

**Figure 7.1:** Effective bond energy distribution

Wertheim's multi - density formalism, and cluster expansions in general, rely on the assumption of pairwise additivity. To include triplet correlations in the form of bond cooperativity we will construct $\Delta c^{(o)}$ in a single chain approximation using generalized Mayer functions. Instead of using TPT*M* to enforce intra - cluster steric constraints, we will employ an analogous perturbation theory to incorporate bond cooperativity. In a chain with the effective bond energy distribution shown in Fig. 7.1, one should expect the product of Mayer functions in a chain of length *s* to be

$$\tilde{f}(1\cdots s) = f_{AB}^{(1)}(12) f_{AB}^{(2)}(23)\cdots f_{AB}^{(2)}(s-1,s) \tag{7.10}$$

Where we have introduced the Mayer functions $f_{AB}^{(j)}(12)$ defined by

$$f_{AB}^{(j)}(12) = \left(\exp\left(\varepsilon^{(j)}/k_B T\right)-1\right)\text{O}_{AB}(12) = f_{AB}^{(j)}\text{O}_{AB}(12) \tag{7.11}$$

Since we are considering the two site case with axial symmetry, steric hindrance between sites is small and the correlation functions $g_{HS}(1\cdots s)$ can be approximated by a linear superposition

$$g_{HS}(1\cdots s)=\prod_{k=1}^{s-1} g_{HS}(k,k+1) \tag{7.12}$$

Combining Eqns. (7.10) and (7.12) one might conclude that the chain integral $I_n$ for this case (similar to Eq. (7.8) but for bond cooperativity) would be given by

$$\frac{1}{\Omega^n}\int \tilde{f}(1\cdots n+1)g_{HS}(12)\cdots g_{HS}(n,n+1)d(2)...d(n+1) \tag{7.13}$$

However, Eq. (7.13) is incorrect due to the fact we did not properly subtract off the lower order contributions to keep from overcounting, as was discussed above in the context of TPT*M*. To account for this we must subtract off all the ways a chain of *l* spheres could be created from chains containing *m* < *l* spheres. Proper accounting results in the following chain integral

$$I_n=\frac{1}{\Omega^n}\int \tilde{F}(1\cdots n+1)g_{HS}(12)\cdots g_{HS}(n,n+1)d(2)...d(n+1) \tag{7.14}$$

Where $\tilde{F}(1\cdots s)$ is defined by the same partition properties of $G_{HS}(1\cdots s)$ with the sole difference being the exchange $g_{HS}(1\cdots s)\rightarrow \tilde{f}(1\cdots s)$. For instance $\tilde{F}(12)=\tilde{f}(12)=f_{AB}^{(1)}O_{AB}(12)$ and

$$\tilde{F}(123)=\tilde{f}(123)-\tilde{f}(12)\tilde{f}(23)=f_{AB}^{(1)}\left(f_{AB}^{(2)}-f_{AB}^{(1)}\right)O_{AB}(12)O_{AB}(23)$$

$$\begin{aligned}\tilde{F}(1234)&=\tilde{f}(1234)-\tilde{f}(123)\tilde{f}(34)-\tilde{f}(12)\tilde{f}(234)+\tilde{f}(12)\tilde{f}(23)\tilde{f}(34)\\&=f_{AB}^{(1)}\left(f_{AB}^{(2)}-f_{AB}^{(1)}\right)^2O_{AB}(12)O_{AB}(23)O_{AB}(34)\end{aligned} \tag{7.15}$$

For the general case $\tilde{F}(1\cdots s)$ we obtain the following relation

$$\tilde{F}(1\cdots s)=f_{AB}^{(1)}\left(f_{AB}^{(2)}-f_{AB}^{(1)}\right)^{s-2}\prod_{k=1}^{s-1}O_{AB}(k,k+1) \tag{7.16}$$

In constructing the chain integrals for the case of bond cooperativity we have essentially made a change of variables as compared to the standard TPT*M*. In its standard form, TPT*M* corrects for multi – body (>2) effects in relation to intra – cluster repulsions by introduction of higher order correlation functions. In the development of Eq. (7.14), we have assumed multi – body effects associated with intra – cluster repulsions are small, and used an analogous TPT$M$ to correct for multi – body effects in the association energies in the form of the association Mayer functions. We have simply switched the roles of intra - cluster attractions and repulsions in reference to higher order corrections to the perturbation theory. The standard TPT$M$ is built on a methodical theoretical footing, where our approach here is intuitive. Further justification of our approach will be given by excellent agreement with simulation predictions.

The form of Eq. (7.16) allows for a complete summation over all chain graphs. Finally, we write the new fundamental graph sum for an $M^{\text{th}}$ order perturbation theory for bond cooperativity and let the order of perturbation become infinitely large $M \to \infty$ to obtain

$$\frac{\Delta c^{(o)}}{V} = \sigma_A \sigma_B \sum_{n=1}^{\infty} \rho_o^{n-1} I_n = \frac{\sigma_A \sigma_B f_{AB}^{(1)} \Delta}{1 - \left( f_{AB}^{(2)} - f_{AB}^{(1)} \right) \rho_o \Delta} \tag{7.17}$$

Where $\Delta = \xi \kappa$ with $\xi$ given by Eq. (4.10) and $\kappa$ by Eq. (2.21). To evaluate the infinite sum in Eq. (7.17) the assumption was made that

$$|\gamma| \leq 1 \quad where \quad \gamma = \left( f_{AB}^{(2)} - f_{AB}^{(1)} \right) \rho_o \Delta \tag{7.18}$$

The validity of this assumption will be discussed in section 7.3. Equation (7.17) is the central result of this chapter and is remarkably simple. We have developed a new perturbation theory where we use perturbations to correct for bond cooperativity. We then allowed the order of perturbation to become infinitely large allowing for a summation over all chain contributions. For the case $\varepsilon^{(2)} = \varepsilon^{(1)}$ the standard first order perturbation theory[37] is recovered. Using Eq. (7.17) we can minimize the free energy Eq. (7.5) with respect to $\sigma_B$ and $\rho_o$ to obtain the mass action equations

$$\frac{\sigma_A}{\rho_o} - 1 = \frac{\sigma_A f_{AB}^{(1)} \Delta}{1 - \left(f_{AB}^{(2)} - f_{AB}^{(1)}\right)\rho_o \Delta} \tag{7.19}$$

and

$$\frac{\rho}{\rho_o} = \left(\frac{\sigma_A}{\rho_o}\right)^2 + \left(f_{AB}^{(2)} - f_{AB}^{(1)}\right) f_{AB}^{(1)} \left(\frac{\sigma_A \Delta}{1 - \left(f_{AB}^{(2)} - f_{AB}^{(1)}\right)\rho_o \Delta}\right)^2 \tag{7.20}$$

where $\sigma_A = \sigma_B$ due to symmetry. Using Eq. (7.19) we can simplify the free energy as

$$\frac{A - A_{HS}}{k_B T V} = \rho \ln \frac{\rho_o}{\rho} - \sigma_A + \rho \tag{7.21}$$

Combining (7.19) and (7.20) we obtain a closed equation for $\rho_o$

$$\frac{\rho}{\rho_o} = 1 + \frac{2\rho_o f_{AB}^{(1)} \Delta}{1 - \rho_o f_{AB}^{(2)} \Delta} + \frac{\left(\rho_o \Delta\right)^2 f_{AB}^{(1)} f_{AB}^{(2)}}{\left(1 - \rho_o f_{AB}^{(2)} \Delta\right)^2} \tag{7.22}$$

Equation (7.22) is similar to the mass action equation obtained by SJ (Eq. (28) of ref[99]), with the only difference being in the theory due to SJ the Mayer functions $f_{AB}^{(j)}$ are replaced by an exponential $f_{AB}^{(j)} \rightarrow e^{\varepsilon^{(j)}/k_B T}$. The advantage of Eq. (7.22) is that the exact non - associating limit $\rho = \rho_o$ is obtained, while in the approach of SJ this limit is obtained approximately.

A very interesting limit of Eq. (7.22) is for the case that the energy of the first bond in a cluster is zero $\varepsilon^{(1)} \rightarrow 0$. For this case we easily find two analytic solutions for the monomer density

$$\rho_o \rightarrow 1 / f_{AB}^{(2)} \Delta \tag{7.23}$$

$$\rho_o \rightarrow \rho \tag{7.24}$$

For the density parameter $\sigma_A$ there is only a single limit

$$\sigma_A \rightarrow \rho_o \tag{7.25}$$

Equation (7.25) shows that the density of spheres bonded once is vanishing in this limit, meaning any chains must be very long since there are few chain ends. Since it is unphysical for the monomer density to be greater than the total density $\rho_o > \rho$, Eq. (7.24) must be the correct solution for the case $\rho f_{AB}^{(2)}\Delta < 1$. For the case $\rho f_{AB}^{(2)}\Delta > 1$, Eq. (7.23) has the lowest free energy meaning it is the correct solution. At $\rho f_{AB}^{(2)}\Delta = 1$ there is a transition point which we solve for as

$$\left(\frac{\varepsilon^{(2)}}{k_B T}\right)_{tr} = \ln\left(\frac{\rho\Delta + 1}{\rho\Delta}\right) \tag{7.26}$$

Below the transition point, $\varepsilon^{(2)} / k_B T < \left(\varepsilon^{(2)} / k_B T\right)_{tr}$, the fluid is composed of only monomers while after the transition the fluid is composed of a mixture of monomers and very long chains. At first it is surprising that the solution Eq. (7.23) would exist. However, it is easy to show the genesis of this solution. For the case of small $\varepsilon^{(1)}$ the function $\tilde{F}(1\cdots s)$ will simplify to $\tilde{F}(1\cdots s) = f_{AB}^{(1)}(12) f_{AB}^{(2)}(23)\cdots f_{AB}^{(2)}(s-1,s)$ where terms which contain products of $f_{AB}^{(1)}$ have vanished. At the point $\varepsilon^{(1)} = 0$ these leading terms in $f_{AB}^{(1)}$ will also vanish with the exception of the case of an infinitely long chain whose contribution to $\Delta c^{(o)}$ is given by

$$\frac{\Delta c_\infty}{V} = \frac{\sigma_A \sigma_B}{\rho_o} \frac{f_{AB}^{(1)}}{f_{AB}^{(2)}} \lim_{n\to\infty}\left(f_{AB}^{(2)}\rho_o\Delta\right)^n \tag{7.27}$$

The limit in Eq. (7.27) is evaluated as

$$\lim_{n\to\infty}\left(f_{AB}^{(2)}\rho_o\Delta\right)^n = \begin{cases} \infty & for \quad f_{AB}^{(2)}\rho_o\Delta > 1 \\ 1 & for \quad f_{AB}^{(2)}\rho_o\Delta = 1 \\ 0 & for \quad f_{AB}^{(2)}\rho_o\Delta < 1 \end{cases} \tag{7.28}$$

Which shows that for $f_{AB}^{(2)}\rho_o\Delta$ being infinitesimally larger than 1, the limit diverges. This diverging limit tames the vanishing $f_{AB}^{(1)}$ resulting in a finite $\Delta c_\infty$ and the solution of the monomer density given by $\rho_o = 1/ f_{AB}^{(2)}\Delta$. As will be shown in section 7.3, the very simple relations given by Eqns. (7.23) – (7.26) are surprisingly accurate.

## 7.2: Simulation methodology

To test the theory we perform new Monte Carlo simulations in the canonical ensemble for molecules which interact with the potential given by Eqns. (7.1) - (7.3). We choose potential parameters $\theta_c = 27^\circ$ and $r_c = 1.1d$ such that each association site is singly bondable and place the sites on opposite poles of the sphere. The simulations are performed using standard methodology[79]. The simulations were allowed to equilibrate for $N \times 10^6$ configurations and averages were taken over another $N \times 10^6$ configurations. A trial configuration was generated by displacing and rotating a sphere. For each simulation we used $N = 864$ associating hard spheres. While in general having triplet contributions to the system energy can significantly increase computation time, for the current potential we simply needed to keep track of the number of spheres bonded twice, which added little computation time as compared to the pairwise additive system.

## 7.3: Numerical results

In this section we compare theoretical and simulation predictions. We begin our discussion with Fig. 7.2 where we compare theoretical and simulation predictions for the fraction of spheres bonded $k$ times $X_k$ and the excess internal energy $E^* = E_{AS} / Nk_BT$. We consider two general cases. In case I we set $\varepsilon^{(1)} = 7k_BT$ and vary $\varepsilon^{(2)}$ and for case II we fix $\varepsilon^{(2)} = 7k_BT$ and vary $\varepsilon^{(1)}$. For each case we use a density of $\rho^* = \rho d^3 = 0.6$. We begin our discussion with case I. For $\varepsilon^{(2)} = 0$, there is no energetic benefit for a sphere to bond twice which results in $X_2 \to 0$. Increasing $\varepsilon^{(2)}$ we see a steady increase in $X_2$ and the fractions $X_1$ and $X_o$ remain nearly constant until $\varepsilon^{(2)} \sim 5k_BT$ at which point they decline sharply. The excess internal energy also remains approximately constant until $\varepsilon^{(2)} \sim 5k_BT$ and then begins to decrease. Theory and simulation are in near perfect agreement.

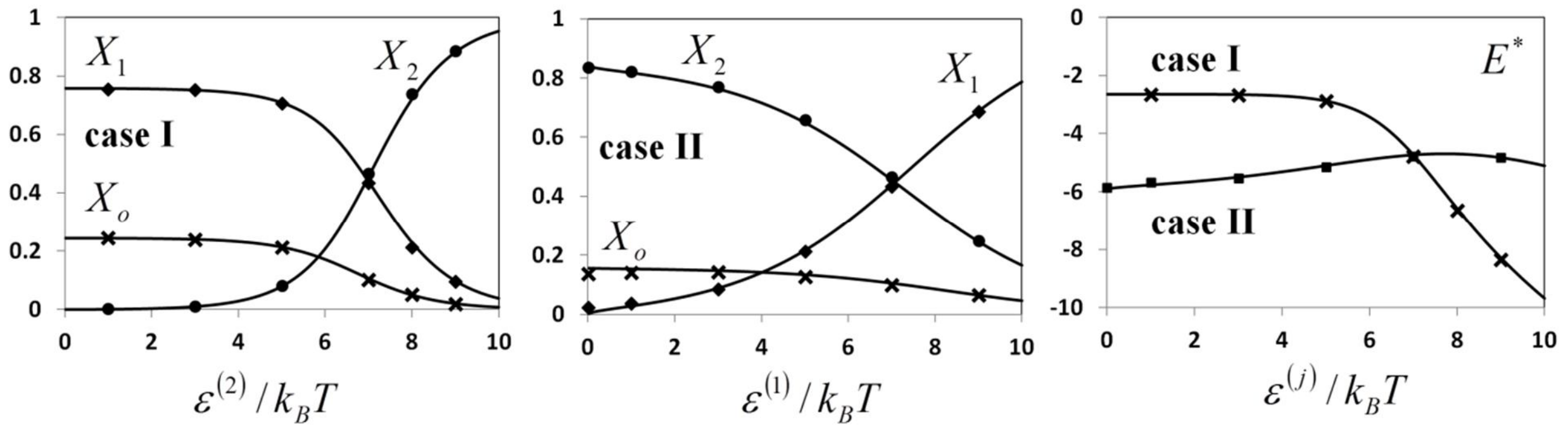

**Figure 7.2:** Comparison of theoretical predictions (curves) and simulation results (symbols) for the fraction of molecules bonded $k$ times $X_k$ for case I – left panel and case II – center panel. The excess internal energy for both cases is given in the right panel. Each case is at a density of $\rho^* = 0.6$. In case I $\varepsilon^{(1)} = 7k_BT$ and $\varepsilon^{(2)}$ is varied while in case II $\varepsilon^{(2)} = 7k_BT$ and $\varepsilon^{(1)}$ is varied. In the right panel $j = 2$ for case I and $j = 1$ for case II

Now considering case II, we set the energy $\varepsilon^{(2)}$ and vary $\varepsilon^{(1)}$. We note the opposite behavior for the fraction $X_2$ as compared to case I. Increasing $\varepsilon^{(1)}$ decreases $X_2$ while increasing $X_1$. This behavior results from the fact that for small $\varepsilon^{(1)}$ the absolute value of the energy per bond is much lower for the first bond in an associated cluster than all remaining bonds in that cluster. For this reason the system minimizes $X_1$. The behavior of $E^*$ is remarkable for this case. The internal energy increases with increasing $\varepsilon^{(1)}$ until $\varepsilon^{(1)} \sim 7.8k_BT$ at which point there is a maximum and $E^*$ begins to decrease. It is counter intuitive that increasing $\varepsilon^{(1)}$ could result in an increase in energy. Again, theory and simulation are in excellent agreement. We also performed calculations for the monomer fraction $X_o$ and excess internal energy $E^*$ using the theory of SJ[99]; these predictions coincided nearly exactly to the calculations performed with the approach presented in this work. No prescription is given by SJ for the calculation of the fractions $X_1$ and $X_2$.

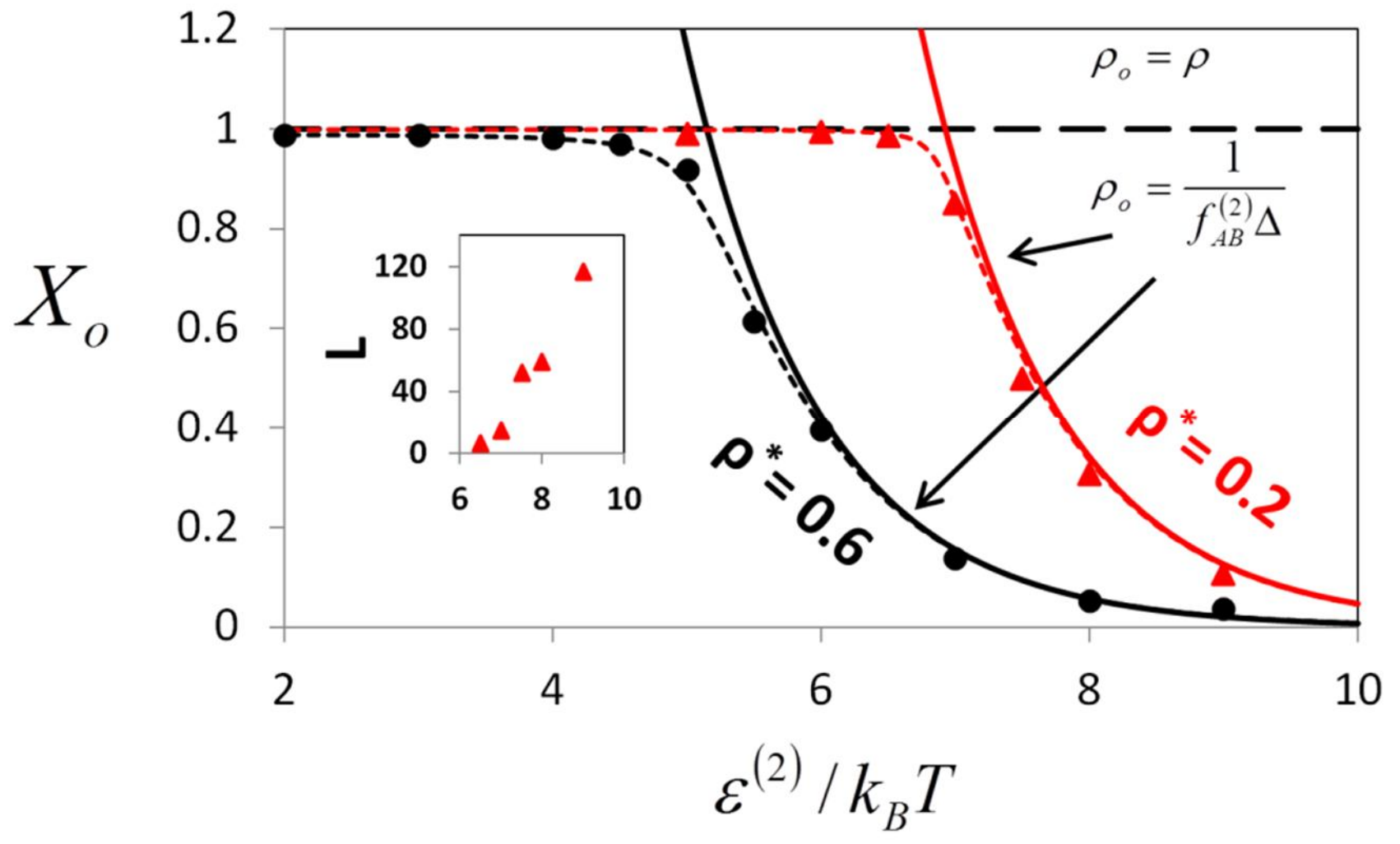

**Figure 7.3:** Monomer fractions versus $\varepsilon^{(2)}/k_BT$ for the case $\varepsilon^{(1)} = 0$. Symbols give simulation results (triangles - $\rho^* = 0.2$ and circles - $\rho^* = 0.6$). Long dashed curve gives the theoretical solutions Eq. (7.24) and solid curves give the solution Eq. (7.23). Short dashed curves give theoretical results from the theory of SJ[99]. The inset gives the simulated values of the average chain length of associated clusters (excluding monomers) versus $\varepsilon^{(2)}/k_BT$ at $\rho^* = 0.2$

Now we focus on the specific case where $\varepsilon^{(1)} = 0$. For this case the monomer fraction is given by Eqns. (7.23) – (7.24). We compare predictions of these simple equations to simulation results for the monomer fractions in Fig. 7.3 at densities $\rho^* = 0.2$ and 0.6. The locations where the solid curves cross the long dashed curve are the transition points predicted by Eq. (7.26). Before the transition the monomer density is given by Eq. (7.24), and after the transition by Eq. (7.23). We also include theoretical predictions from the approach of SJ[99]. As can be seen, both simulation and SJ predict a smooth transition while the transition predicted by the current approach is sharp. This discrepancy is at the heart of the difference between our approach and the approach of SJ. In Wertheim's theory the exact non – associating limit $\rho_o = \rho$ is obtained. That is, if the association energy is zero there is no association. The exception to this rule, is for the case presented in this chapter for bond cooperativity where $\varepsilon^{(1)} = 0$ and the monomer density is given by Eq. (7.23). The theory of SJ does not obey this non – associating limit. In their approach associated chains of spheres can exist even when all association energies are zero due to the fact that association has a purely geometric definition, meaning the exact reference equation of state is not recovered. Molecular simulation also uses this strictly geometric definition of bonding. For these reasons, both the theory of SJ and molecular simulation show a smooth transition while the current approach shows a sharp transition.

Another prediction of the current approach was that after the transition the fluid would be composed of a mixture of monomers and very long chains. The inset of Fig. 7.3 shows simulation results for the average chain length of associated chains (not including monomers) $L$ as a function of $\varepsilon^{(2)} / k_B T$ and a density $\rho^* = 0.2$. As can be seen, after the transition, $L$ grows rapidly as $\varepsilon^{(2)} / k_B T$ is increased. At an energy $\varepsilon^{(2)} = 9 k_B T$ the monomer fraction is found to be $X_o \approx 0.1$ with the fraction bonded once as $X_1 \approx 0.016$ giving an average chain length of associated clusters as $L \approx 117$. This shows that the fluid is primarily composed of monomers and long chains as predicted by the theory.

We conclude this section with a brief discussion of the condition given by Eq. (7.18) that $|\gamma| \leq 1$ for the evaluation of the infinite sum given by Eq. (7.17). For all cases with positive bond cooperativity $\varepsilon^{(2)} \geq \varepsilon^{(1)}$ we find that condition Eq. (7.18) is easily satisfied. However, for the calculated cases with strong negative bond cooperativity $\varepsilon^{(1)} < \varepsilon^{(2)} - k_B T$ , the condition given by Eq. (7.18) can be violated . For instance, in Fig. 7.2, $|\gamma| > 1$ when $\varepsilon^{(2)} < 5.4 k_B T$ for case I and $\varepsilon^{(1)} > 8 k_B T$ for case II. Of course, as can be seen in Fig. 7.2, the theory is in excellent agreement with simulation even when $|\gamma| > 1$. This shows that the final equations derived assuming $|\gamma| \le 1$ are also accurate for $|\gamma| > 1$. Furthermore, in nature bond cooperativity arises from the fact that when a polyfunctional hydrogen bonding molecule forms multiple hydrogen bonds, the polarization of the molecule is increased.[66] This necessarily results in positive bond cooperativity only, for which the condition $|\gamma| \le 1$ always holds.

## 7.4: Summary and conclusions

We have extended Wertheim's theory to account for bond cooperativity in two site associating fluids by the development of a new perturbation theory. This was accomplished in an intuitive manner, guided by the structure of Wertheim's TPT*M*. The new theory was tested against Monte Carlo simulation results and was found to be accurate. For the case that the first bond of an associated cluster had zero energy $\varepsilon^{(1)} = 0$, a simplified form of the theory was obtained predicting that a sharp transition existed between a fluid composed of only monomers and a fluid composed of monomers and very long chains. This transition was also observed by molecular simulation; however, the simulated transition was smooth. The discrepancy between theory and simulation is attributable to the different definitions of association.

The theory of SJ[99] was also found to be accurate. This was the first time this approach had been compared to molecular simulation. An advantage of the approach developed here is that the exact non – associating limit is obtained, while in the approach of SJ this approach is obtained approximately. That said, the two theories given nearly identical predictions when all association energies are finite.

Taking the logic used in this paper further, we can consider the case where the $k^{th}$ bond in a chain receives an energy $-\varepsilon^{(k)}$ and is assigned a Mayer function $f_{AB}^{(k)}$. For this case the functions $\tilde{f}(1\cdots s)$ would be given by

$$\tilde{f}(1\cdots s)=\prod_{k=1}^{s-1} f_{AB}^{(k)}\mathrm{O}(k,k+1) \tag{7.29}$$

This would be a more realistic model for HF[65] than the model considered in this paper, due to the fact that the bonding energy changes as a function of chain length beyond what is given by Eqns. (7.1) and (7.2). Combining Eqns. (7.14) and (7.29) gives a generalized theory for bond cooperativity in chain formation for two site associating fluids.

Lastly, Kalyuzhnyi and Stell[75] developed a cluster expansion similar to Wertheim's, except for the densities are defined in a way which is convenient for molecules which interact with spherically symmetric association interactions (e. g. ions). This approach has also been applied in the development of an equation of state for dipolar hard spheres[100] as well as patchy colloids with patches which bond twice[101]. It should be possible to include bond cooperativity in these theories also.

# CHAPTER 8

# Association in orienting fields

Throughout this dissertation it has been assumed that the fluids were homogeneous in both position and orientation. For bulk fluids, this treatment is sufficient; however, for interfacial / confined fluids or fluids which are in the presence of some external field (electric, magnetic etc…) this treatment may be insufficient. One of the distinct advantages of Wertheim's theory is that it is naturally valid for inhomogeneous systems. This is evidenced by the fact that the Helmholtz free energy Eq. (1.39) is a functional of the inhomogeneous density $\rho(1)$. This was first noted by Chapman,[41] and has since been widely exploited by researchers in the development of accurate density functional theories for associating fluids.[44, 47, 55, 102]

Each of these approaches has considered inhomogeneities in position only (confined fluids, interfaces etc…) and have neglected the possibility of inhomogeneities in orientation. Of course, there are many cases in which associating fluids are placed in an external field which acts on orientation. First and foremost, due to anisotropy of the pair potential, any inhomogeneity in position is also an inhomogeneity in orientation. Other examples include associating dipolar molecules[103], patchy colloids in uniform electric fields and magnetic Janus particles in magnetic fields[104]. With the wide ranging success of Wertheim's theory in describing bulk associating fluids and associating fluids which exhibit spatial inhomogeneities, it seems timely to extend the approach to include the effects of orientating external fields.

In this chapter we develop a new density functional theory for associating fluids in orienting fields. For simplicity, we will assume that the system is homogeneous in position with the only inhomogeneities being in orientation. Using density functional theory (DFT) in the canonical ensemble, we derive an exact orientational distribution function (ODF) for associating fluids in orienting fields. It is then shown that significant simplification results if it is assumed that the reference fluid is isotropic. Finally, we apply the theory at the level of first order perturbation theory (TPT1) to study the effect of a linear orienting field on the association of two site associating spheres.

## 8.1: General theory

In this section the new DFT for associating fluids in orienting external fields will be developed. We consider fluids composed of molecules / colloids whose intermolecular potential is given by Eq. (1.1). We allow for inhomogeneities in orientation, which result from the application of an external field external field $\phi_E(\Omega)$. For this case the exact Helmholtz free energy is obtained from Eq. (1.33)

$$\frac{A}{Vk_BT} = \int \left(\rho(\Omega)\ln\left(\rho_o(\Omega)\Lambda^3\right) + Q(\Omega)\right)d\Omega - \frac{c^{(o)}}{V} \tag{8.1}$$

Here $\rho(\Omega)$ is the density as a function of orientation with the total system density (bulk density) given by

$$\rho = \int \rho(\Omega) d\Omega \tag{8.2}$$

Alternatively, $\rho(\Omega)$ can be written in terms of the ODF given by

$$\tau(\Omega) = \frac{\rho(\Omega)}{\rho} \tag{8.3}$$

Combining Eqns. (8.2) – (8.3) we obtain the normalization condition for the ODF

$$\int \tau(\Omega) d\Omega = 1 \tag{8.4}$$

To obtain the orientational density profile, or equivalently the ODF, we employ classical density functional theory in the canonical ensemble. We construct the appropriate functional by adding the field contribution to the intrinsic free energy and adding a Lagrange multiplier term to enforce a constant number of spheres to obtain

$$\frac{A}{k_B TV} + \int d\Omega \rho(\Omega) \frac{\phi_E(\Omega)}{k_B T} - \lambda \left( \int \rho(\Omega) d\Omega - \rho \right) \tag{8.5}$$

where $\lambda$ is the Lagrange multiplier which we use to enforce the condition (8.2). Note, that when Eq. (8.2) is satisfied, the contribution in Eq. (8.5) which contains the Lagrange multiplier vanishes identically.

To obtain a relation for $\rho(\Omega)$ we now minimize Eq. (8.5) with respect to $\rho(\Omega')$

$$\ln\left(\rho_o(\Omega')\Lambda^3\right) - \frac{c_o(\Omega')}{V} + \frac{\phi_E(\Omega')}{k_B T} - \lambda = 0 \tag{8.6}$$

where we used the fact that $Q(\Omega)$ is independent of the total density.[35] The function $c_o(\Omega')$ in Eq. (8.6) is given by the functional derivative

$$c_o(\Omega') = \frac{\delta c^{(o)}}{\delta \rho(\Omega')} = -\frac{\delta}{\delta \rho(\Omega)}\left(\frac{A_{ref}^{EX}}{k_B T}\right) + \frac{\delta \Delta c^{(o)}}{\delta \rho(\Omega)} \tag{8.7}$$

where $A_{ref}^{EX} = -k_B T c_{ref}$ is the reference system excess Helmholtz free energy functional. Adding and subtracting $\ln\left(\rho(\Omega')\Lambda^3\right)$ in Eq. (8.6) we obtain the relation

$$\rho(\Omega') = \frac{1}{\Lambda^3 X_o(\Omega')} \exp\left(-\frac{\phi_E(\Omega')}{k_B T} + \frac{c_o(\Omega')}{V} + \lambda\right) \tag{8.8}$$

where $X_o(\Omega) = \rho_o(\Omega)/\rho(\Omega)$ is the fraction of molecules with orientation $\Omega$ which are unbonded. Employing Eq. (8.2) to eliminate the Lagrange multiplier $\lambda$ we obtain the ODF

$$\tau(\Omega) = \frac{\rho(\Omega)}{\rho} = \frac{\dfrac{1}{X_o(\Omega)} \exp\left(-\dfrac{\phi_E(\Omega)}{k_B T} + \dfrac{c_o(\Omega)}{V}\right)}{\displaystyle\int \frac{1}{X_o(\Omega')} \exp\left(-\frac{\phi_E(\Omega')}{k_B T} + \frac{c_o(\Omega')}{V}\right) d\Omega'} \tag{8.9}$$

Equation (8.9) is the formally exact ODF for associating fluids which interact with the pair potential given by Eq. (1.1). The orientational average of any quantity $Q$ is then given by $\langle Q \rangle = \int \tau(\Omega) Q(\Omega) d\Omega$. At this point, no restriction has been placed on the reference system potential $\phi_{ref}$. If it is assumed that $\phi_{ref}$ is isotropic $\phi_{ref}(12) = \phi_{ref}(r_{12})$ then the term $c_o(\Omega)$ becomes independent of orientation and Eq. (8.9) simplifies to

$$\tau(\Omega) = \frac{\dfrac{1}{X_o(\Omega)} \exp\left(-\dfrac{\phi_E(\Omega)}{k_B T}\right)}{\displaystyle\int \frac{1}{X_o(\Omega')} \exp\left(-\frac{\phi_E(\Omega')}{k_B T}\right) d\Omega'} \tag{8.10}$$

Equation (8.10) is the exact ODF for associating fluids with an isotropic reference system. The prefactor $1/X_o(\Omega)$ modifies the Boltzmann distribution favoring orientations which maximize association. Both Eqns. (8.9) and (8.10) give formally exact results for the ODF; however, both of these equations rely on the knowledge of the monomer fraction $X_o(\Omega)$ which must be approximated. At the level of first order perturbation theory (TPT1) for molecules with a set of association sites $\Gamma = \{A, B, C, \cdots, Z\}$ the monomer fraction is given by

$$X_o(\Omega) = \prod_{A \in \Gamma} X_A(\Omega) \tag{8.11}$$

where $X_A(\Omega)$ is the fraction of molecules with orientation $\Omega$ which are not bonded at association site *A* and is obtained from Eq. (1.44) as

$$X_A(\Omega_1) = \frac{\sigma_{\Gamma - A}(\Omega_1)}{\rho(\Omega_1)} = \frac{1}{1 + \rho \int \tau(\Omega_2) g_{ref}(\Omega_1, \Omega_2, r_{12}) \sum_{B \in \Gamma} f_{AB}(\Omega_1, \Omega_2, r_{12}) X_B(\Omega_2) d\vec{r}_{12} d\Omega_2} \tag{8.12}$$

If we assume the reference potential is isotropic, the reference system pair correlation function becomes independent of orientation $g_{ref}(\Omega_1, \Omega_2, r_{12}) = g_{ref}(r_{12})$.

## 8.2: Application to the two site *AB* case

In this section we specialize the general results of section 8.1 to the two site *AB* case considered in Fig. 5.1 **a**, where there are only *AB* attractions, and the angle between the centers of association sites is taken to be $180^{\circ}$. We consider an isotropic hard sphere reference system with the association potential $\phi_{AB}$ given by the conical sites discussed in section 2.1. We will restrict our attention to an external potential which has the form of a dipole in an electric field

$$\frac{\phi_E(\Omega)}{k_B T} = \frac{\phi_E(\cos\gamma)}{k_B T} = -\frac{\mu E}{k_B T}\cos\gamma = -C^* \cos\gamma \tag{8.13}$$

where $\gamma$ is the angle between the orientation vector and the field $\vec{E}$ (see Fig. 8.1) and $\mu$ is the dipole moment. Here we define the orientation vector of sphere *j* as the vector which passes through the center of the *A* patch $\vec{r}_A^{(j)}$ (hydrogen association site).

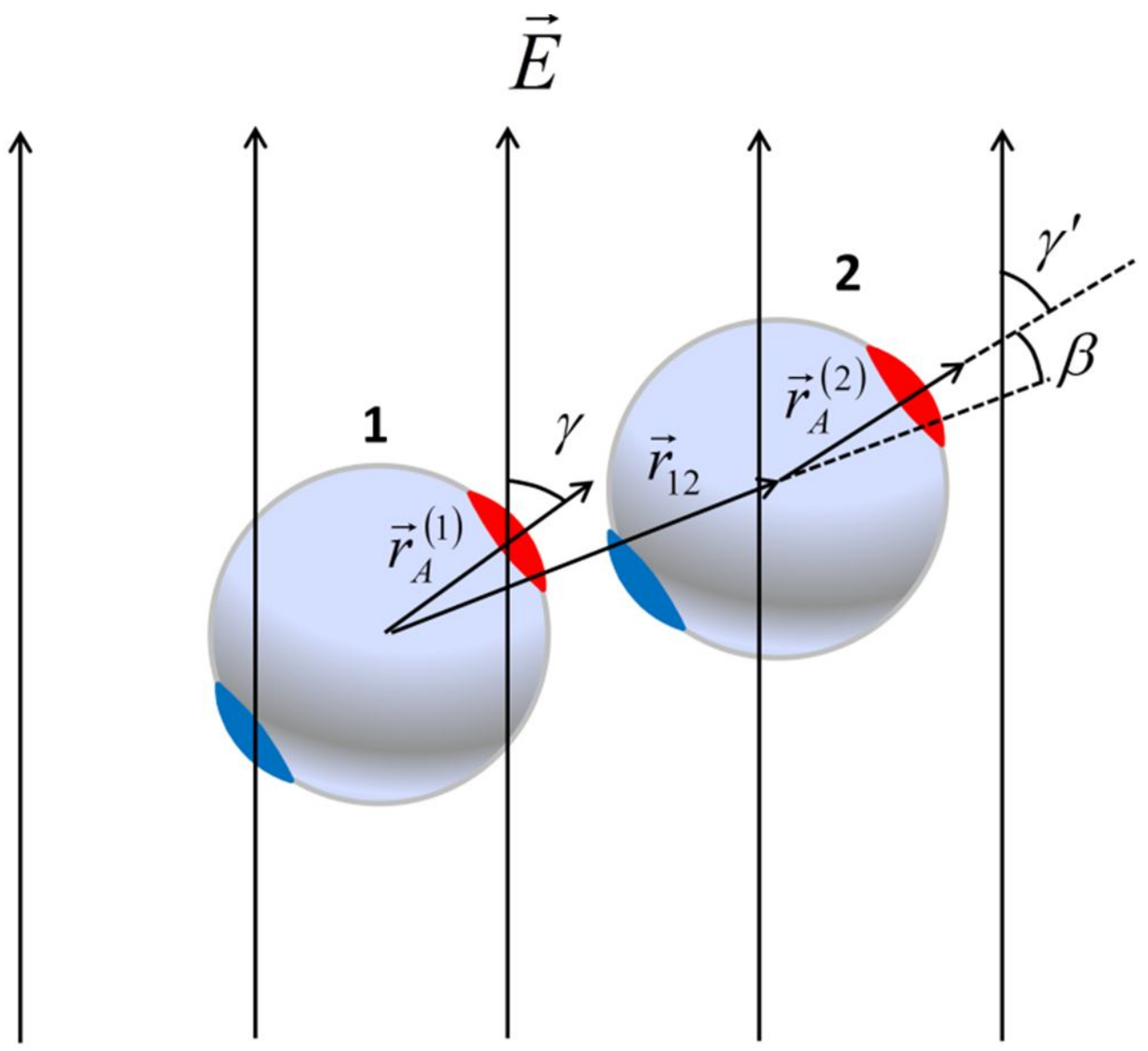

**Figure 8.1:** Diagram of associating spheres in an external field $\vec{E}$ . The orientation of sphere $j$ is defined as the orientation of the center of patch *A*, denoted by the site orientation vector $\vec{r}_A^{(j)}$

To complete the theory we must carry out the integration in Eq. (8.12). To perform this integral in a rigorous manner a numerical 4D integral (two spatial angles and two orientation angles) would need to be performed. A more computationally efficient method, although not exact, is to carry out the orientational integration *I* for sphere 2 in an orientational reference frame centered on the site *A* orientation vector $\vec{r}_A^{(1)}$ of sphere 1 ( $\vec{r}_{12} \, // \, r_A^{(1)}$ in Fig. 8.1), and then assume that the total integral in Eq. (8.12) is given by this integral *I* multiplied by the solid angle of the patch. Following this approach only a double integral must be evaluated, which is much more computationally convenient. Carrying out this process we obtain

$$\frac{1}{\sqrt{X_o(\cos\gamma)}} = 1 + \frac{(1-\cos\theta_c)}{2} \xi \, f_{AB} \rho \int_{\cos\theta_c}^{1} \int_{0}^{2\pi} \tau(\cos\gamma') \sqrt{X_o(\cos\gamma')} \, d\cos\beta \, d\alpha \tag{8.14}$$

Where $f_{AB} = \exp(\varepsilon_{AB} / k_B T) - 1$ and $\xi = 4\pi \int_d^{r_c} g_{HS}(r) r^2 dr$ which is given by Eq. (4.10). *Note, the definition of* $\xi$ *differs from that of the original publication*[105] *by a factor of 4π.* The constraint angle is given by $\cos\gamma' = \cos\gamma\cos\beta - \sin\alpha\sin\beta\sin\gamma$ where $\beta$ and α are the polar and azimuthal angles $\vec{r}_A^{(2)}$ makes in a coordinate system whose *z* axis is parallel to $\vec{r}_{12}$. In Eq. (8.14) we have employed Eq. (8.11) with the result $X_A(\cos\gamma) = X_B(\cos\gamma)$ to eliminate $X_A$ and $X_B$ in favor of $X_o$. For this case the ODF is given by

$$\tau(\cos\gamma) = \frac{\frac{1}{X_o(\cos\gamma)}\exp\left(-\frac{\phi_E(\cos\gamma)}{k_B T}\right)}{2\pi\int_{-1}^{1}\frac{1}{X_o(x)}\exp\left(-\frac{\phi_E(x)}{k_B T}\right)dx} \tag{8.15}$$

Combining Eq. (8.14) and (8.15) a closed integral equation for the monomer fraction is obtained, which is solved using a simple Picard iteration. The approach is computationally efficient, with a typical calculation taking mere seconds on a modern laptop.

## 8.3: Numerical calculations

To test the theory we perform *NVT* Monte Carlo simulations to measure the average orientation of spheres with respect to the field $\langle\cos\gamma\rangle$ and the average chain length of associated chains $\langle L_C\rangle = 1/\langle X_A\rangle$. Simulations were allowed to equilibrate for ~$10^9$ trial moves and averages were taken over an additional ~$10^9$ trial moves. We used $N$ = 864 spheres for each simulation; however, additional simulations were performed using 2048 spheres to verify the absence of any system size effects. In this work we use the potential parameters $r_c = 1.1d$ and $\theta_c = 27^\circ$ such that sites are only singly bondable.

Figure 8.2 compares theoretical predictions and simulations for both $\langle\cos\gamma\rangle$ and $\langle L_C\rangle$ as a function of $\varepsilon^* = \varepsilon_{AB}/k_B T$ for various reduced field strengths $C^* = \mu E/k_B T$ at both low $\rho^* = \rho d^3$ = 0.2 and high $\rho^*$ = 0.6 density cases. The general trend (excluding the zero field case) is that increasing the reduced association energy $\varepsilon^*$ increases $\langle \cos\gamma \rangle$ showing that association increases order in the system. For instance, for $\rho^*$ = 0.2 and $C^*$ = 0.5 increasing $\varepsilon^*$ from 0 to 10 results in a threefold increase in $\langle \cos\gamma \rangle$. We also note that increasing the field strength $C^*$ while holding $\varepsilon^*$ constant results in a significant increase in the average chain length $\langle L_C \rangle$.

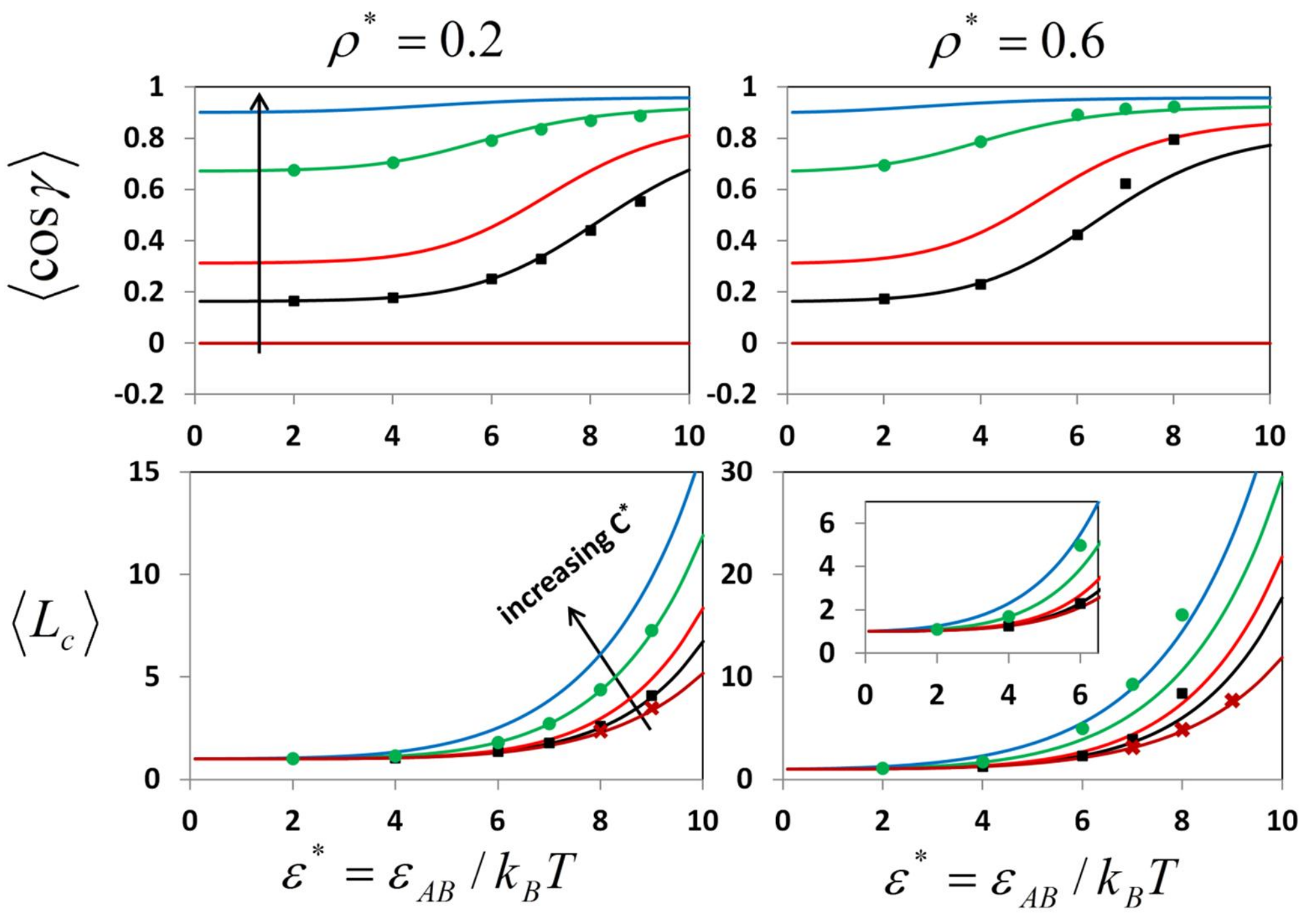

**Figure 8.2:** Average orientation with respect to the field <cos$\gamma$> and average chain length <$L_C$> versus reduced association energy $\varepsilon^*$ for $\rho^*$ = 0.2(left) and $\rho^*$ = 0.6 (right). Curves give theoretical predictions for, from bottom to top, $C^*$ = 0, 0.5, 1, 3 and 10. Symbols give simulation results for $C^*$= 0 (crosses), $C^*$= 0.5 (squares) and $C^*$= 3 (circles). Inset on bottom right panel shows results for <$L_C$> at $\rho^*$ = 0.6 and $\varepsilon^*$ < 6.5

The increase in $\langle L_C \rangle$ with increasing $C^*$ is due to the fact that the field pre-orients the spheres thereby decreasing the penalty paid in decreased orientational entropy when a bond is formed. The relationship between $\langle \cos\gamma \rangle$ and $\varepsilon^*$ seems to be a synergistic effect between the decrease in energy obtained for both bond formation and orientational alignment with the field.

For the low density case $\rho^*$ = 0.2, theory and simulation are in excellent agreement over the full range of $\varepsilon^*$. For the high density case $\rho^*$ = 0.6, theory is in good agreement with simulation for $\varepsilon^* < 6$ and becomes less accurate for larger $\varepsilon^*$. The disagreement between simulation and theory for this case is likely due to the fact that in the simulations the oriented chains create a nematic type phase where long ranged order assists in the formation of chains by restricting associated chains to reptating in a quasi 1D tube formed by neighboring associated chains. The 1D nature of the tube assists in guiding chain ends together to form longer chains. This type of effect cannot be captured in the current theory due to the fact that Eq. (8.14) is obtained in Wertheim's single chain approximation, Eq. (1.41), which neglects interactions between associated clusters beyond that of the reference fluid.[37] This order can be observed in the simulation snapshots given in Fig. 8.3. As can be seen, the field orders and lengthens the associated clusters in the direction of the field.

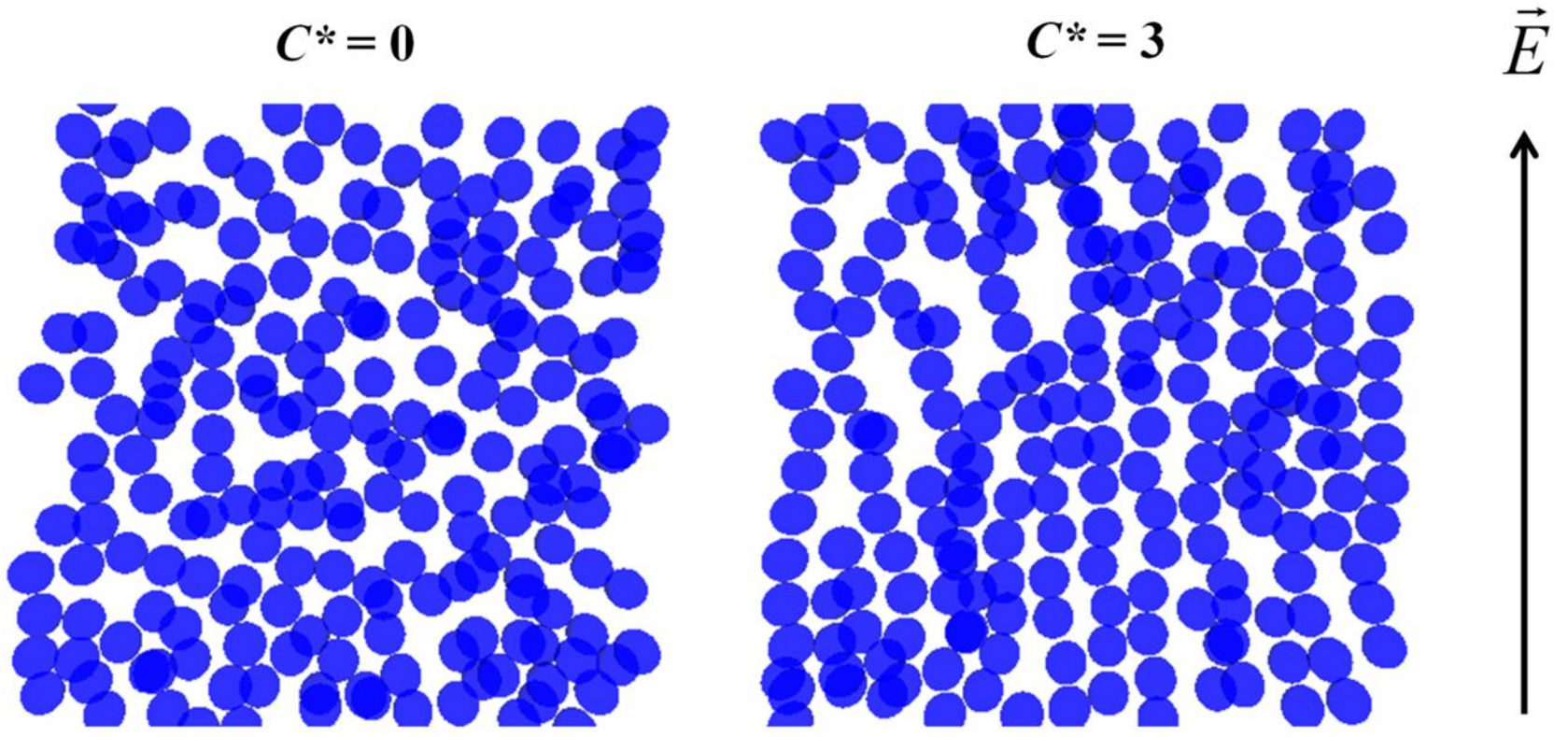

**Figure 8.3:** Simulation snapshots at a density of $\rho^*$ = 0.6 and association energy of $\varepsilon^*$ = 8 with no field $C^*$ = 0 (left) and with field $C^*$ = 3 (right). Snapshots are taken by displaying spheres located in a layer of 1.5*d* at the edge of the simulation cell. Due to periodic boundary conditions and neglect of spheres not in the layer, some clusters are longer than they appear.

Interestingly, for $\rho^* = 0.6$ the theory again becomes accurate for the quantity $\langle \cos\gamma \rangle$ at large $\varepsilon^*$ and $C^*$ = 3. For the case $C^*$ = 0.5 the nematic ordering effect, which is not captured in the theory, assist in orienting the spheres resulting in the underprediction of $\langle \cos\gamma \rangle$; however, for $C^*$ = 3 the orientation of the spheres is completely dominated by the field. It is in this realm the theory again becomes accurate for $\langle \cos\gamma \rangle$ at large $\varepsilon^*$. However, due to the ordering of clusters, the error in $\langle L_C \rangle$ persist.

## 8.4: Summary and conclusions

In this chapter we have applied Wertheim's multi – density formalism to derive the ODF for associating fluids in orienting external fields. For the general case that the reference potential is anisotropic, Eq. (8.9) gives the formally exact ODF. For the special case that the reference system is isotropic, Eq. (8.10) gives the formally exact ODF. To apply the theory, the required monomer fraction was obtained using first order perturbation theory. The broken orientational symmetry, as a result of the external field, allows for the enforcement of bond angle constraints in the theory, which is not possible in TPT1 otherwise.[106] This fact alone hints that the theory should be accurate for the case where there is more than two association sites. Also, we note that the axially symmetric case studied in the chapter is probably the most difficult case to study theoretically due to the fact that, in this case, association induces long range order of the clusters. Possible extensions of this work include incorporation of the results of this chapter with a bulk equation of state for dipolar molecular fluids[107] allowing for the study of the effect of electric fields on phase equilibria. Also, using the multi – density approach of Kalyuzhnyi and Stell[75], Kalyuzhnyi et *al*.[100] developed an equation of state for dipolar hard spheres. Combining their approach with the approach developed in this paper may allow for the development of a density functional theory for dipolar hard spheres in electric fields.

# CHAPTER 9

# Concluding remarks

This dissertation presented numerous extensions of Wertheim's thermodynamic perturbation theory (TPT) to account for various association interactions beyond that treated in first order perturbation theory (TPT1). While simple in form and remarkably successful as an equation of state for hydrogen bonding fluids, as well as patchy colloids, a number of simplifying assumptions (listed in section 1.5) limit the applicability of TPT1.

Chapter 2 extended TPT to account for the possibility of multiple bonds per association site. TPT1 assumes that each association site is singly bondable. While usually accurate for hydrogen bonding molecules, this assumption can fail for patchy colloids with patch sizes that allow for multiple bonds per patch. Here we extended TPT to account for the possibility of two bonds per patch, including contributions to the free energy and for chains of doubly bonded patches as well as triatomic rings of doubly bonded patches. The theory was found to be highly accurate in comparison to simulation data. Our treatment here was limited to the one patch case; however, our approach to ring formation has since been implemented in a theory which accounts for doubly bonded patches in colloids with multiple patches[108]. Also, following a similar approach to the one presented here,[109] we developed a classical density function theory (DFT) for single patch colloids with a doubly bondable patch.[102]

In chapter 3 we extended TPT to mixtures of spherically symmetric colloids (*s* colloids) and single patch colloids (*p* colloids). This model had the key features of the *s* and *p* colloid mixture synthesized by Feng et *al.*[8]. Through the introduction of cluster partition functions, the intrinsic higher order nature of the theory was tamed. The theory was shown to be highly accurate in the prediction of the self - assembly and thermodynamics of this mixture. It was shown that the arm number distribution, fraction of *p* colloids bonded, internal energy and pressure could be manipulated as a function of temperature, density and composition.

In chapter 4 we extended the theory developed in chapter 3 to the case that the *p* colloids can have multiple patches. The theory was then applied to the case of a binary mixture of bi-functional *p* colloids which have an *A* and *B* type patch with only *AB* attractions and *s* colloids which are not attracted to other *s* colloids and are attracted to only the patch *A* on the *p* colloids. Since there are no *AA* or *BB* attractions, and the s colloids are only attracted to the *A* patch of the *p* colloid, no associated cluster can involve more than one *s* colloid. This mixture reversibly self assembles into both colloidal star molecules, where the *s* colloid is the articulation segment and the *p* colloids form the arms, and free chains composed of only *p* colloids. It was shown that temperature, density, composition and relative attractive strengths can all be varied to manipulate the number of arms per colloidal star molecule, average arm length and the fraction of chains (free chains + arms) which are star arms.

An interesting application of the theory developed in chapters 3 – 4 is the following. Mixtures of *p* colloids have been shown to be excellent candidates for network fluids and low density gels.[12] This stems from the limited valence of the *p* colloids. Patchy colloids which contain only two patches can only form linear chains which results in the absence of a liquid / vapor transition. However, if one introduces another species of *p* colloid with more than two patches, these colloids can act as junctions to seed network formation. With this type of approach, very low density gels can be produced. What has not been explored is the case where the mixture consists of *p* colloids and *s* colloids. Like mixtures which consist of two species of *p* colloids, we should expect that mixtures of *s* colloids and two patch *p* colloids produce network forming fluids and low density gels where the *s* colloids act as network junctions (Fig. 9.1).

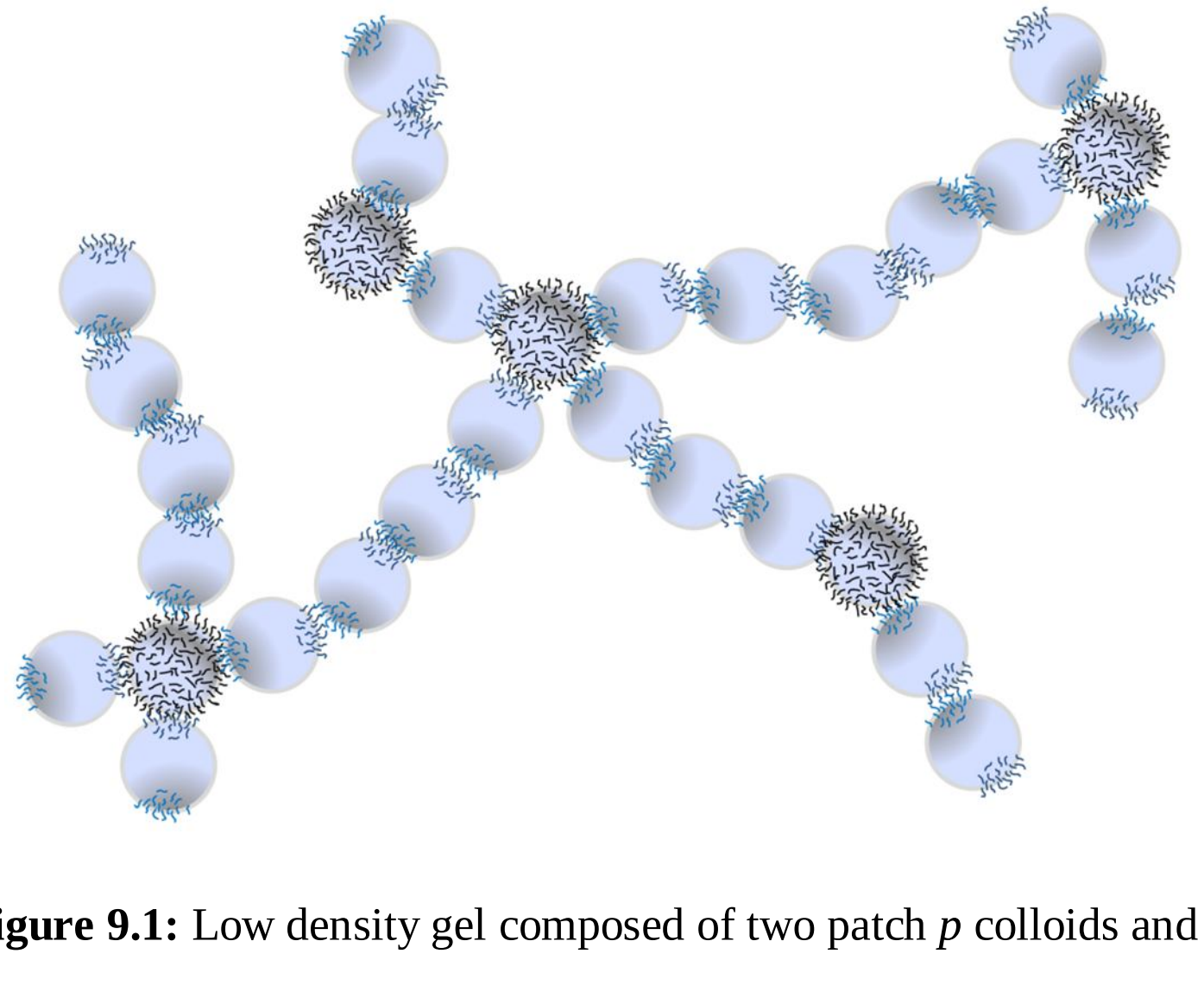

**Figure 9.1:** Low density gel composed of two patch *p* colloids and *s* colloids.

Where now, unlike the case studied in chapter 5, we consider the case that that there are *AA* and *BB* attractions and both patches are attracted to the *s* colloid. For this case, there can be multiple *s* colloids in an associated cluster acting as network junctions in low density gels. Unlike the case where multi - patch (>2) *p* colloids are used as network junctions, *s* colloids do not have a fixed valence. This fact will offer more temperature dependent control over the network fluid properties. Also, as stated by Feng et al.[8], it is possible to produce the two patch colloids as depicted in Fig. 9.1 in relatively large quantities. In addition, synthesis of spherically symmetric colloids is commonplace with large quantities available. For these reasons gel phases composed of these mixtures are practical and realizable. The theory developed in this dissertation could easily be applied to study mixtures of this type.

A significant limitation of TPT1 is the fact that the equation of state is independent of the relative locations of the association sites. For the case of associating spheres with two association sites the angle between the centers of the two association sites (we call this the bond angle) provides sufficient information to describe the site geometry. In chapter 5 we extended TPT to account for bond angle for this two site case. To develop a theory valid over all possible bond angles: chain formation, ring formation, double bonding as well as steric hindrance between association sites all had to be accounted for. The resulting theory was shown to be highly accurate for the prediction of the distribution of associated clusters as well as the equation of state. This was the first time the effect of bond angle had been included in TPT for associating fluids. In chapter 6 we extended these results by allowing for additional association sites, but treating the additional sites in first order. That is, the theory only depends on the bond angle between the *A* and *B* association sites. This simplifying assumption allowed for the logical and relatively simple extension of the results of chapter 5. As a test, the theory was compared to simulation results for the case of an associating sphere with three association sites. The theory was shown to be accurate.

A limitation of the approach presented in chapter 6 is that it can only account for the bond angle between a single pair of association sites with the remaining patches treated independently. Of course, it is easy to imagine situations where small bond angle effects for more than two sites would need to be accounted for.  For instance, considering patchy colloids, one could have four patch colloids where two pairs of patches have small bond angles, with each patch pair being essentially independent of the other. Another case would be three patch colloids where the patches are located in close proximity to each other on the surface of the colloid such that steric hindrance, ring formation, double bonding and triple bonding can occur. Figure 9.2 illustrates these possibilities. A generalized theoretical model which could account for these complex possibilities could be extensively employed by synthetic chemists to design, from the bottom up, patchy colloids which self – assemble into predetermined structures in a temperature dependent and reversible manner.  Extension of TPT to incorporate these complex features is an important avenue for future research.

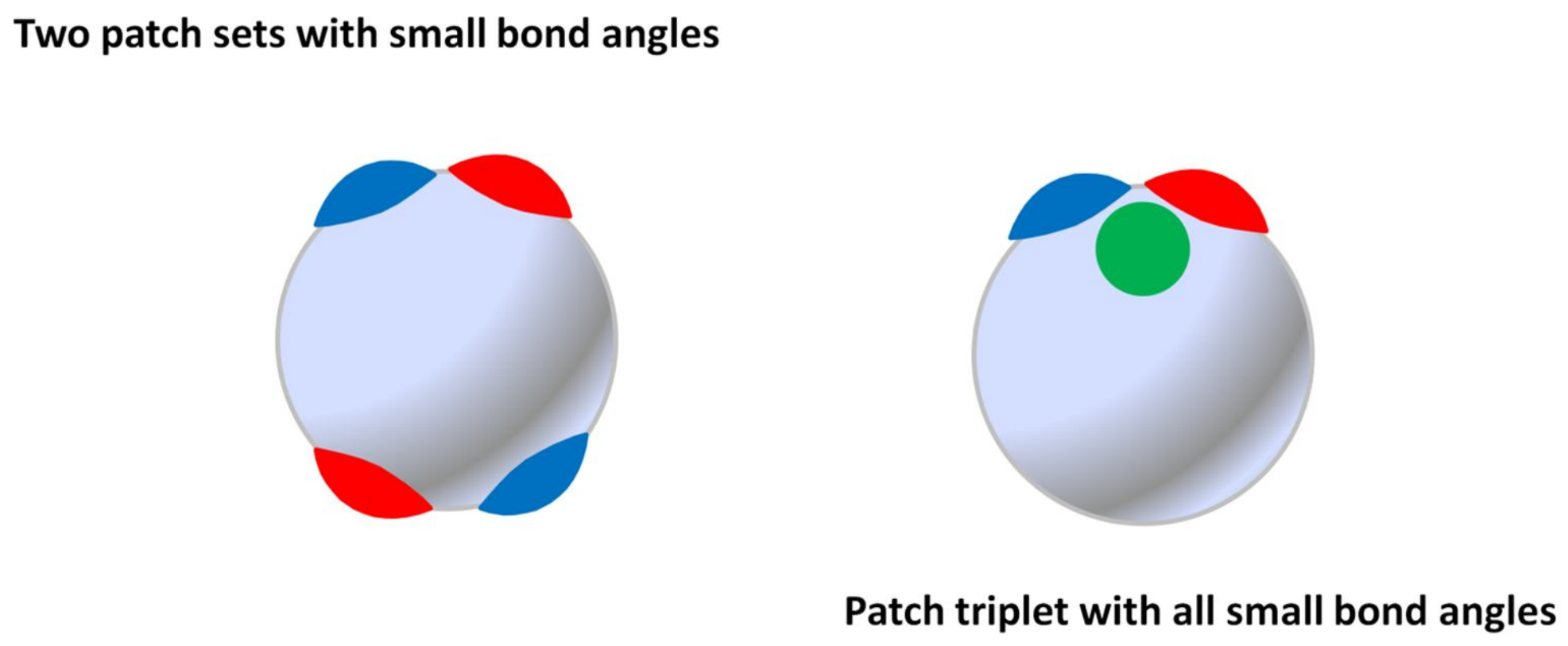

**Figure 9.2:** Sample scenarios of when multiple small bond angle effects would need to be accounted for In chapter 7 we extended TPT to account for bond cooperativity in two site associating fluids. This seems like an improbable extension due to the fact that Wertheim's entire multi – density formalism is constructed around the assumption of pairwise additivity. However, we were able to show that when bond cooperativity is treated as a perturbation at infinite order, the theory seems to naturally incorporate the effect of bond cooperativity. The new theory was tested against new Monte Carlo simulation data and found to be highly accurate. This represents a significant advance due to the fact that many multi – functional hydrogen bonding molecules exhibit some degree of bond cooperativity. In the development of this theory it was assumed that the association sites were located on opposite poles of the sphere (bond angle of $180^{\circ}$) such that the possibility of ring formation and double bonding as well as steric hindrance between association sites did not need to be accounted for. An important extension of this theory is the inclusion of ring formation and steric hindrance as was done in chapter 5 in two site associating fluids which exhibit bond cooperativity. We have recently completed and published this extension.[110]

Finally, in chapter 8 we developed a new classical density functional theory (DFT) in the canonical ensemble for the case of an associating fluid in a spatially uniform orienting external field. Using this DFT, exact results were obtained for the orientational distribution function (ODF). These exact solutions were dependent on the orientation dependent monomer fractions which could not be evaluated exactly and were approximated in TPT1. The theory was tested against Monte Carlo simulation data for the case of hard spheres with two association sites. The theory was shown to be highly accurate at low densities and reasonably accurate at high densities. The theory is relatively simple and computationally cheap which allows for broad applicability. An obvious application of this theory is the study of the effects of electric fields on the phase equilibria of hydrogen bonding multi – polar molecules. Also, the theory could be extended to include translational inhomogeneities, in addition to orientational, to study ordering of hydrogen bonding multi – polar molecules near charged surfaces or in the vicinity of ions.